\documentclass[11pt, onecolumn, preprintnumbers, amsmath, amssymb, aps, prd, nofootinbib, superscriptaddress, showpacs, showkeys]{revtex4-2}

\usepackage[utf8]{inputenc}
\usepackage{graphicx}
\usepackage{dcolumn}
\usepackage{bm}
\usepackage{color}
\usepackage[colorlinks=true, linkcolor=blue, citecolor=blue]{hyperref}
\usepackage{amsfonts}

\begin{document}

\title{Optical Appearance of the Kerr-Bertotti-Robinson Black Hole with a Magnetically Driven Synchrotron Emissivity Model}

\author{Zeng-Yi Zhang}
\author{Xiang-Qian Li}
\email{lixiangqian@tyut.edu.cn}
\author{Hao-Peng Yan}
\email{yanhaopeng@tyut.edu.cn}
\author{Xiao-Jun Yue}

\affiliation{College of Physics and Optoelectronic Engineering, Taiyuan University of Technology, Taiyuan 030024, China}

\begin{abstract}
We investigate the optical appearance of a Kerr-Bertotti-Robinson (Kerr-BR) black hole illuminated by a geometrically and optically thin accretion disk. Instead of using a phenomenological power-law emissivity, we adopt a magnetically driven synchrotron emissivity proxy coupled to the local electromagnetic environment. With a backward ray-tracing framework, we examine the effects of the spin $a$, magnetic parameter $B$, and observer inclination $\theta_O$ on the ray-classification maps, redshift distributions, and specific-intensity images. We show that the ISCO position is modified by both $a$ and $B$, and that rapidly rotating prograde configurations can develop an additional model-dependent inner cutoff when the magnetically dominated approximation underlying the emissivity prescription ceases to be applicable. High-resolution one-dimensional intensity profiles further separate the direct image, the $n=1$ lensing-ring contribution, and the higher-order $n\geq 2$ photon-ring subimages, while quantifying the Doppler-induced brightness asymmetry. Retrograde disks exhibit a wider emission-depleted central region because of the outwardly shifted ISCO, making the higher-order lensed components more clearly distinguishable from the direct emission. These results show that the disk inner boundary and the magnetic-field-dependent emissivity can substantially influence the observable appearance of Kerr-BR black holes.
\end{abstract}

% PACS numbers (Physics and Astronomy Classification Scheme)

% Keywords (optional but recommended)

\maketitle

% 如果你确实希望正文从第二页开始，请取消下面这行代码的注释。
% 否则，建议遵循期刊默认排版，不要加 \newpage。
\newpage

\section{Introduction}
\label{sec:intro}

The Event Horizon Telescope (EHT) images of M87* \cite{EHT_M87_1} and Sgr A* \cite{EHT_SgrA_1} have provided direct observational access to the near-horizon regions of supermassive black holes. These observations reveal a compact central brightness depression surrounded by a bright emission ring, offering a new way to study strong-field gravity and accretion physics. The observed image is determined by two closely related ingredients: the gravitational lensing structure of the background spacetime, which controls the black hole shadow and the hierarchy of higher-order images \cite{Gralla2019,Johnson2020}, and the radiative properties of the emitting plasma, which determine how the lensed photon trajectories are illuminated \cite{Narayan:2019imo}. This has motivated extensive studies of black hole shadows, photon rings, and disk images in both geometric-optics and wave-optics settings \cite{Luminet1979,Falcke2000,Kanai:2013jfc,Gralla:2020pra,Gralla:2020srx,Guo2024,Zeng:2020vsj,Zeng:2021dlj,He:2021htq,Hu:2022rsv,Hu:2023xgy,Zhang:2023hxd,Zeng:2023gcr,Li:2024lpo,Hu:2024qwo,Hu:2025ycf,He:2026dyn}.

In theoretical studies of black hole images, geometrically and optically thin disks with simple radial emissivity profiles, such as $r^{-2}$ or $r^{-3}$, are often used to isolate relativistic lensing effects \cite{Luminet1979,Bambi2019,Guo2024,Hu:2024qwo}. Such prescriptions are useful because they provide clean and controllable models for comparing different spacetimes. However, the millimeter emission relevant to EHT observations is generally associated with synchrotron radiation from relativistic electrons in a magnetized plasma. The local magnetic field, plasma density, electron pressure, and velocity field can therefore affect both the spatial distribution and the observed intensity of the radiation. Recent studies of magnetized accretion disks and thick-disk imaging further indicate that electromagnetic fields, disk geometry, and polarization effects can leave measurable imprints on black hole images and spectra \cite{Moscibrodzka:2009ph,Moscibrodzka:2014wpa,Porth:2019rfi,EHT_M87_5,Gelles:2021wgp,Vincent:2022fwj,Hu:2025rbo,Yang:2025xsy,Yang:2026ybh}. This motivates the use of emissivity prescriptions that retain at least part of the dependence on the local electromagnetic environment, especially when the background solution itself contains a nontrivial magnetic field.

Black holes immersed in external magnetic fields have long been studied in the context of exact Einstein-Maxwell solutions. A well-known example is the Ernst-Wild magnetized Kerr solution, often referred to as the Kerr-Melvin spacetime \cite{Ernst1976}. Although this class of solutions provides an important theoretical framework, Melvin-type rotating solutions may exhibit features such as asymptotic ergoregions and nontrivial geodesic behavior at large distances. Recently, Podolsk\'{y} and Ovcharenko constructed an exact type-D rotating black hole solution immersed in a uniform Bertotti-Robinson (BR) magnetic field \cite{Podolsky2025_KBR}. This Kerr-BR spacetime offers a useful setting in which the influence of a magnetic background on photon propagation and accretion-disk observables can be investigated analytically and numerically.

The Kerr-BR solution and related geometries have motivated several recent studies. These include analyses of the exact solution structure and its generalizations \cite{Ovcharenko:2025qov, Podolsky:2025zlm, Astorino:2025lih, Ovcharenko:2025cpm, Ovcharenko:2026pow, Astorino:2026kuv, Barrientos:2026shy, Ovcharenko:2026byw, Barrientos:2026kdl, Ahmed:2025ril}, hidden symmetries and geodesic integrability \cite{Gray:2025lwy, Wang:2025vsx, Lu:2026kcm}, and thermodynamic or Kerr/CFT-related aspects \cite{Siahaan:2025ngu, Siahaan:2026tuf, Hu:2026slp, Al-Badawi:2026whl}. Astrophysical applications have also been discussed, including particle motion, extreme mass-ratio inspirals, gravitational-wave-related signatures \cite{Li:2025rtf, Zhang:2025ole, Wang:2026czl, Xamidov:2026kqs, Mustafa:2026gly, Xu:2026tgb}, energetic processes such as the magnetic Penrose process, magnetic reconnection, and quasi-periodic oscillations \cite{Mirkhaydarov:2026fyn, Zeng:2025olq, Rehman:2026rzq}, and observational signatures related to lensing, photon rings, and black hole shadows \cite{Vachher:2025jsq, Ali:2025beh, Zeng:2025tji, Wang:2025vsx, Wan:2026lca, Sharipov:2026syz}. In particular, these works show that the magnetic parameter can modify the photon region, shadow size, lensing-ring structure, and potential observational constraints of Kerr-BR black holes \cite{Zeng:2025tji,Wang:2025vsx,Wan:2026lca}.

An important feature of the Kerr-BR spacetime is that the magnetic parameter $B$ affects not only the horizon and ergoregion structure but also the timelike circular orbits. In particular, the location of the innermost stable circular orbit (ISCO) depends on both the spin parameter $a$ and the magnetic parameter $B$ \cite{Wang2025_ISCO}. Since the ISCO is commonly adopted as the inner edge of a geometrically thin accretion disk, its dependence on the spacetime parameters can directly influence the size and morphology of the emitting region. In addition, for an emissivity model based on a magnetically dominated synchrotron prescription, the domain of validity of the magnetic-dominance condition can provide an additional constraint on the inner emission boundary. Thus, the observable image can be sensitive not only to the geodesic structure, but also to the physical prescription used to determine where the disk emission is terminated.

In this work, we study the optical appearance of a Kerr-BR black hole surrounded by a geometrically and optically thin equatorial accretion disk. The disk emission is described by a synchrotron emissivity proxy coupled to the local electromagnetic field. We use a backward ray-tracing scheme to compute photon trajectories from the observer's screen to the disk and examine how the image changes with the black hole spin $a$, the magnetic parameter $B$, and the observer inclination $\theta_O$. Particular attention is paid to the relation between the ISCO, the validity range of the magnetically dominated emissivity prescription, and the resulting inner edge of the emitting region. In rapidly rotating prograde configurations, the adopted prescription can cease to be physically meaningful before the formal ISCO is reached; in our model, this radius is therefore treated as an additional inner cutoff for the synchrotron-emitting region.

To complement the two-dimensional intensity maps, we also extract one-dimensional intensity profiles along the horizontal and vertical directions on the observer's screen. These profiles allow us to separate the broad direct image from the narrower higher-order structures, including the $n=1$ lensing ring and the $n\geq 2$ photon-ring subimages. They also provide a useful diagnostic of the Doppler-induced brightness asymmetry at high inclination. Finally, we compare prograde and retrograde accretion disks. In the retrograde case, the ISCO is shifted outward, leading to a larger central brightness depression and a clearer separation between the direct emission and the higher-order lensed components.

The paper is organized as follows. In Section \ref{sec:spacetime}, we review the Kerr-BR spacetime and the ray-tracing setup. In Section \ref{sec:model}, we introduce the magnetically driven synchrotron emissivity model and the radiative-transfer prescription. Section \ref{sec:results} presents the numerical images and one-dimensional intensity profiles for prograde and retrograde disks. Finally, Section \ref{sec:conclusion} summarizes our main results. We use geometrized units $G=c=M=1$ throughout.

\section{The Kerr-BR Spacetime and Ray-Tracing Geometry}
\label{sec:spacetime}

In this section, we summarize the Kerr-Bertotti-Robinson (Kerr-BR) geometry and the ray-tracing setup used to construct the accretion-disk images. We first introduce the metric functions, then describe the mapping between photon constants of motion and the observer's screen, and finally define the ray-classification scheme and the ISCO-based disk boundary used as the baseline inner edge of the emitting region.

\subsection{The Metric of the Kerr-BR Black Hole}

The Kerr-BR spacetime, derived by Podolsk\'{y} and Ovcharenko \cite{Podolsky2025_KBR}, is an exact type-D solution of the Einstein-Maxwell equations. It describes a rotating black hole with mass parameter $m$ and spin parameter $a$ embedded in a Bertotti-Robinson-type magnetic background characterized by the parameter $B$. In Boyer-Lindquist-like coordinates $(t,r,\theta,\phi)$, the line element can be written as
\begin{equation}
    ds^2 = g_{tt}dt^2 + 2g_{t\phi}dtd\phi
    + g_{\phi\phi}d\phi^2 + g_{rr}dr^2 + g_{\theta\theta}d\theta^2 .
\end{equation}
The non-vanishing covariant components are
\begin{align}
    g_{tt} &= -\frac{Q-a^2P\sin^2\theta}{\rho^2\Omega^2}, \label{eq:gtt} \\
    g_{t\phi} &= \frac{a\sin^2\theta\,[P(r^2+a^2)-Q]}{\rho^2\Omega^2}, \\
    g_{\phi\phi} &= \frac{\sin^2\theta\,[P(r^2+a^2)^2-Qa^2\sin^2\theta]}{\rho^2\Omega^2}, \\
    g_{rr} &= \frac{\rho^2}{\Omega^2 Q}, \\
    g_{\theta\theta} &= \frac{\rho^2}{\Omega^2 P}. \label{eq:gthth}
\end{align}
The metric functions are given by
\begin{align}
    \rho^2 &= r^2+a^2\cos^2\theta, \\
    \Delta &= \left(1-B^2m^2\frac{I_2}{I_1^2}\right)r^2
    -2m\frac{I_2}{I_1}r+a^2, \label{eq:delta} \\
    \Omega^2 &= 1+B^2r^2-B^2\Delta\cos^2\theta, \\
    Q &= (1+B^2r^2)\Delta, \\
    P &= 1+B^2\left(m^2\frac{I_2}{I_1^2}-a^2\right)\cos^2\theta,
\end{align}
where
\begin{equation}
    I_1=1-\frac{1}{2}B^2a^2,
    \qquad
    I_2=1-B^2a^2 .
\end{equation}
In the limit $B\rightarrow 0$, one has $\Omega^2\rightarrow 1$, $P\rightarrow 1$, $Q\rightarrow \Delta$, $I_1\rightarrow 1$, and $I_2\rightarrow 1$, so that the metric reduces smoothly to the standard Kerr solution.

\subsection{Null Geodesics and Observer's Sky}

Photon trajectories are governed by the null geodesic equations in the Kerr-BR spacetime. Because the metric is stationary and axisymmetric, the photon energy and axial angular momentum,
\begin{equation}
    E=-p_t,
    \qquad
    L_z=p_\phi ,
\end{equation}
are conserved along the ray. The Hamilton-Jacobi equation is separable in this geometry, giving rise to an additional Carter-like constant $\mathcal{Q}$. We therefore introduce the dimensionless impact parameters
\begin{equation}
    \xi \equiv \frac{L_z}{E},
    \qquad
    \eta \equiv \frac{\mathcal{Q}}{E^2},
\end{equation}
which are independent of the overall photon energy.

The observer is placed at a finite radius $r_O$ and inclination angle $\theta_O$. We define a local zero-angular-momentum observer (ZAMO) tetrad at the observer's position and use the corresponding local photon momentum to construct the screen coordinates $(\alpha,\beta)$. To avoid coordinate singularities near the rotation axis, we use the reduced metric components
\begin{equation}
    \tilde{g}_{t\phi}=\frac{g_{t\phi}}{\sin^2\theta},
    \qquad
    \tilde{g}_{\phi\phi}=\frac{g_{\phi\phi}}{\sin^2\theta}.
\end{equation}
For each point on the observer's screen, the constants of motion are obtained from the tetrad projection as
\begin{align}
    \xi &= \frac{p_\phi}{E}, \\
    \eta &= P\left(\frac{\sqrt{g_{\theta\theta}}\,p^{(\theta)}}{E}\right)^2
    +\frac{(\xi/\sin\theta_O-a\sin\theta_O)^2}{P}
    -(a-\xi)^2 .
\end{align}
Here $p^{(\theta)}$ denotes the locally measured tetrad component of the photon momentum in the polar direction. Once $(\xi,\eta)$ are specified for a given screen pixel, the corresponding null geodesic is integrated backward from the observer toward the black hole and the accretion disk. In the numerical calculation, we use a fourth-order Runge-Kutta scheme with adaptive step-size control to resolve the rapid variation of the geodesics in the strong-field region.

\subsection{Ray Classification and ISCO}

We classify photon trajectories according to the number of intersections with the equatorial plane, where the geometrically thin disk is located. This classification follows the spirit of the topological approach used in studies of black hole photon rings \cite{Gralla2019}, while here the index $n$ labels the order of the corresponding subimage. Specifically, we use the following convention:
\begin{itemize}
    \item \textbf{Direct image ($n=0$):} rays that intersect the equatorial plane once before reaching the observer. These rays usually form the broad primary image of the disk.

    \item \textbf{Lensing ring ($n=1$):} rays that intersect the equatorial plane twice. They undergo stronger gravitational deflection and generate a narrower, demagnified lensed image of the disk.

    \item \textbf{Higher-order photon-ring subimages ($n\geq 2$):} rays that intersect the equatorial plane three or more times. These rays pass close to the critical photon region and form increasingly narrow higher-order subimages.
\end{itemize}

The inner boundary of the emitting region is another important ingredient in the image construction. As a baseline prescription for a geometrically thin disk, we take the innermost stable circular orbit (ISCO) as the inner edge of the disk. In the Kerr-BR spacetime, the ISCO position depends on both the spin parameter $a$ and the magnetic parameter $B$. The roots of the modified horizon equation $\Delta=0$ are \cite{Podolsky2025_KBR, Wang2025_ISCO}
\begin{equation}
    r_{\pm}
    =
    \frac{m I_2 \pm \sqrt{m^2 I_2-a^2I_1^2}}
    {I_1^2-B^2m^2I_2}\,I_1 .
\end{equation}
Introducing
\begin{equation}
    \lambda=\frac{2\sqrt{r_+r_-}}{r_++r_-},
\end{equation}
the ISCO radius can be expressed as \cite{Wang2025_ISCO}
\begin{equation}
    r_{\rm ISCO}
    =
    \frac{r_++r_-}{2}
    \left[
    3+Z_2
    \mp
    \sqrt{(3-Z_1)(Z_1+2Z_2+3)}
    \right],
\end{equation}
where
\begin{align}
    Z_1 &= 1+(1-\lambda^2)^{1/3}
    \left[(1+\lambda)^{1/3}+(1-\lambda)^{1/3}\right], \\
    Z_2 &= \sqrt{3\lambda^2+Z_1^2}.
\end{align}
The upper sign corresponds to prograde orbits, while the lower sign corresponds to retrograde orbits.

In the ray-tracing calculation, an equatorial crossing contributes to the observed intensity only when it lies within the emitting disk. For the purely ISCO-truncated disk, this condition is $r\geq r_{\rm ISCO}$. In the magnetically driven synchrotron model introduced in the next section, an additional validity condition associated with magnetic dominance can further restrict the inner emitting region. The ray classification described above then allows us to separate the direct image, the lensing-ring contribution, and the higher-order photon-ring subimages in the resulting intensity maps.

\section{Magnetically Driven Synchrotron Emission Model}
\label{sec:model}

In this section, we introduce the emission prescription used to illuminate the Kerr-BR spacetime. Instead of using a purely phenomenological radial power-law emissivity, we adopt a synchrotron emissivity proxy motivated by GRMHD studies of radiatively inefficient accretion flows (RIAFs) \cite{Chael2019, Chael2021}. The purpose of this prescription is not to replace a full GRMHD radiative-transfer calculation, but to incorporate, in a controlled analytic way, the dependence of the emission on the local electromagnetic environment.

\subsection{Local Magnetic Field and Emissivity Profile}
\label{subsec:emissivity}

In low-luminosity accretion systems such as M87*, the millimeter emission is commonly associated with synchrotron radiation from relativistic electrons in a magnetized plasma. For a thermal Maxwell-J\"{u}ttner electron distribution, a useful time-averaged emissivity proxy can be written as \cite{Chael2021}
\begin{equation}
    j_{\rm sim} \propto
    \frac{\rho^3}{p_e^2}
    \exp \left[
    -C
    \left(
    \frac{\rho^2}{|B_{\rm local}|p_e^2}
    \right)^{1/3}
    \right],
    \label{eq:jsim}
\end{equation}
where $\rho$ is the proper plasma density, $p_e$ is the electron pressure, $C$ is a dimensionless fitting parameter, and $|B_{\rm local}|$ denotes the magnetic-field strength measured in the local plasma frame. For 230 GHz emission, a typical value is $C\simeq 0.2$ \cite{Chael2021}.

It is important to distinguish the metric parameter $B$ from the local magnetic field entering the synchrotron emissivity. The former characterizes the strength of the Bertotti-Robinson magnetic background in the Kerr-BR solution, whereas the latter is the magnetic-field strength relevant to the local plasma emission. In a general frame, the electric and magnetic fields are related to the Faraday invariant through
\begin{equation}
    |B_{\rm local}|^2
    =
    \frac{1}{2}F_{\mu\nu}F^{\mu\nu}
    +
    |E_{\rm local}|^2 ,
    \label{eq:B_local_def}
\end{equation}
where $|E_{\rm local}|$ is the electric-field strength measured in the same local frame.

Motivated by magnetically dominated accretion flows, we approximate the local magnetic energy by the electromagnetic invariant in the region where
\begin{equation}
    |B_{\rm local}|^2 \gg |E_{\rm local}|^2 .
\end{equation}
Under this assumption,
\begin{equation}
    |B_{\rm local}|^2
    \simeq
    \frac{1}{2}F_{\mu\nu}F^{\mu\nu}.
    \label{eq:mag_dominated_proxy}
\end{equation}
This relation should be understood as a model assumption for the emissivity prescription rather than as a full plasma solution.

Using the Newman-Penrose form of the electromagnetic field for the Kerr-BR solution \cite{Podolsky2025_KBR}, and restricting the emission to the equatorial plane $\theta=\pi/2$, this approximation gives
\begin{equation}
    |B_{\rm local}|
    \simeq
    B
    \sqrt{
    I_2
    \left(
    1-\frac{B^2m^2}{I_1^2}
    \right)
    -
    \frac{2m}{r}
    \frac{I_2}{I_1}
    }
    \equiv
    B\,\mathcal{M}(r).
    \label{eq:B_local_analytical}
\end{equation}
Here $\mathcal{M}(r)$ is a dimensionless factor that encodes the radial modification of the effective local magnetic field in the equatorial plane. The above expression is used only where the quantity inside the square root is non-negative.

To obtain an analytically tractable radial emissivity, we parameterize the thermodynamic variables by power-law RIAF-like profiles,
\begin{equation}
    \rho(r)=\rho_0 r^{-p},
    \qquad
    p_e(r)=P_0 r^{-q}.
\end{equation}
For the fiducial model, we take $p=1$ and $q=2$, corresponding to a wind-modified density profile and a virial-type pressure scaling. Defining
\begin{equation}
    j_0\equiv \frac{\rho_0^3}{P_0^2},
    \qquad
    \alpha\equiv C\left(\frac{\rho_0^2}{P_0^2}\right)^{1/3},
\end{equation}
and substituting Eq.~(\ref{eq:B_local_analytical}) into Eq.~(\ref{eq:jsim}), we obtain the radial emissivity model
\begin{equation}
    \mathcal{E}_\nu(r)
    =
    j_0\,r
    \exp\left[
    -\alpha
    \left(
    \frac{r^2}{B\,\mathcal{M}(r)}
    \right)^{1/3}
    \right].
    \label{eq:final_emissivity}
\end{equation}
In the numerical images, $j_0$ fixes only the overall intensity normalization, whereas $\alpha$ controls the steepness of the exponential suppression.

Equation~(\ref{eq:final_emissivity}) provides a simple way to incorporate the influence of the magnetic background into the disk emissivity. Its domain of validity, however, is limited by the requirement that the magnetic-dominance approximation in Eq.~(\ref{eq:mag_dominated_proxy}) remains meaningful. We therefore define the magnetic-dominance boundary $r_{\rm md}$ by
\begin{equation}
    \mathcal{M}^2(r_{\rm md})=0,
    \qquad
    \mathcal{M}^2(r)>0
    \quad
    \text{for}
    \quad
    r>r_{\rm md}.
    \label{eq:rmd_def}
\end{equation}
For radii where $\mathcal{M}^2(r)<0$, the invariant satisfies $F_{\mu\nu}F^{\mu\nu}<0$, indicating that the field is no longer magnetically dominated in the sense assumed above. In this region, the synchrotron emissivity proxy in Eq.~(\ref{eq:final_emissivity}) cannot be consistently continued as a real-valued magnetic-field-dependent model.

The effective inner edge of the emitting disk is then taken to be
\begin{equation}
    r_{\rm in}
    =
    \max\left(r_{\rm ISCO},r_{\rm md}\right).
    \label{eq:rin_def}
\end{equation}
Thus, for configurations in which $r_{\rm md}<r_{\rm ISCO}$, the disk is truncated at the ISCO, as in the standard thin-disk prescription. For rapidly rotating prograde configurations, however, the formal ISCO can move inside the region where the adopted magnetically dominated emissivity prescription is no longer applicable. In that case, $r_{\rm md}$ provides an additional model-dependent inner cutoff. This cutoff should not be interpreted as a proof that all physical radiation vanishes there; rather, it marks the inner limit of validity of the synchrotron emissivity model used in this work.

\begin{figure*}[htbp]
\centering
\begin{tabular}{cc}
\includegraphics[width=0.48\textwidth]{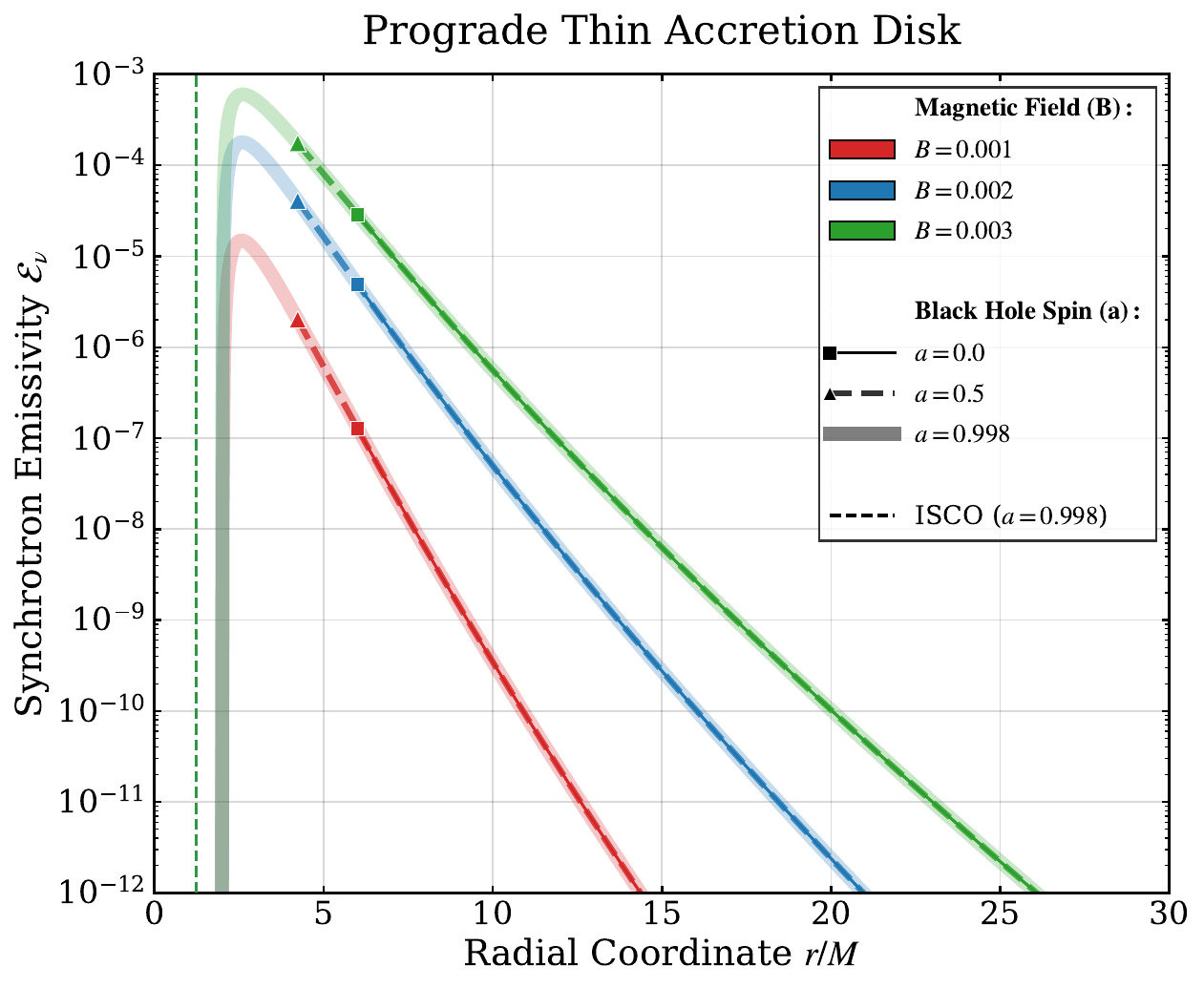} &
\includegraphics[width=0.48\textwidth]{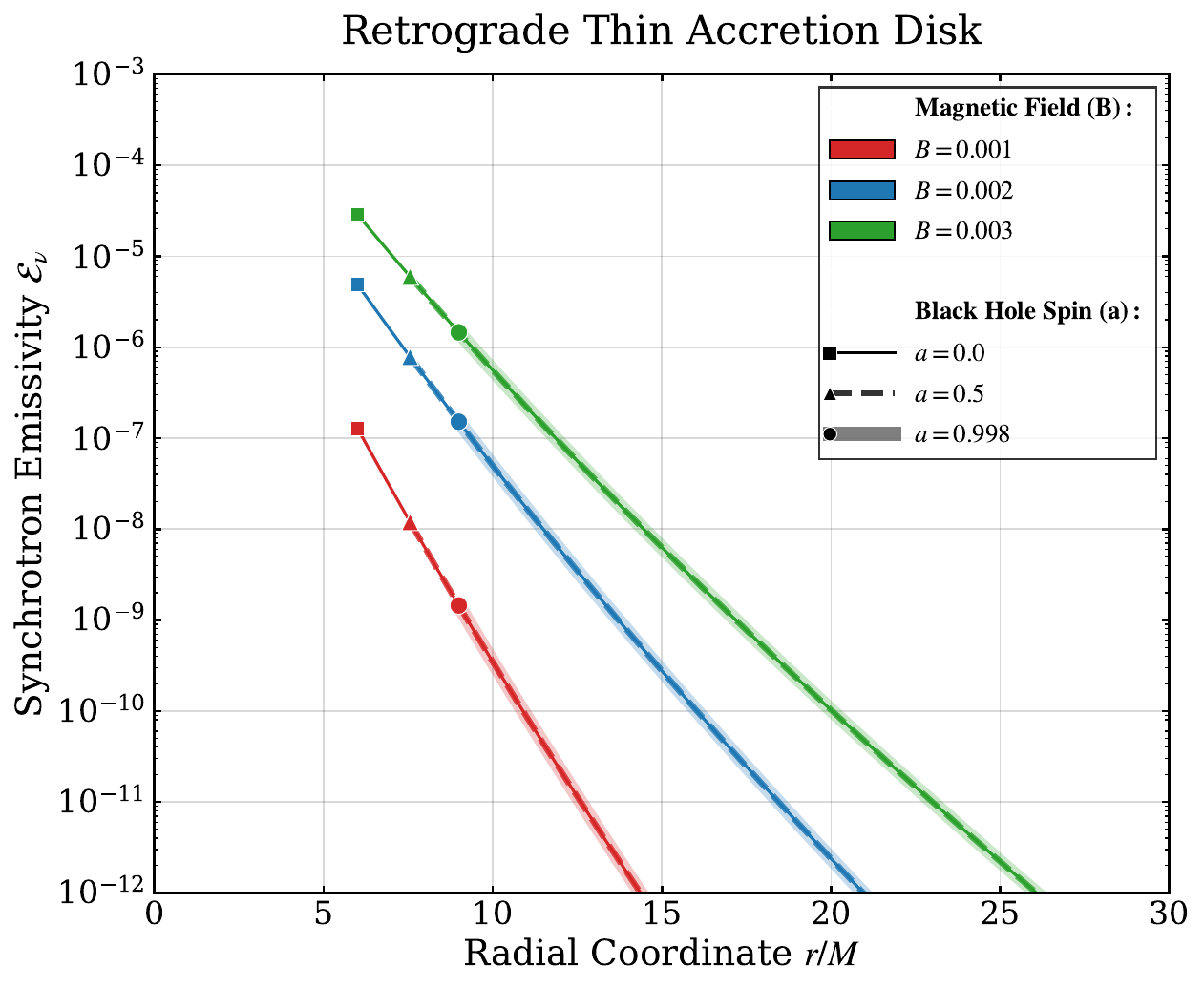} \\
\end{tabular}
\caption{
Radial profiles of the synchrotron emissivity $\mathcal{E}_\nu(r)$ for the Kerr-BR black hole with different values of the spin parameter $a$ and magnetic parameter $B$. The colors denote different magnetic parameters, while the line styles and markers denote different spin values.
\textbf{Left:} Prograde thin accretion disk. Increasing the spin moves the ISCO inward, allowing the emitting region to extend closer to the black hole. For the near-extremal case $a=0.998$, the magnetically dominated condition underlying the emissivity prescription can break down before the formal ISCO is reached. The corresponding inner edge is therefore set by the validity boundary of the adopted emissivity model, producing a sharp model-dependent cutoff in the plotted profile.
\textbf{Right:} Retrograde thin accretion disk. The ISCO is shifted to larger radii, so the emitting region is truncated by the orbital inner edge rather than by the magnetic-dominance boundary. In this case, the emissivity profiles remain regular over the displayed disk region.
}
\label{fig:emissivity}
\end{figure*}

Figure~\ref{fig:emissivity} illustrates the resulting radial emissivity profiles. For prograde disks, increasing the spin generally moves the ISCO inward. In the extreme-spin case, the additional cutoff associated with $r_{\rm md}$ can become relevant before the formal ISCO is reached. For retrograde disks, the ISCO is shifted to larger radii, so the inner boundary is typically set by orbital stability rather than by the magnetic-dominance condition.

\subsection{Plasma Kinematics and Relativistic Redshift}
\label{subsec:kinematics}

The observed photon frequency is affected by both gravitational redshift and the Doppler shift associated with the orbital motion of the emitting plasma. We assume that the radiating gas follows equatorial circular geodesic motion in the emitting region, $r\geq r_{\rm in}$. The plasma 4-velocity is written as
\begin{equation}
    u^\mu_{\rm em}
    =
    u^t(1,0,0,\Omega),
\end{equation}
where $\Omega=d\phi/dt$ is the angular velocity of the circular orbit.

For a stationary and axisymmetric metric, the angular velocity of equatorial circular geodesics is
\begin{equation}
    \Omega_\pm
    =
    \frac{
    -g_{t\phi,r}
    \pm
    \sqrt{
    (g_{t\phi,r})^2
    -
    g_{tt,r}g_{\phi\phi,r}
    }
    }
    {g_{\phi\phi,r}},
    \label{eq:Omega_pm}
\end{equation}
where the comma denotes differentiation with respect to $r$. The plus and minus signs correspond to the two possible orbital orientations. In the following, we use the branch that is consistent with the chosen prograde or retrograde disk. The temporal component is fixed by the normalization condition $g_{\mu\nu}u^\mu u^\nu=-1$,
\begin{equation}
    u^t
    =
    \frac{1}
    {
    \sqrt{
    -g_{tt}
    -2\Omega g_{t\phi}
    -\Omega^2 g_{\phi\phi}
    }
    } .
    \label{eq:ut}
\end{equation}

For an observer at the screen with 4-velocity $u^\mu_{\rm O}$, the frequency shift is
\begin{equation}
    g
    \equiv
    \frac{\nu_{\rm obs}}{\nu_{\rm em}}
    =
    \frac{-p_\mu u^\mu_{\rm O}}
    {-p_\mu u^\mu_{\rm em}} .
\end{equation}
For the static observer used in our ray-tracing setup, this reduces to
\begin{equation}
    g
    =
    \frac{1}{u^t(1-\Omega\xi)},
    \label{eq:g_factor}
\end{equation}
where $\xi=L_z/E$ is the impact parameter of the photon. This factor contains both the gravitational redshift associated with the Kerr-BR metric and the Doppler shift due to the orbital motion of the emitting plasma.

\subsection{Radiative Transfer and Intensity Integration}
\label{subsec:radiative_transfer}

We now describe how the local disk emission is converted into the observed intensity on the screen. The accretion disk is assumed to be geometrically thin and optically thin, so that absorption and multiple scattering are neglected. Liouville's theorem implies that $I_\nu/\nu^3$ is conserved along a photon trajectory. Therefore, the contribution from an equatorial crossing at radius $r^{(n)}$ is weighted by $g_{(n)}^3$.

For a given screen position $(\alpha,\beta)$, a backward-integrated photon trajectory may intersect the equatorial plane multiple times. The observed specific intensity is then approximated by the discrete sum
\begin{equation}
    I_{\nu_{\rm obs}}(\alpha,\beta)
    =
    \sum_n
    g_{(n)}^3
    \mathcal{E}_{\nu_{\rm em}}\left(r^{(n)}\right)
    \Theta\!\left(r^{(n)}-r_{\rm in}\right),
    \label{eq:intensity_sum}
\end{equation}
where $g_{(n)}$ and $r^{(n)}$ are the redshift factor and radial coordinate at the $n$-th equatorial crossing, respectively. The step function enforces the adopted inner boundary of the emitting disk. In practice, the sum includes the direct image ($n=0$), the lensing-ring contribution ($n=1$), and the leading higher-order photon-ring subimages ($n\geq 2$). This prescription allows us to connect the ray-classification maps directly with the final intensity images.

\section{Numerical Results and Image Analysis}
\label{sec:results}

We now present the ray-tracing results for a geometrically and optically thin accretion disk around the Kerr-BR black hole. The analysis is organized into three parts. In Section \ref{subsec:prograde}, we examine the two-dimensional (2D) appearance of prograde disks and discuss how the magnetic parameter $B$, the spin parameter $a$, and the observer inclination $\theta_O$ affect the ray-classification maps, redshift distributions, and specific-intensity images. In Section \ref{subsec:1D_profiles}, we extract one-dimensional (1D) intensity profiles along the horizontal and vertical directions on the observer's screen in order to quantify the higher-order subimage structures and the Doppler-induced brightness asymmetry. Finally, in Section \ref{subsec:retrograde}, we compare these results with retrograde disks, for which the outward displacement of the ISCO leads to a substantially wider emission-depleted central region.

\subsection{Optical Appearance of the Prograde Disk}
\label{subsec:prograde}

The observed appearance of an accretion disk is controlled by both the lensing properties of the background spacetime and the spatial distribution of the emitting plasma. For a prograde thin disk, the ISCO provides the standard baseline inner edge. As discussed in Section \ref{sec:model}, however, in rapidly rotating configurations the formal ISCO may lie inside the region where the magnetically dominated synchrotron emissivity prescription ceases to be applicable. In such cases, the effective inner edge of the emitting region is determined by Eq.~(\ref{eq:rin_def}), rather than by the ISCO alone. This model-dependent cutoff can modify the central brightness depression and the relative visibility of the higher-order lensed components.

\subsubsection{Effect of the Magnetic Parameter $B$}

We first study the dependence of the optical appearance on the magnetic parameter $B$. The observer is placed at $r_O=1000M$, and the field of view is fixed to $3^\circ$. The emissivity parameter is set to $\alpha=0.5$. To isolate the role of $B$, we fix the spin at $a=0.998$ and consider three representative values,
\begin{equation}
    B\in\{0.001,0.002,0.003\},
\end{equation}
for four observer inclinations,
\begin{equation}
    \theta_O\in\{0^\circ,17^\circ,53^\circ,80^\circ\}.
\end{equation}

\begin{figure*}[htbp]
    \centering
    \begin{tabular}{cccc}
        \includegraphics[width=0.23\textwidth]{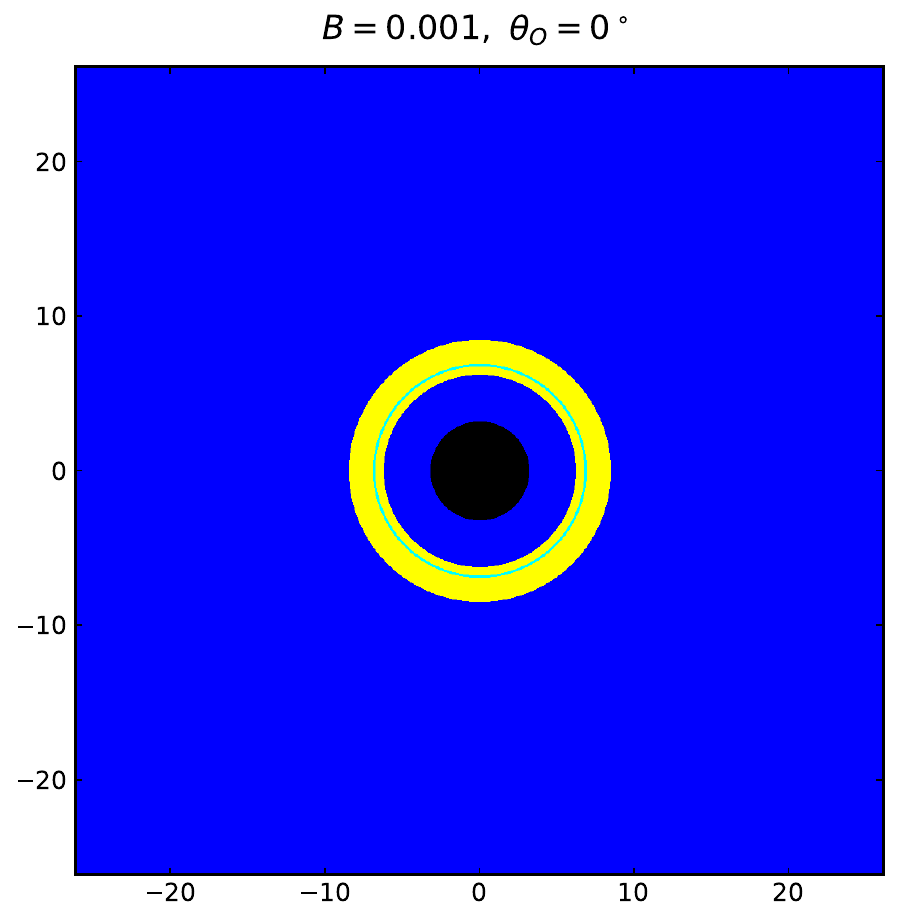} &
        \includegraphics[width=0.23\textwidth]{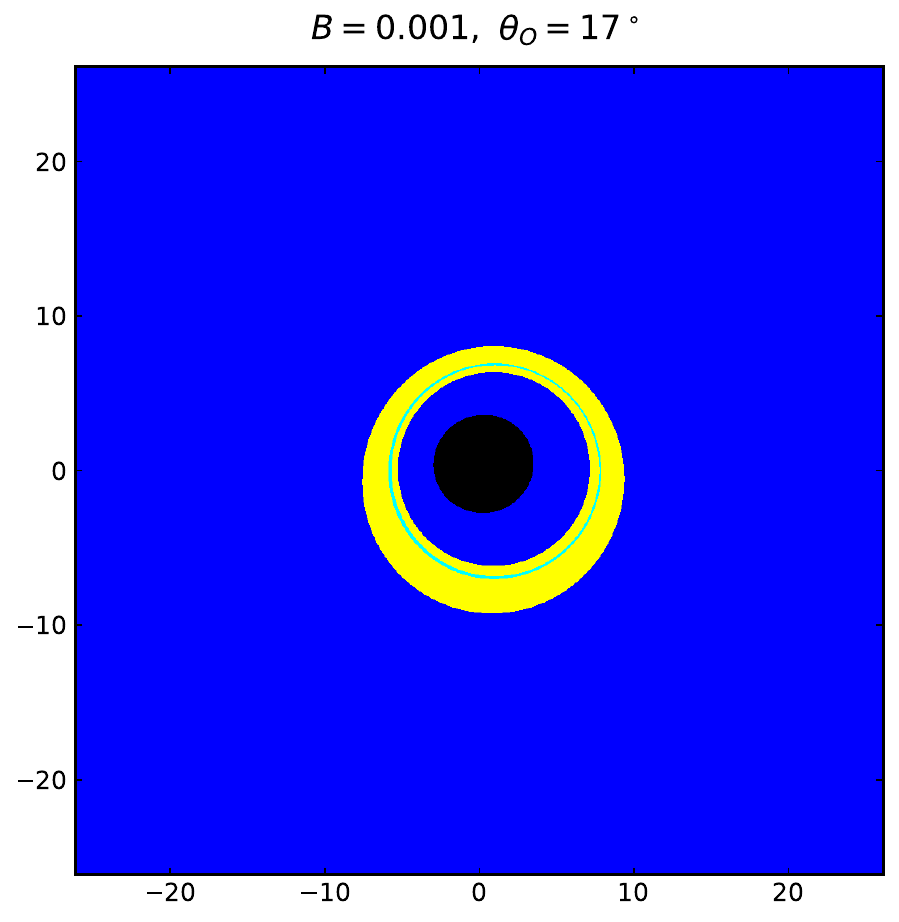} &
        \includegraphics[width=0.23\textwidth]{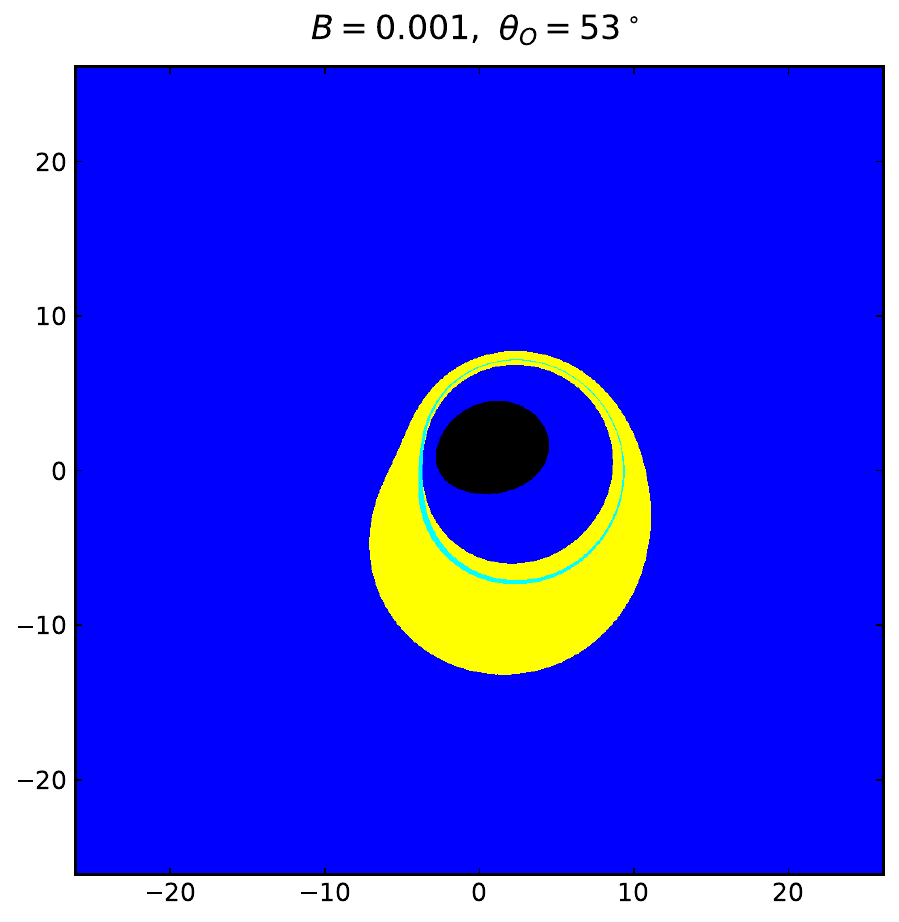} &
        \includegraphics[width=0.23\textwidth]{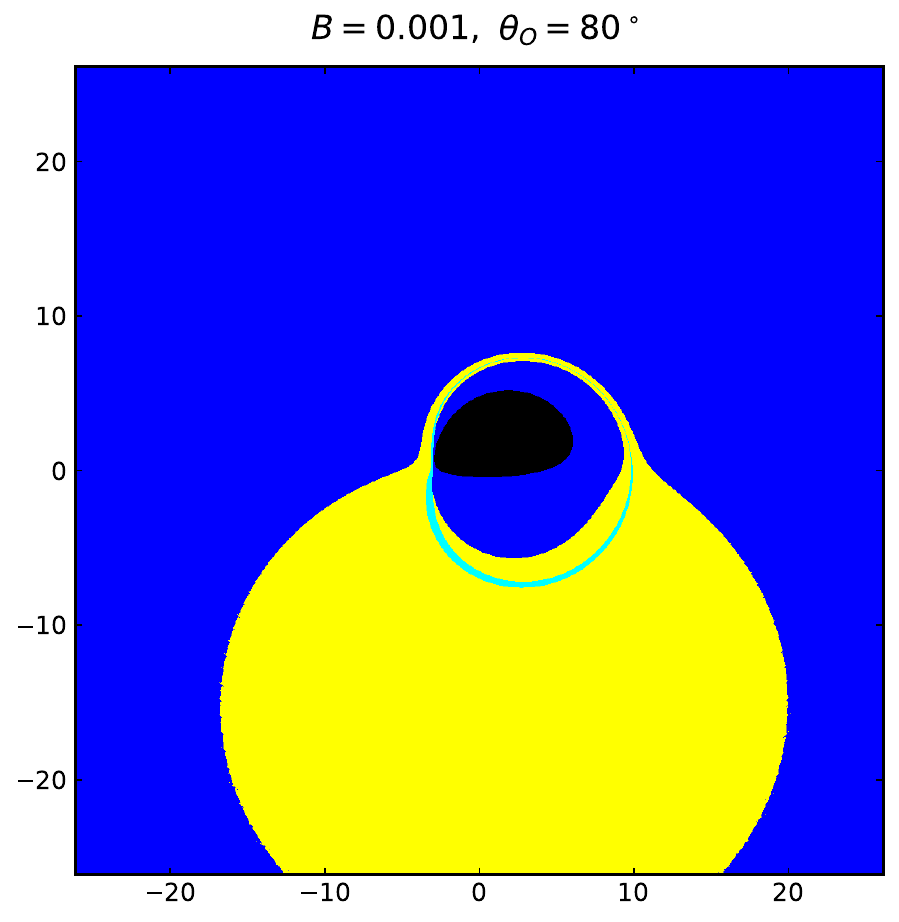} \\
        \includegraphics[width=0.23\textwidth]{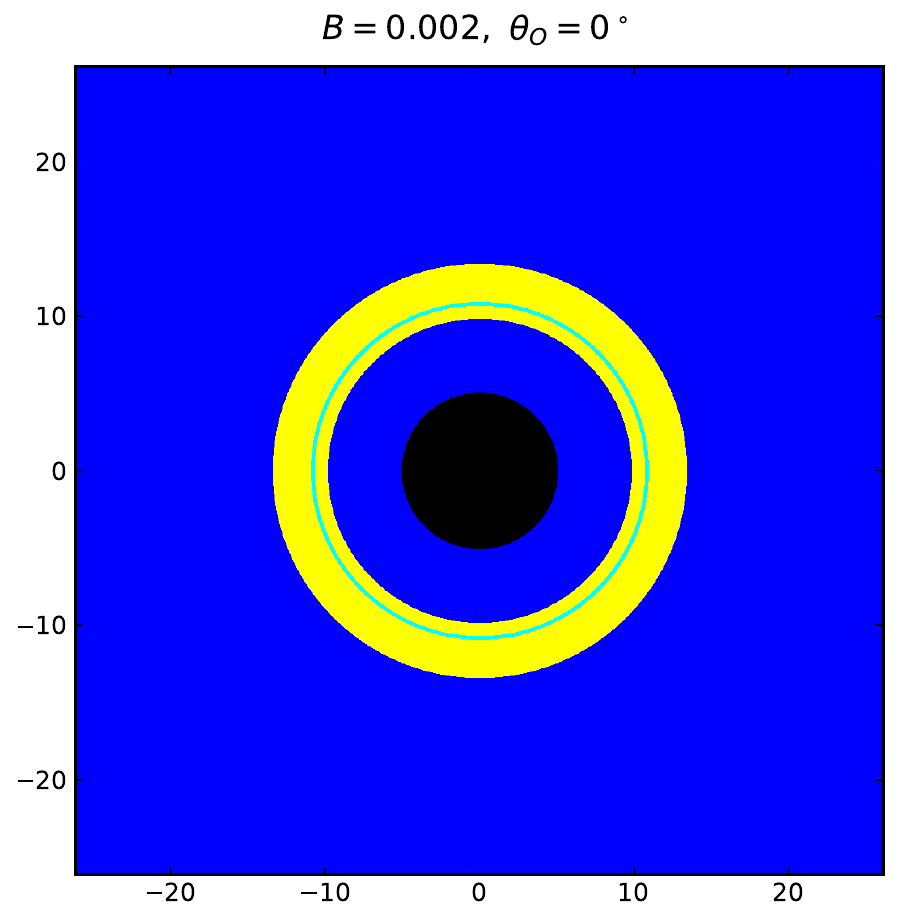} &
        \includegraphics[width=0.23\textwidth]{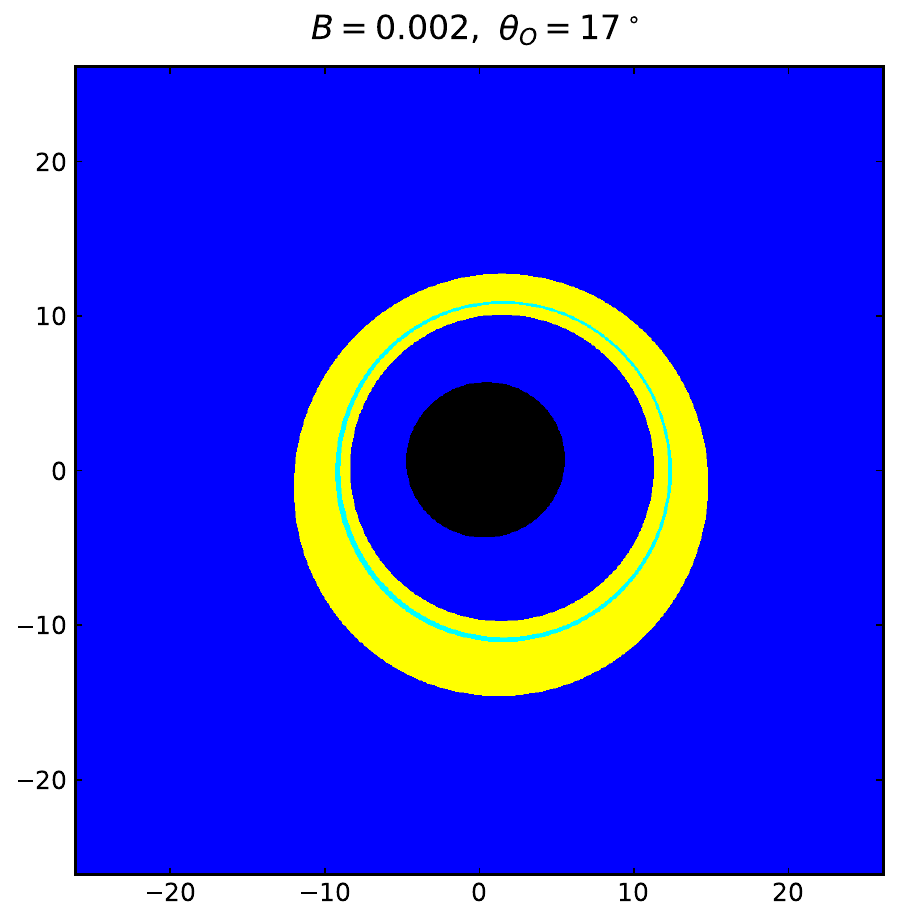} &
        \includegraphics[width=0.23\textwidth]{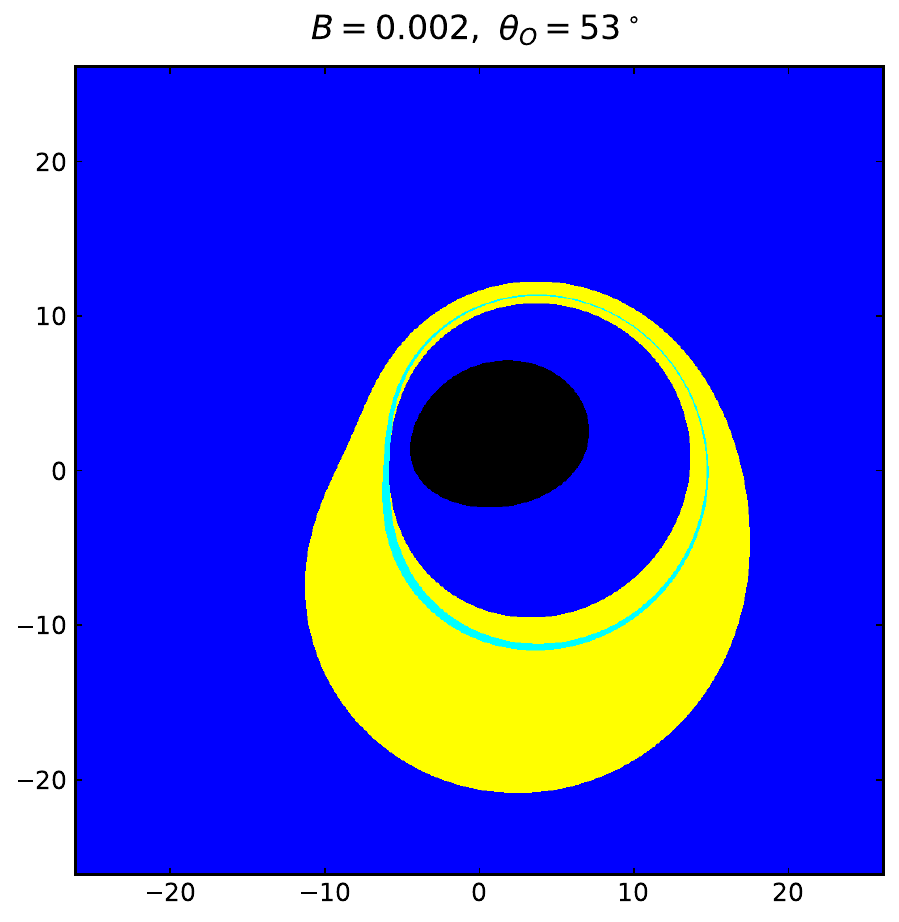} &
        \includegraphics[width=0.23\textwidth]{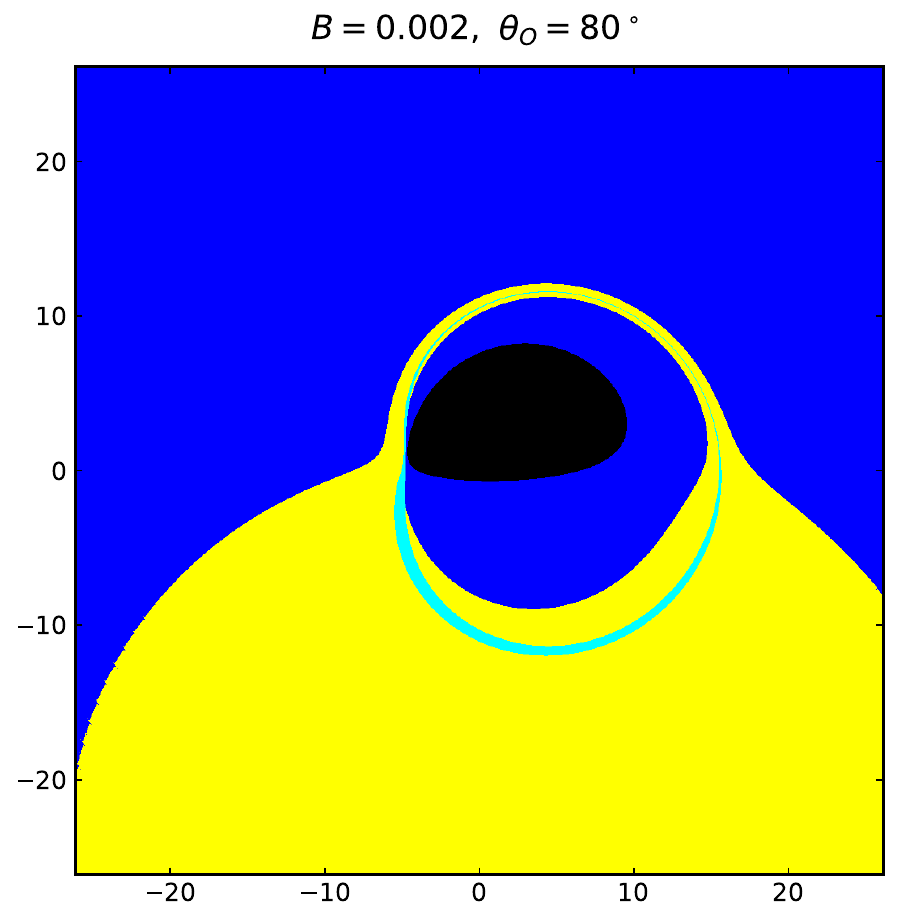} \\
        \includegraphics[width=0.23\textwidth]{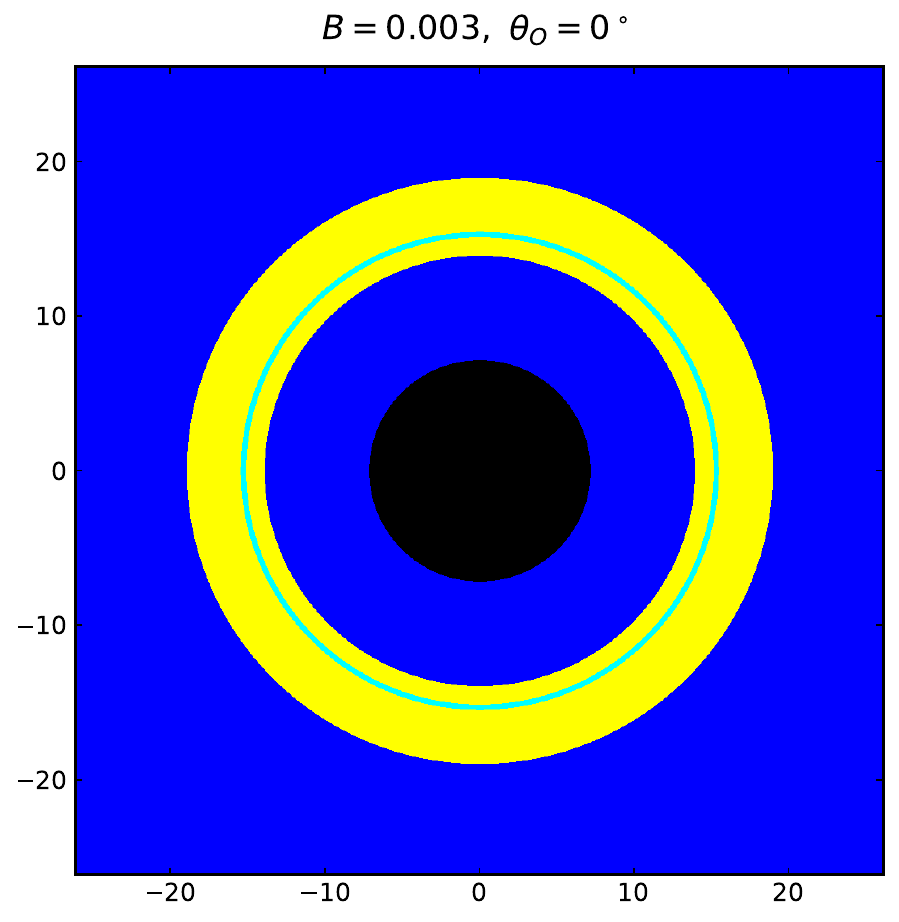} &
        \includegraphics[width=0.23\textwidth]{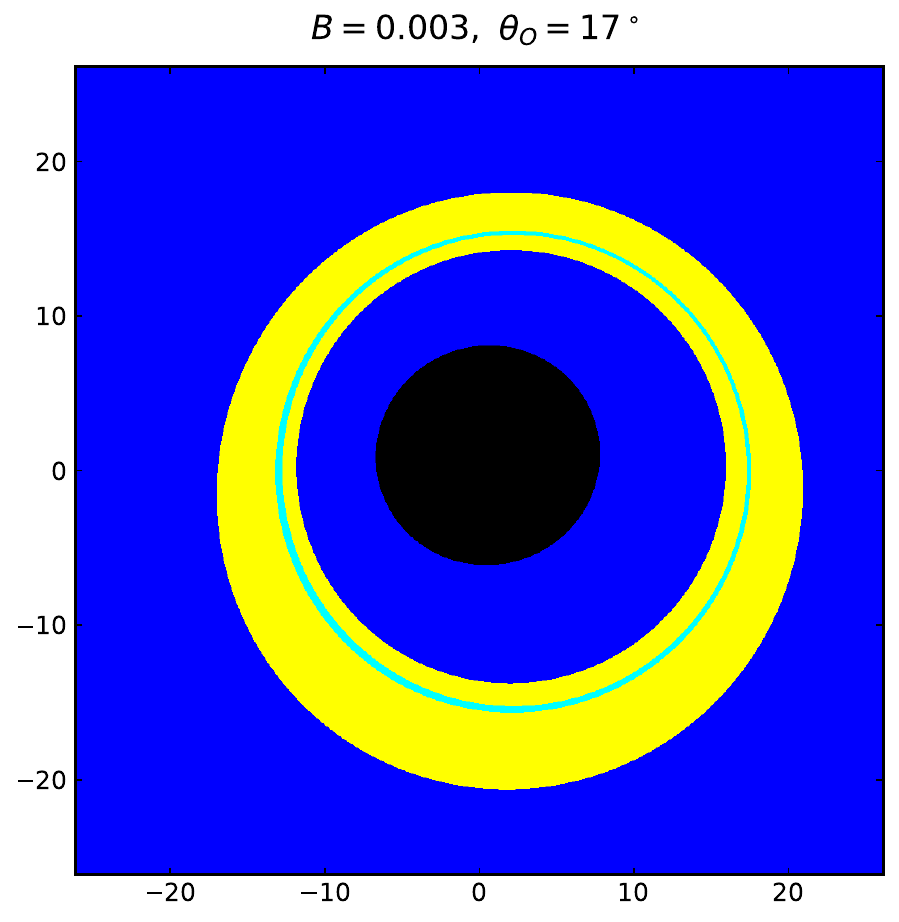} &
        \includegraphics[width=0.23\textwidth]{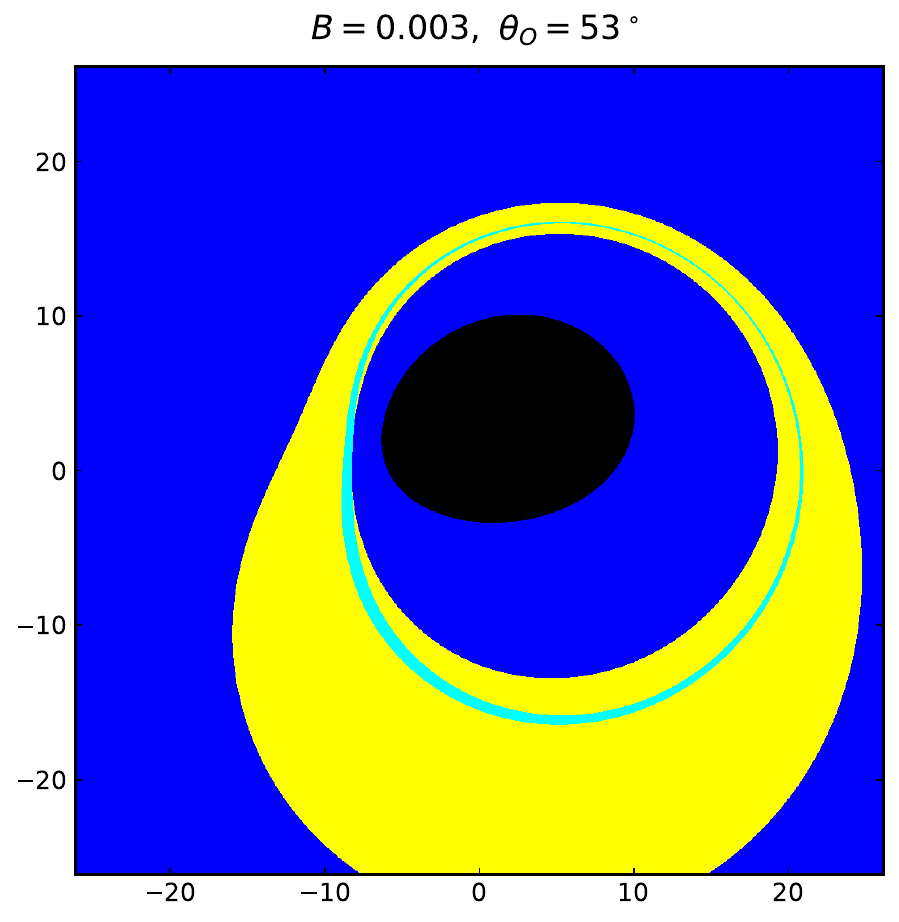} &
        \includegraphics[width=0.23\textwidth]{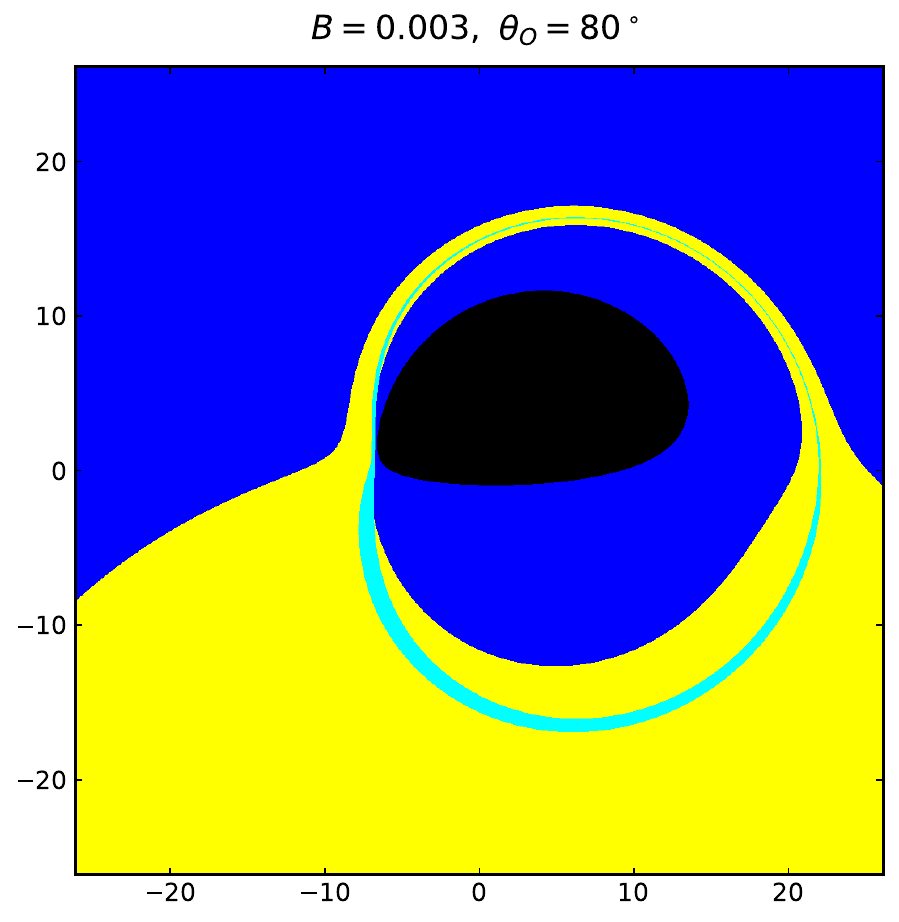} \\
    \end{tabular}
    \caption{Ray-classification maps for prograde disks with fixed spin $a=0.998$. Rows correspond to $B=0.001,0.002,0.003$ from top to bottom, and columns correspond to $\theta_O=0^\circ,17^\circ,53^\circ,80^\circ$ from left to right. The colors denote the subimage order: direct image ($n=0$), lensing-ring contribution ($n=1$), and higher-order photon-ring subimages ($n\geq 2$). The black region represents screen directions that do not receive emission from the disk, including both photon capture and the non-emitting region inside the adopted inner disk boundary.}
    \label{fig:class_varyB}
\end{figure*}

Figure \ref{fig:class_varyB} shows the corresponding ray-classification maps. The colors denote the subimage order defined in Section \ref{sec:spacetime}: $n=0$ for the direct image, $n=1$ for the lensing-ring contribution, and $n\geq 2$ for the higher-order photon-ring subimages. The black region denotes screen directions that do not receive emission from the disk. This region includes rays captured by the black hole as well as rays that pass through the non-emitting region inside the adopted inner disk boundary. It should therefore be interpreted as an emission-depleted central region rather than solely as the geometrical black hole shadow.

As the inclination increases from the face-on case to the nearly edge-on case, the image becomes increasingly asymmetric because of projection effects, frame dragging, and the different path lengths of rays reaching the observer. The higher-order subimages are compressed into narrow structures near the boundary of the central dark region. Varying $B$ changes both the Kerr-BR geodesic structure and the effective emitting region through its influence on the ISCO and on the magnetic-dominance cutoff. As a result, the size of the central emission-depleted region and the widths of the lensed components vary with $B$.

\begin{figure*}[htbp]
    \centering
    \begin{tabular}{cccc}
        \includegraphics[width=0.23\textwidth]{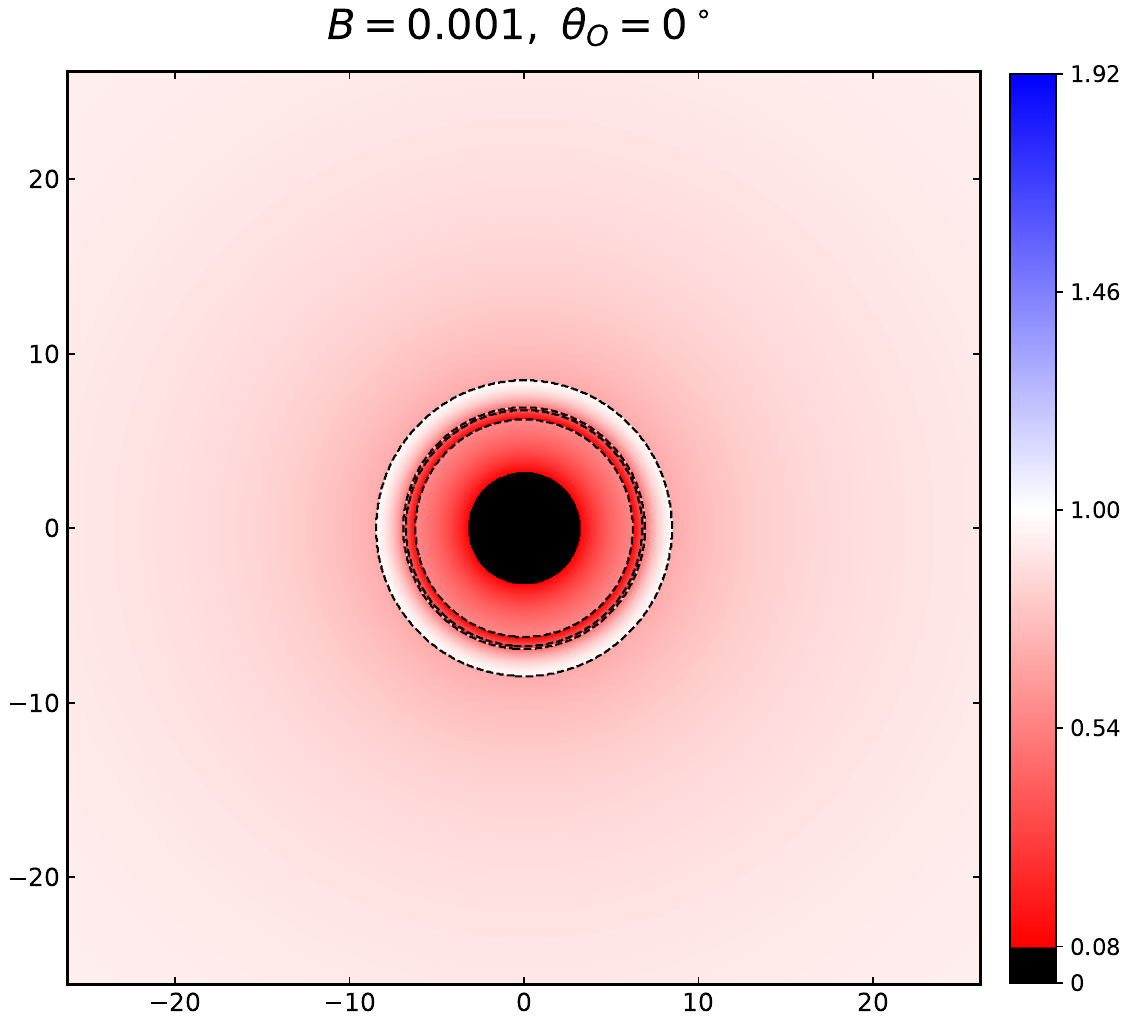} &
        \includegraphics[width=0.23\textwidth]{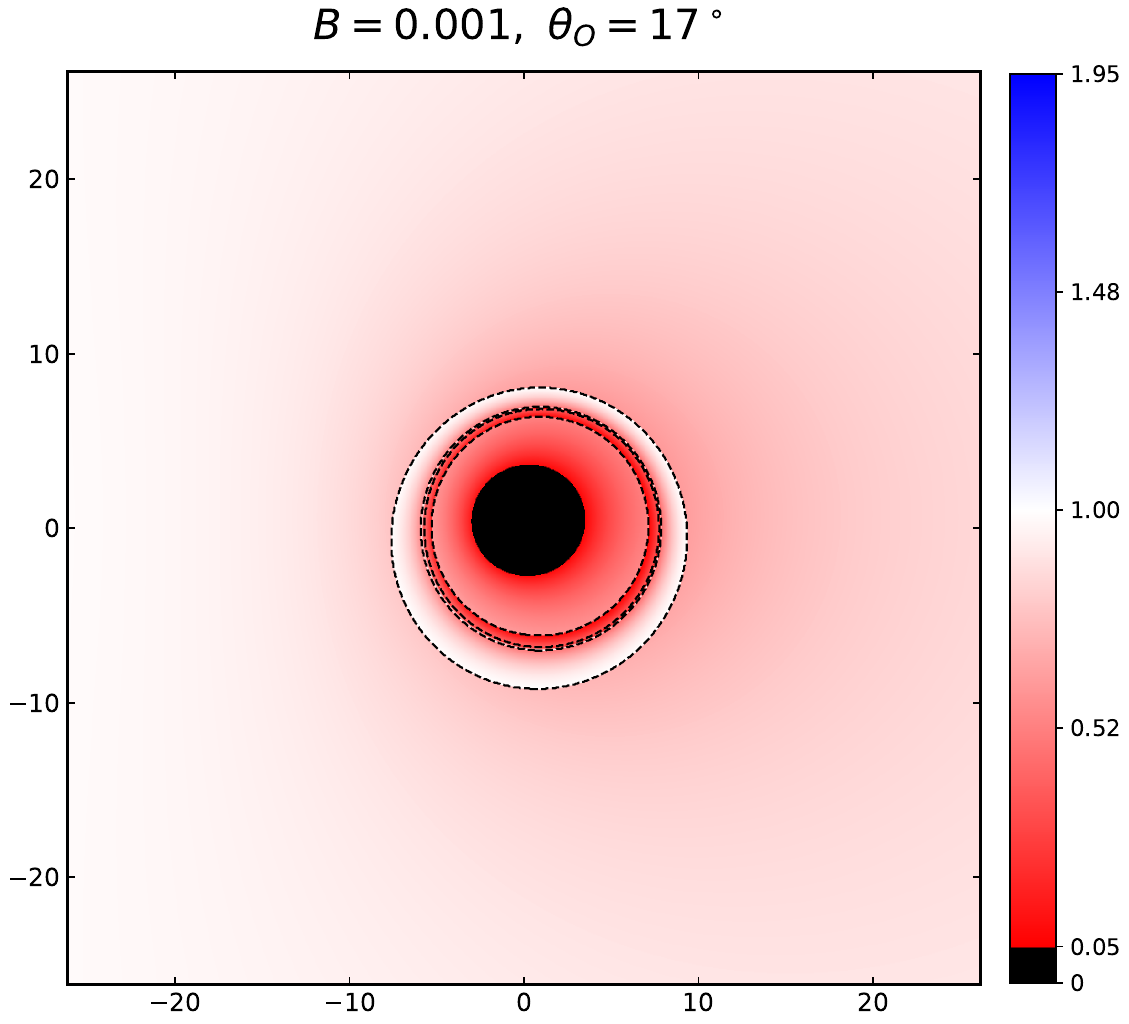} &
        \includegraphics[width=0.23\textwidth]{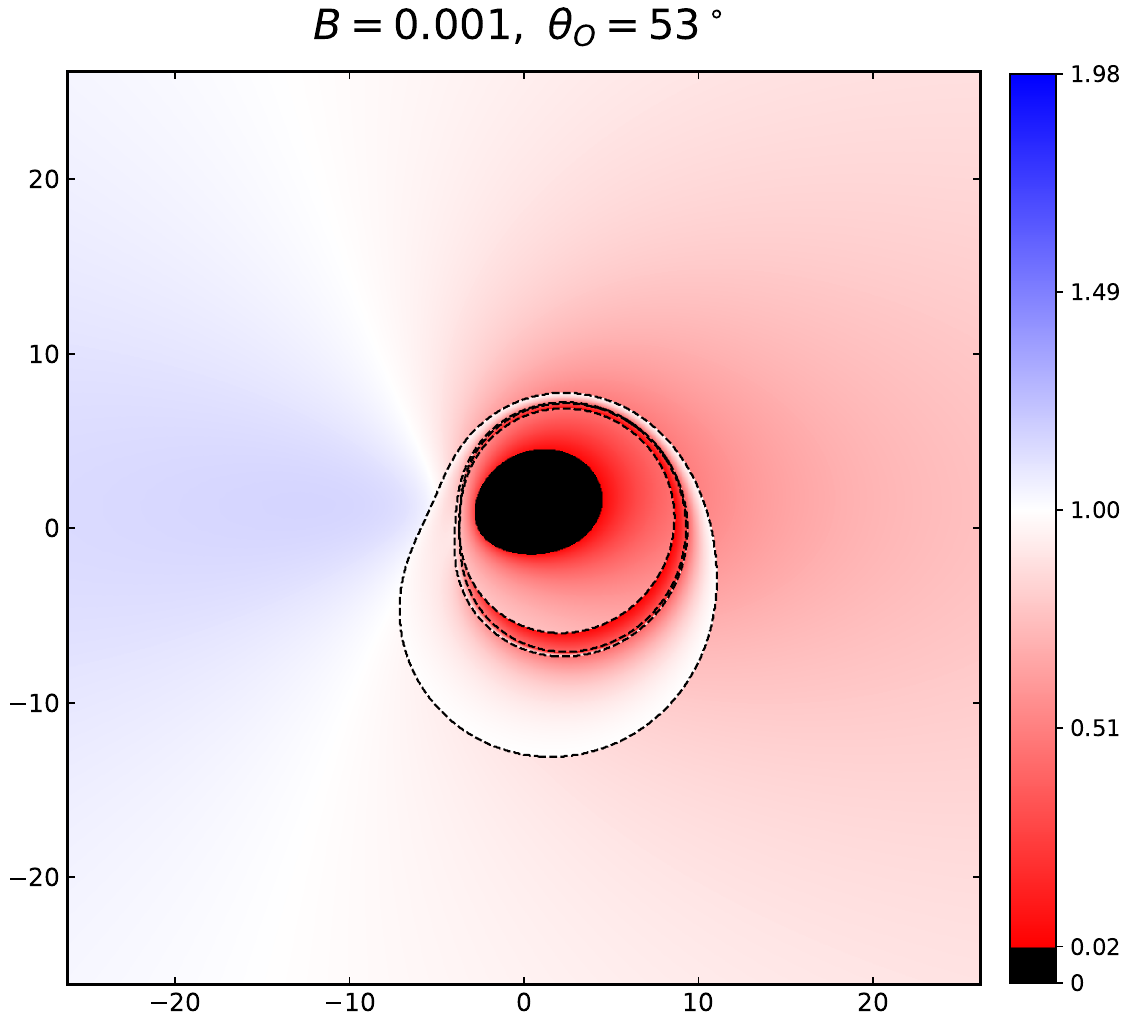} &
        \includegraphics[width=0.23\textwidth]{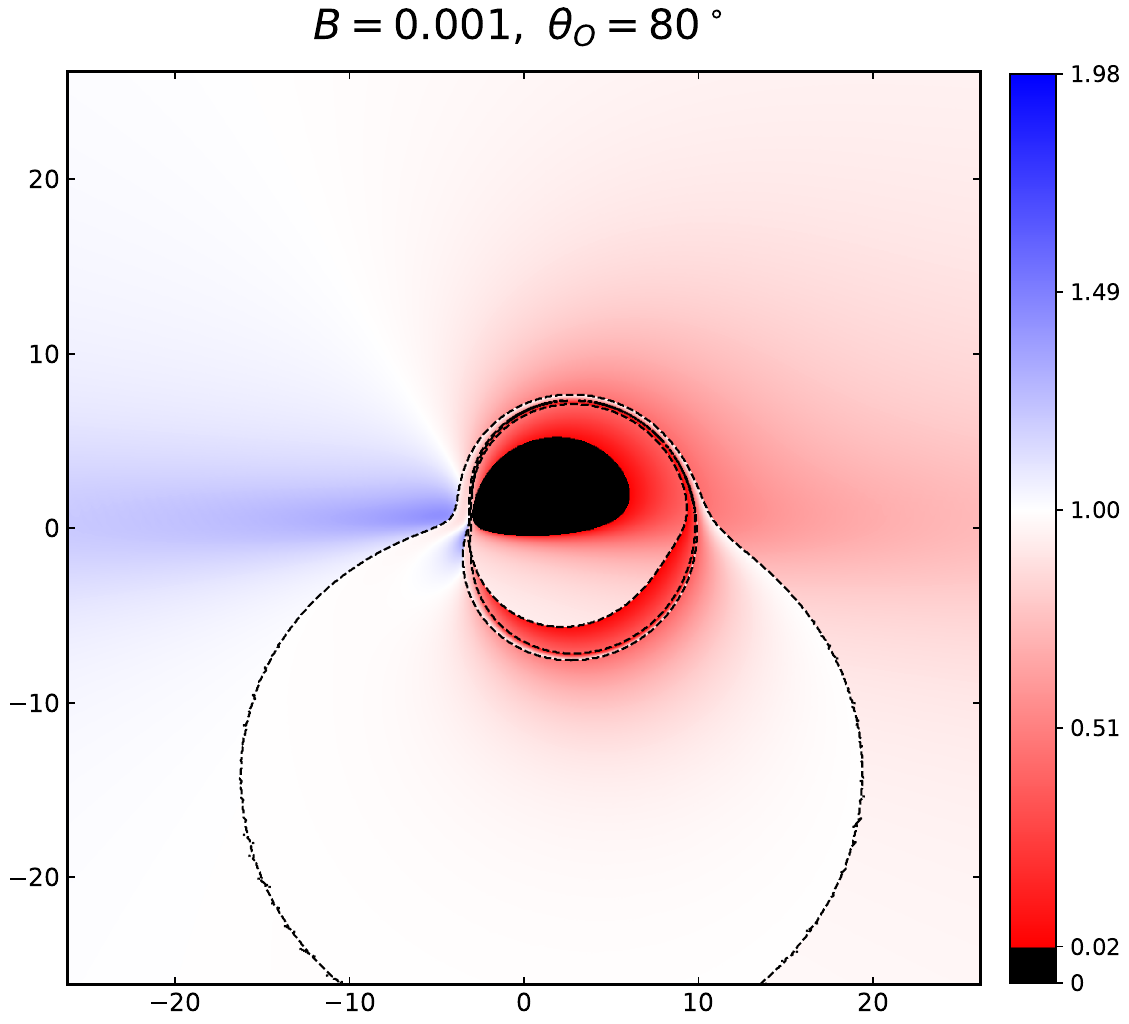} \\
        \includegraphics[width=0.23\textwidth]{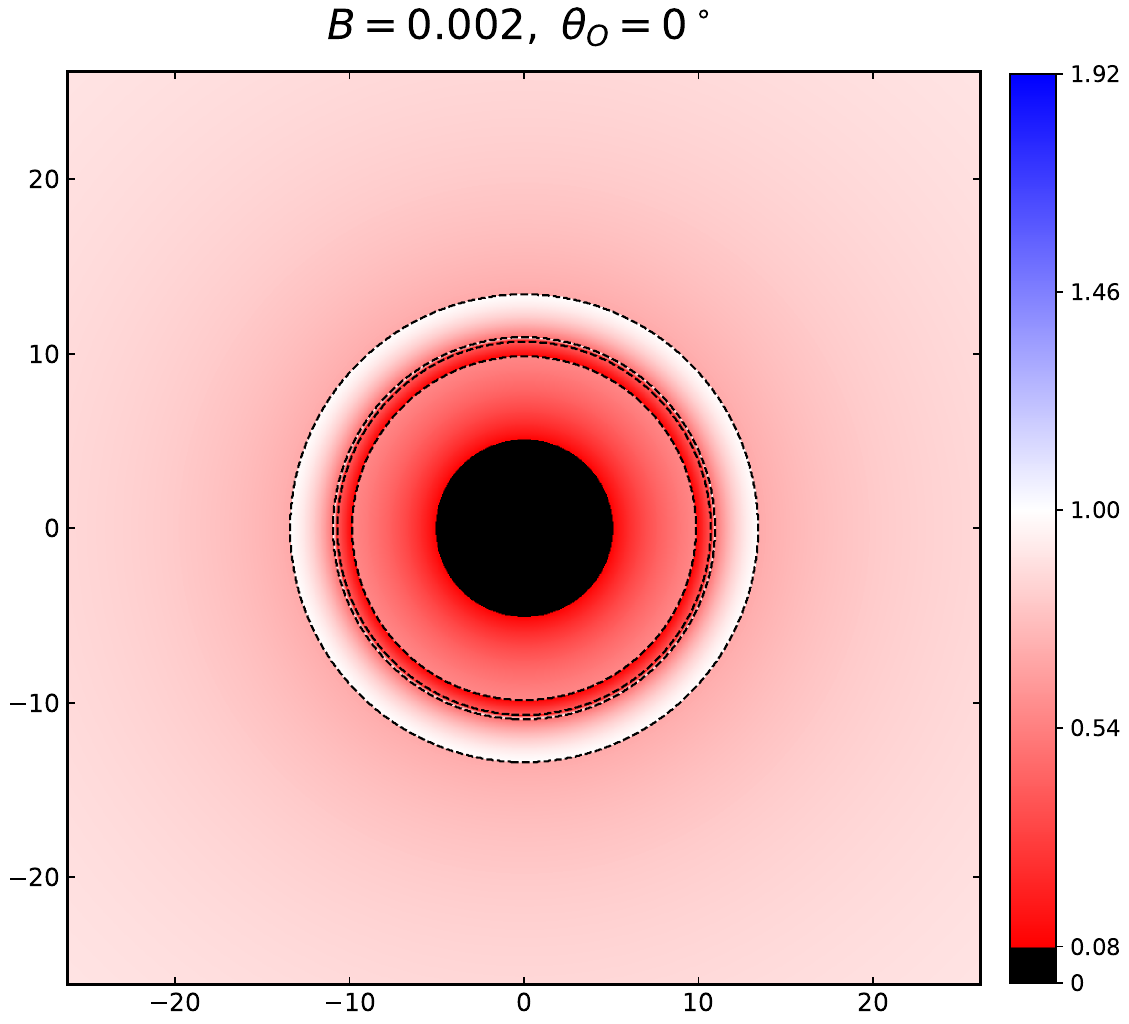} &
        \includegraphics[width=0.23\textwidth]{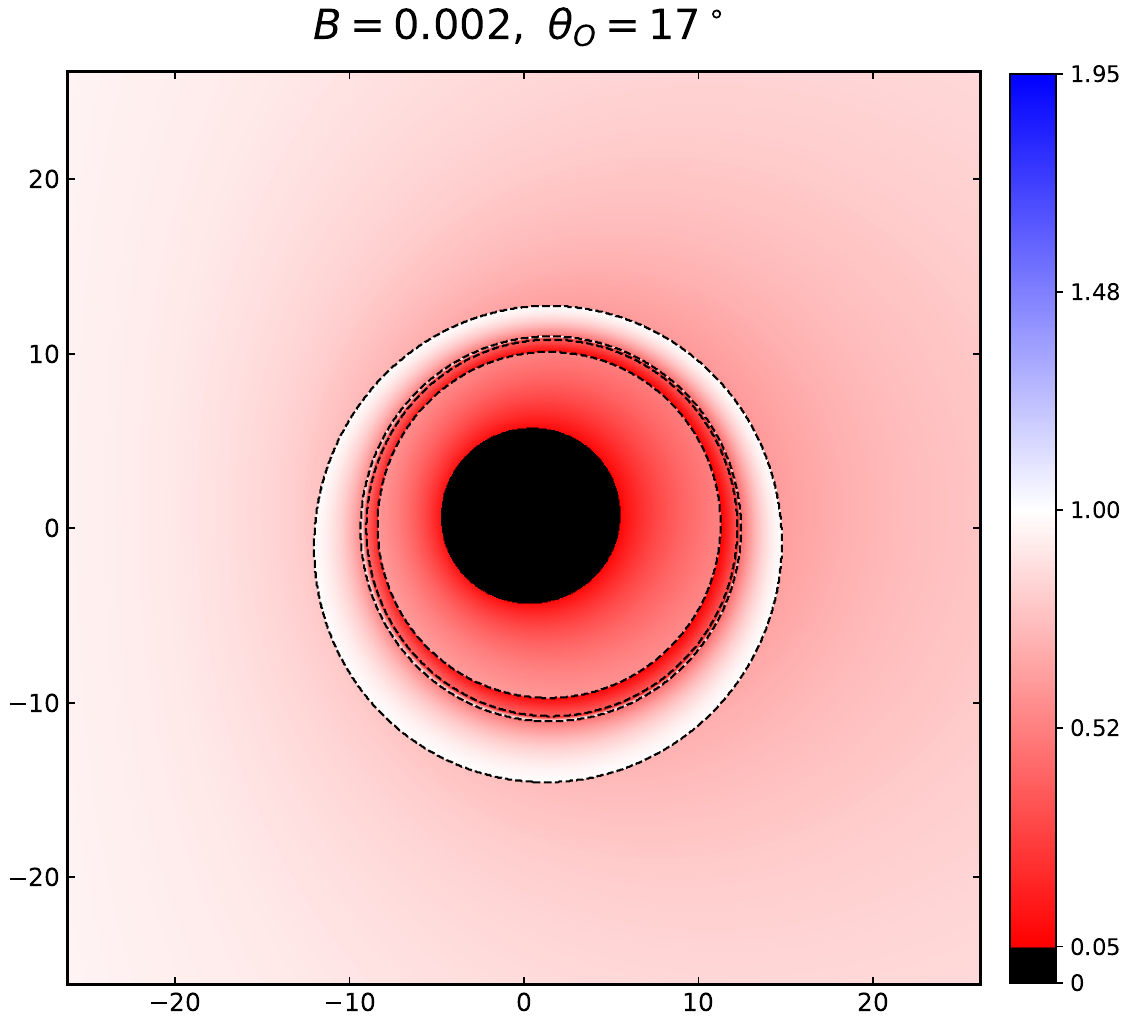} &
        \includegraphics[width=0.23\textwidth]{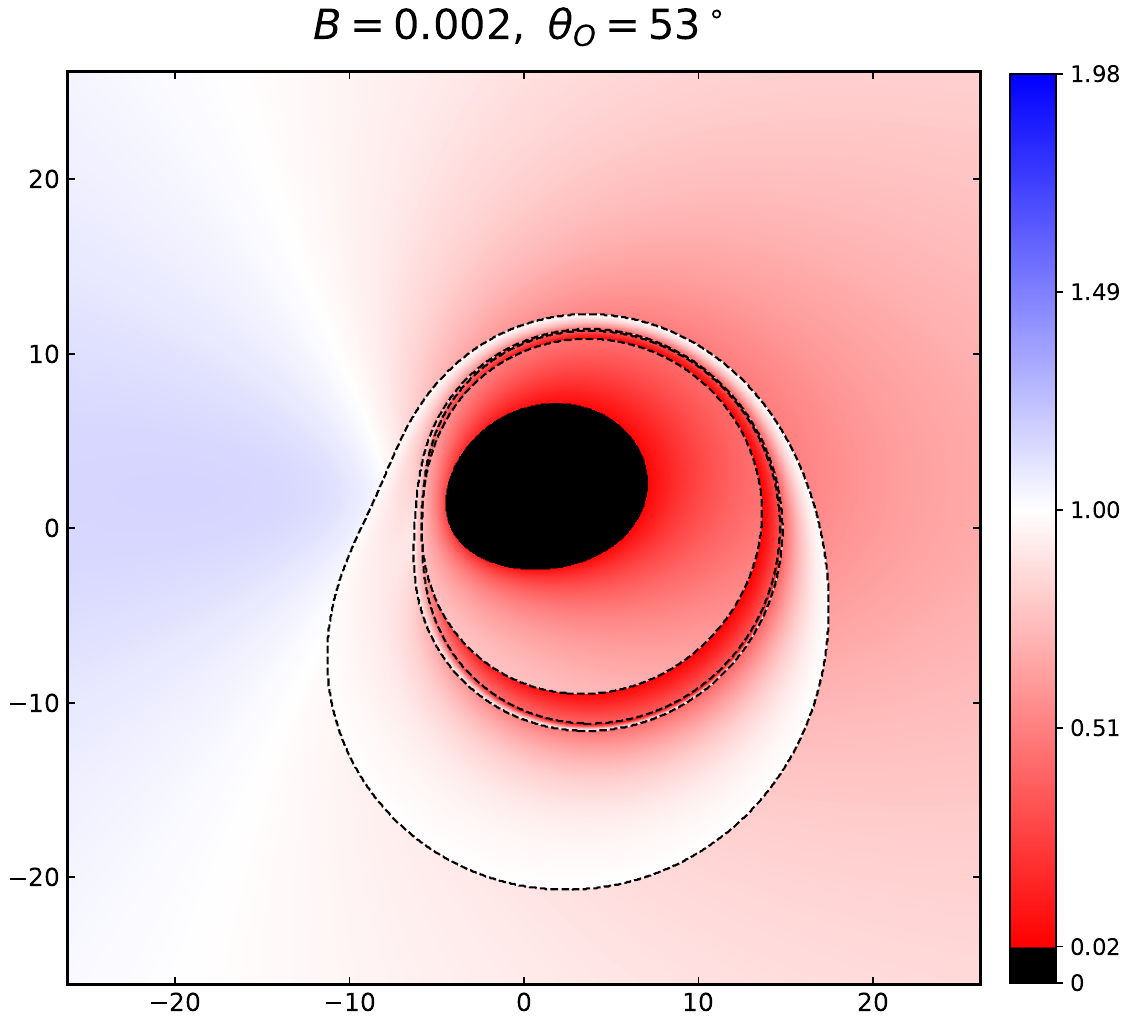} &
        \includegraphics[width=0.23\textwidth]{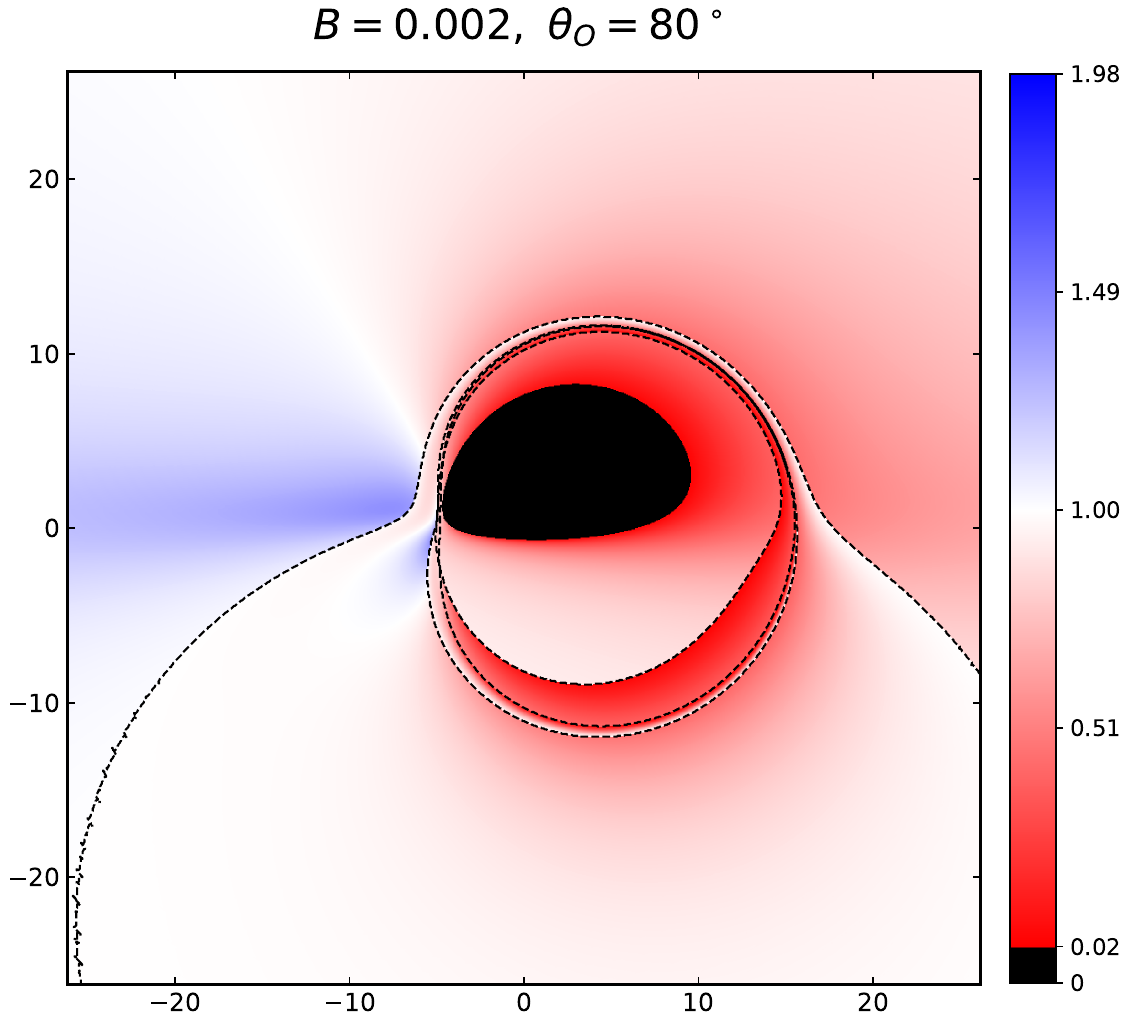} \\
        \includegraphics[width=0.23\textwidth]{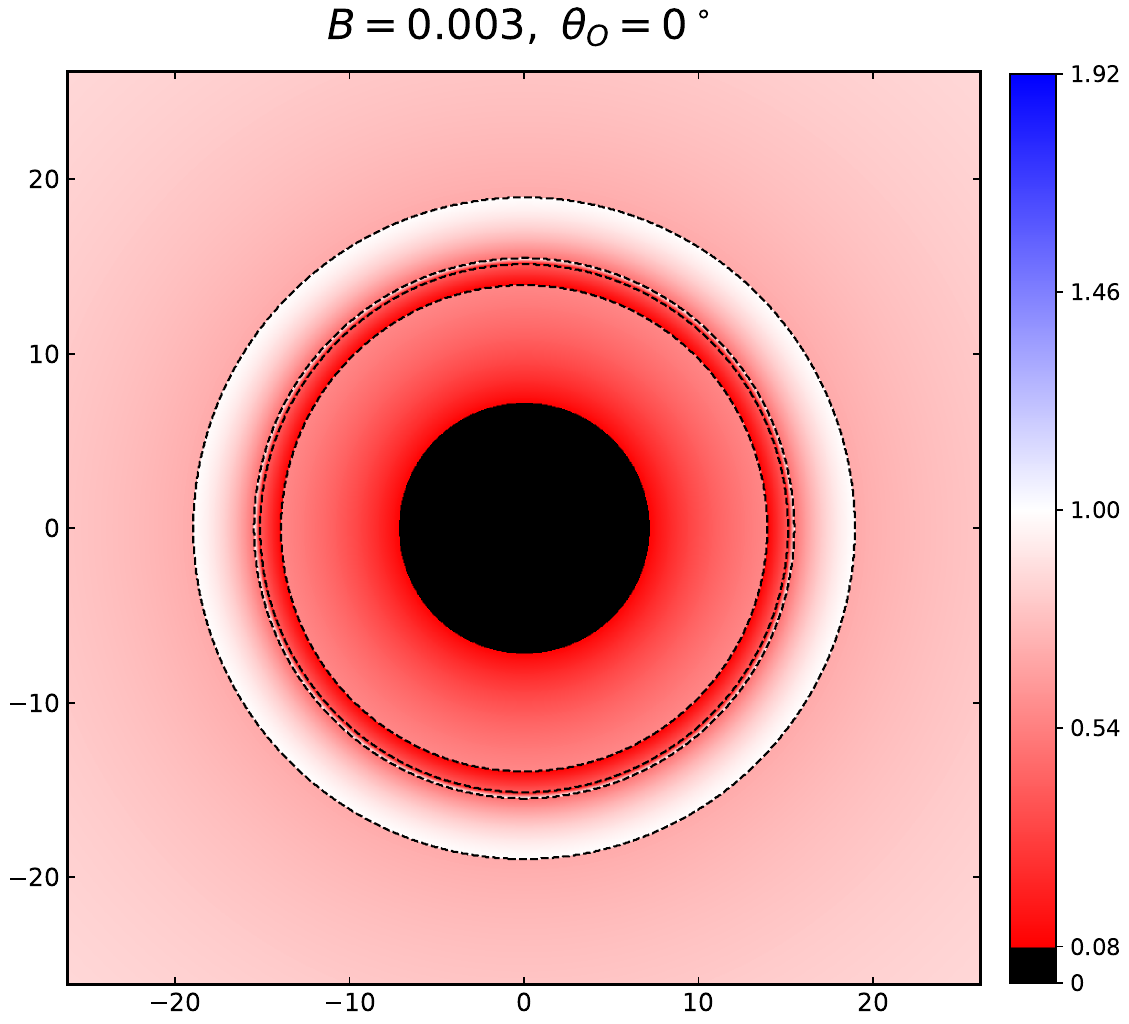} &
        \includegraphics[width=0.23\textwidth]{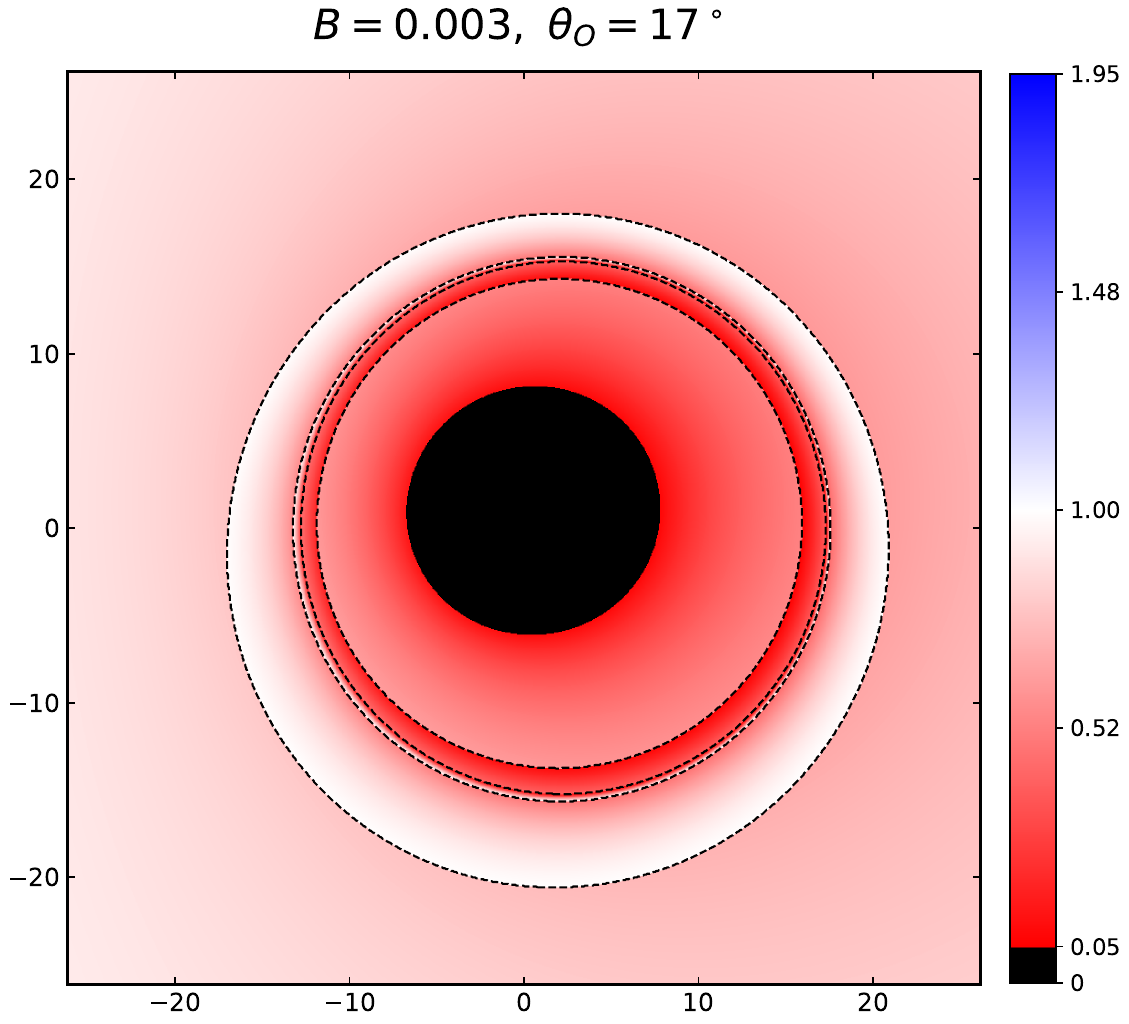} &
        \includegraphics[width=0.23\textwidth]{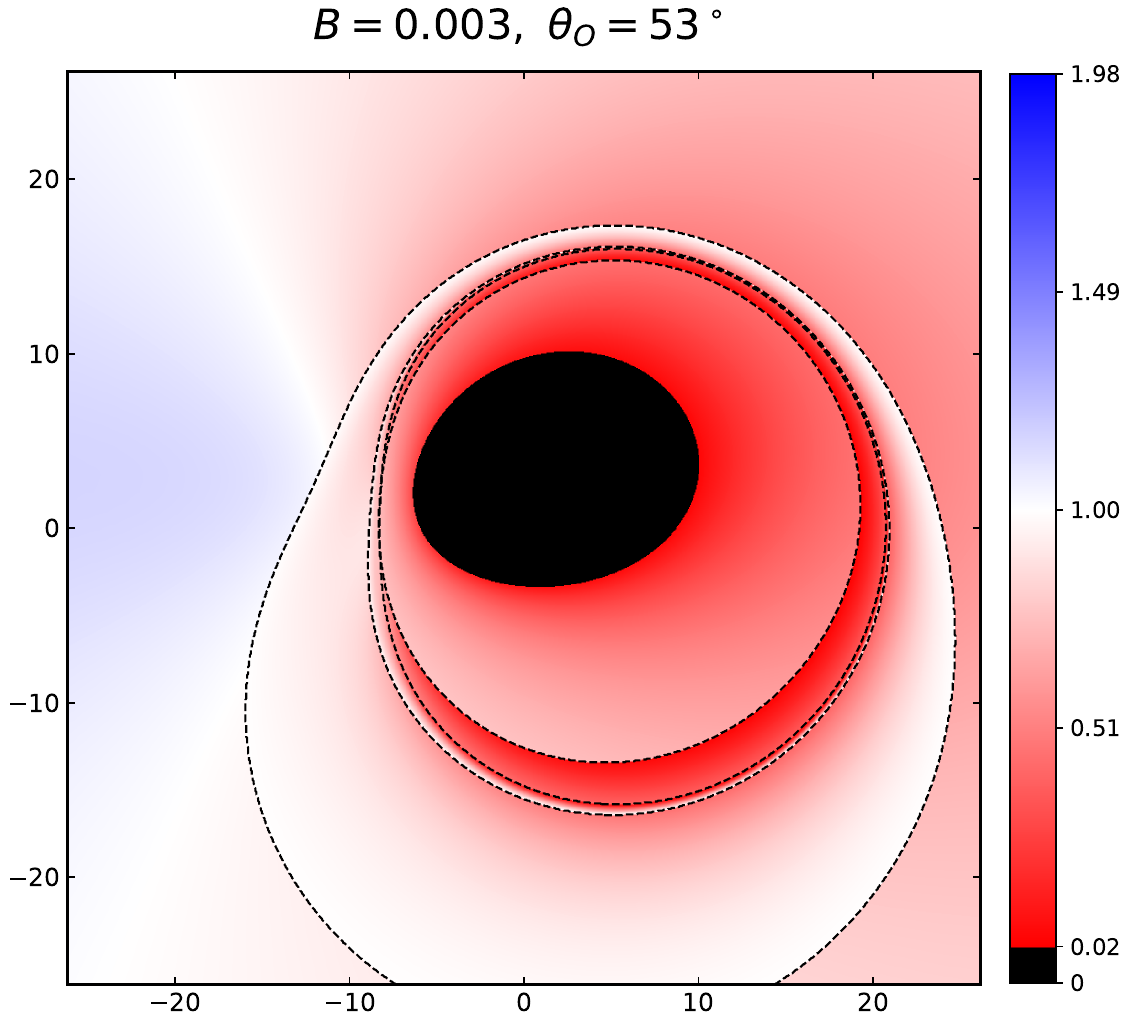} &
        \includegraphics[width=0.23\textwidth]{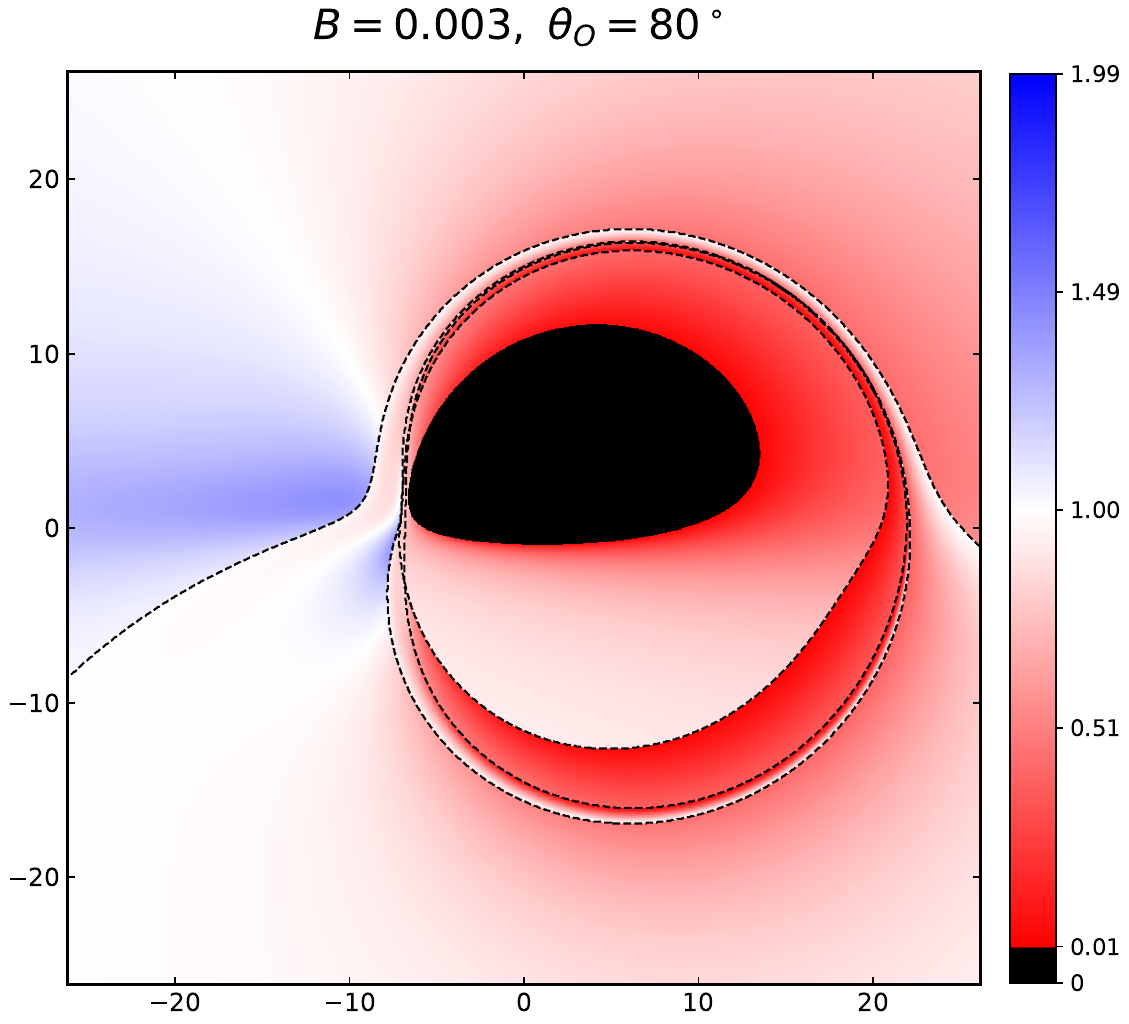} \\
    \end{tabular}
    \caption{Redshift-factor maps for prograde disks with $a=0.998$. Rows correspond to different magnetic parameters $B$, and columns correspond to different observer inclinations $\theta_O$. The dashed curves mark the boundaries between the direct image ($n=0$), the lensing-ring contribution ($n=1$), and the higher-order photon-ring subimages ($n\geq 2$). The left-right asymmetry at large inclination is mainly produced by relativistic Doppler boosting.}
    \label{fig:redshift_varyB}
\end{figure*}

Figure \ref{fig:redshift_varyB} displays the redshift factor $g$ on the observer's screen. Since $g$ depends on the photon momentum and on the orbital velocity of the emitter, it reflects both the gravitational redshift and the Doppler shift, but it is independent of the overall normalization of the emissivity. The dashed curves indicate the boundaries between the ray-classification regions extracted from the corresponding crossing masks. At high inclination, the approaching side of the disk is blueshifted, while the receding side is redshifted. This produces the characteristic left-right asymmetry in the redshift map. Increasing $B$ modifies the metric functions and the circular-orbit angular velocity, leading to a modest change in the extent and location of the blueshifted region.

\begin{figure*}[htbp]
    \centering
    \begin{tabular}{cccc}
        \includegraphics[width=0.23\textwidth]{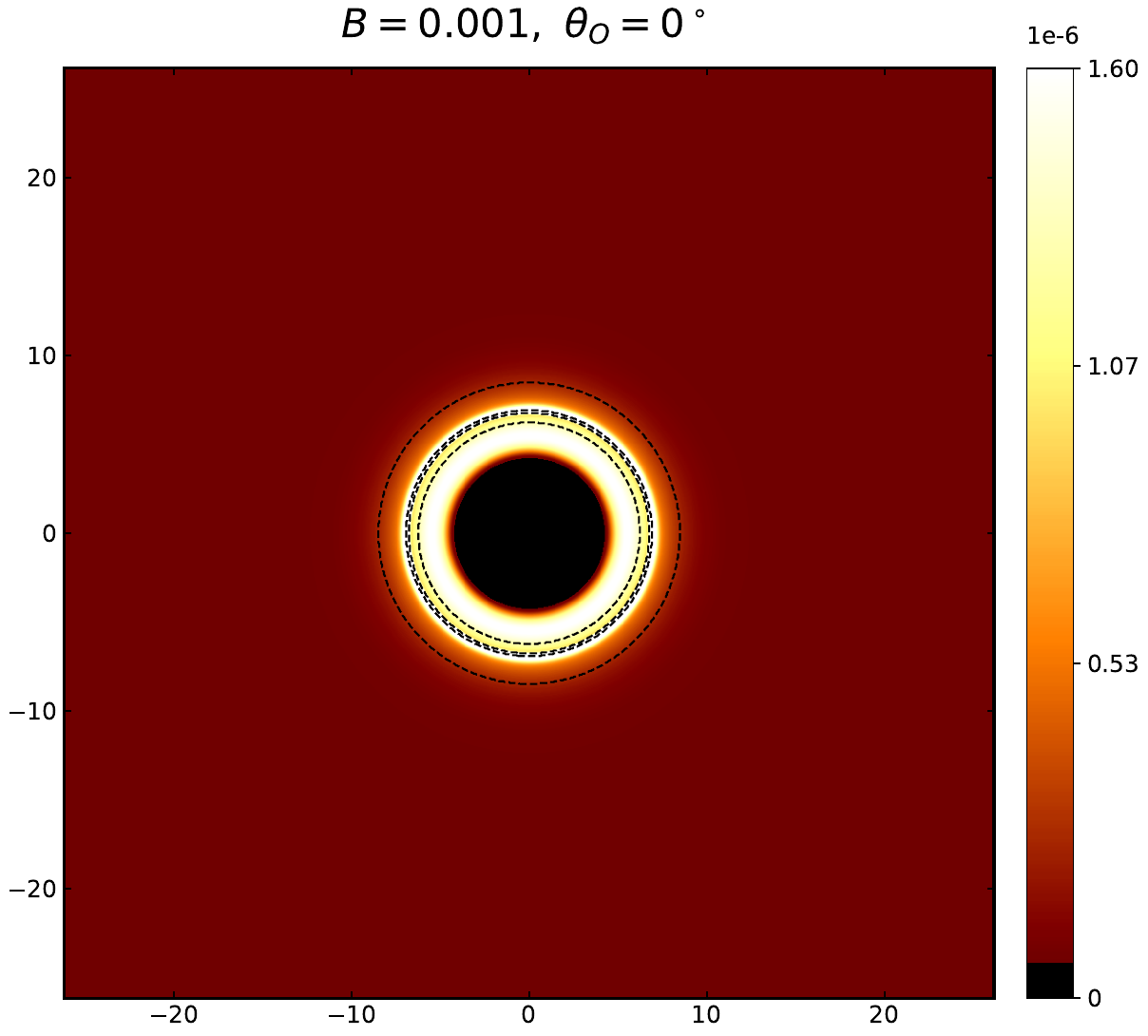} &
        \includegraphics[width=0.23\textwidth]{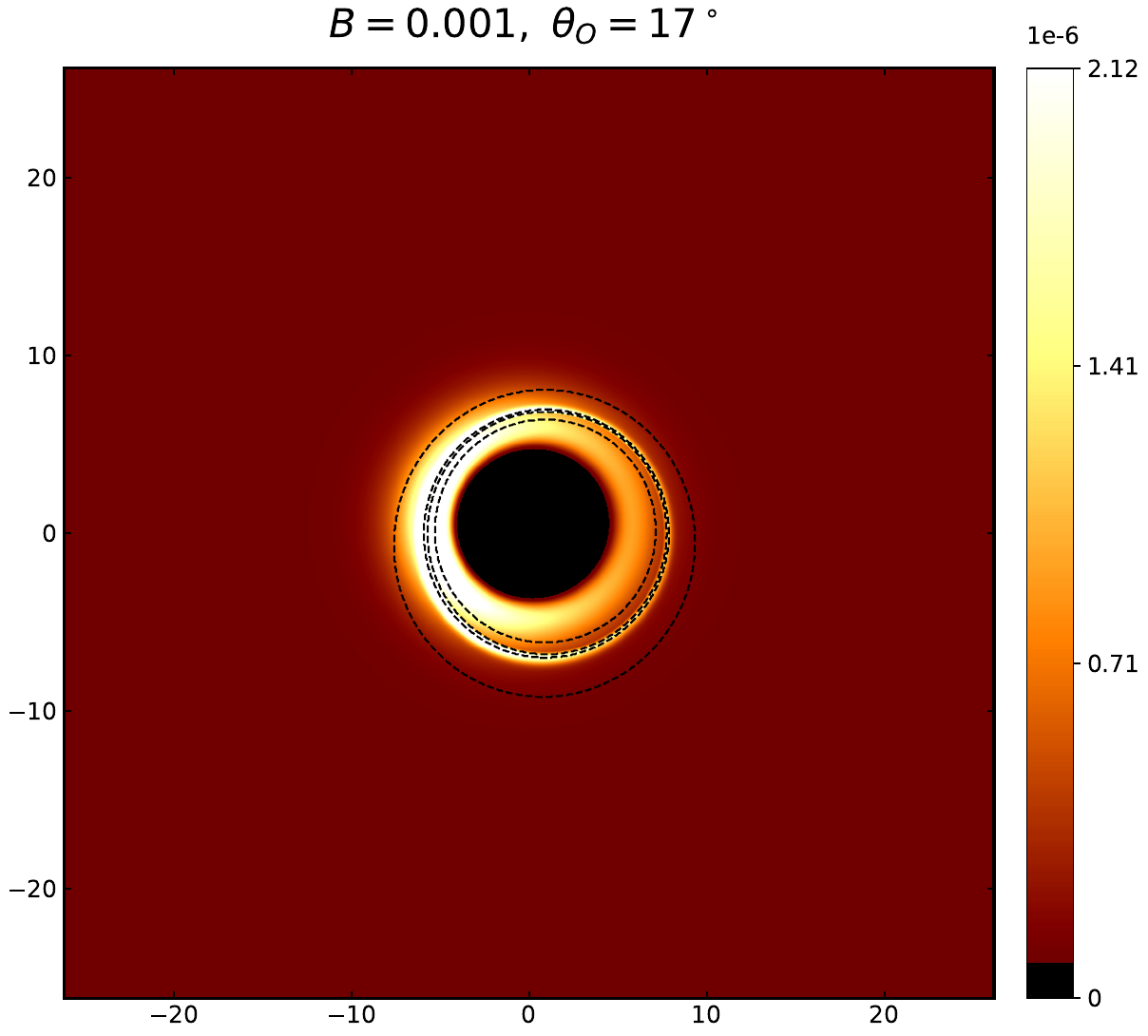} &
        \includegraphics[width=0.23\textwidth]{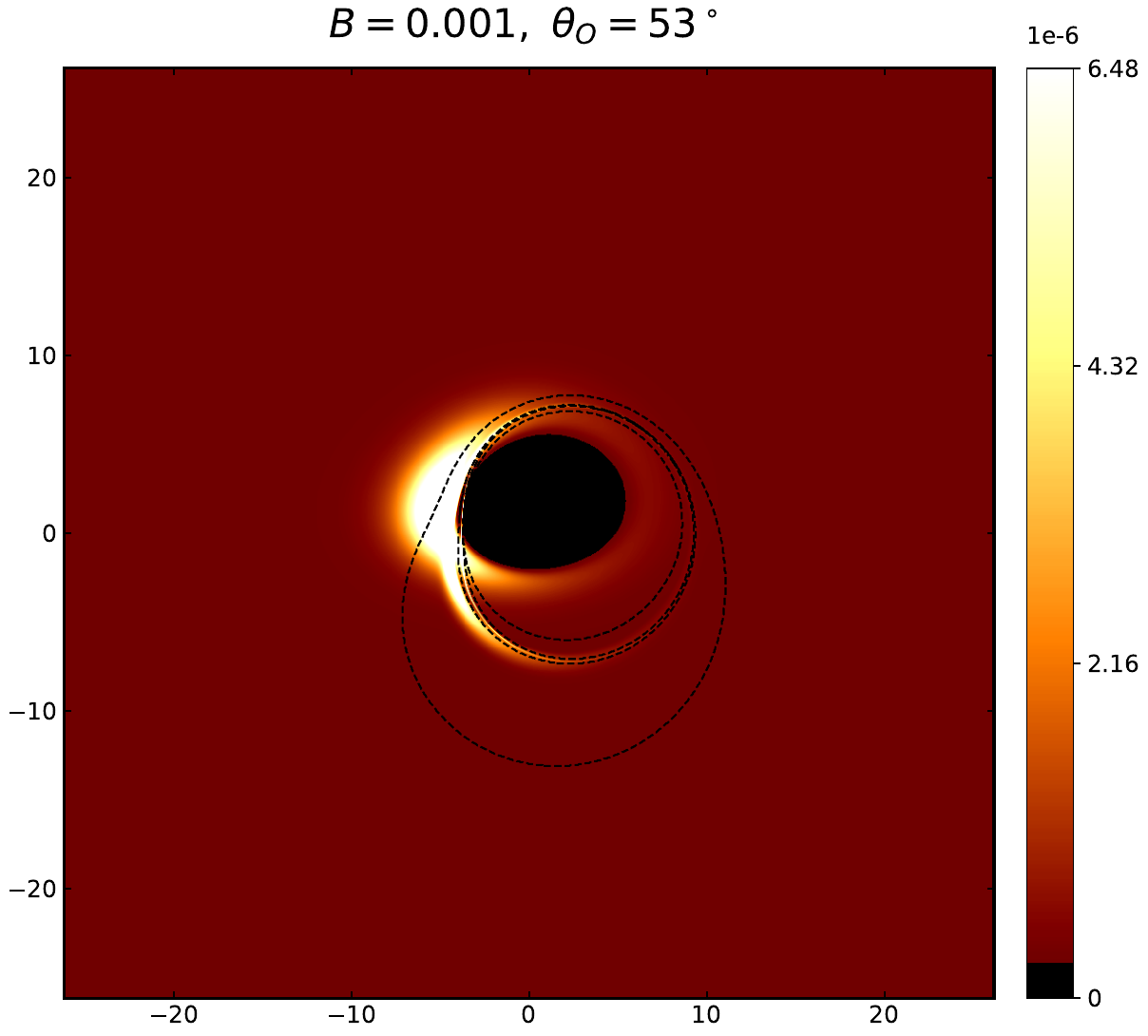} &
        \includegraphics[width=0.23\textwidth]{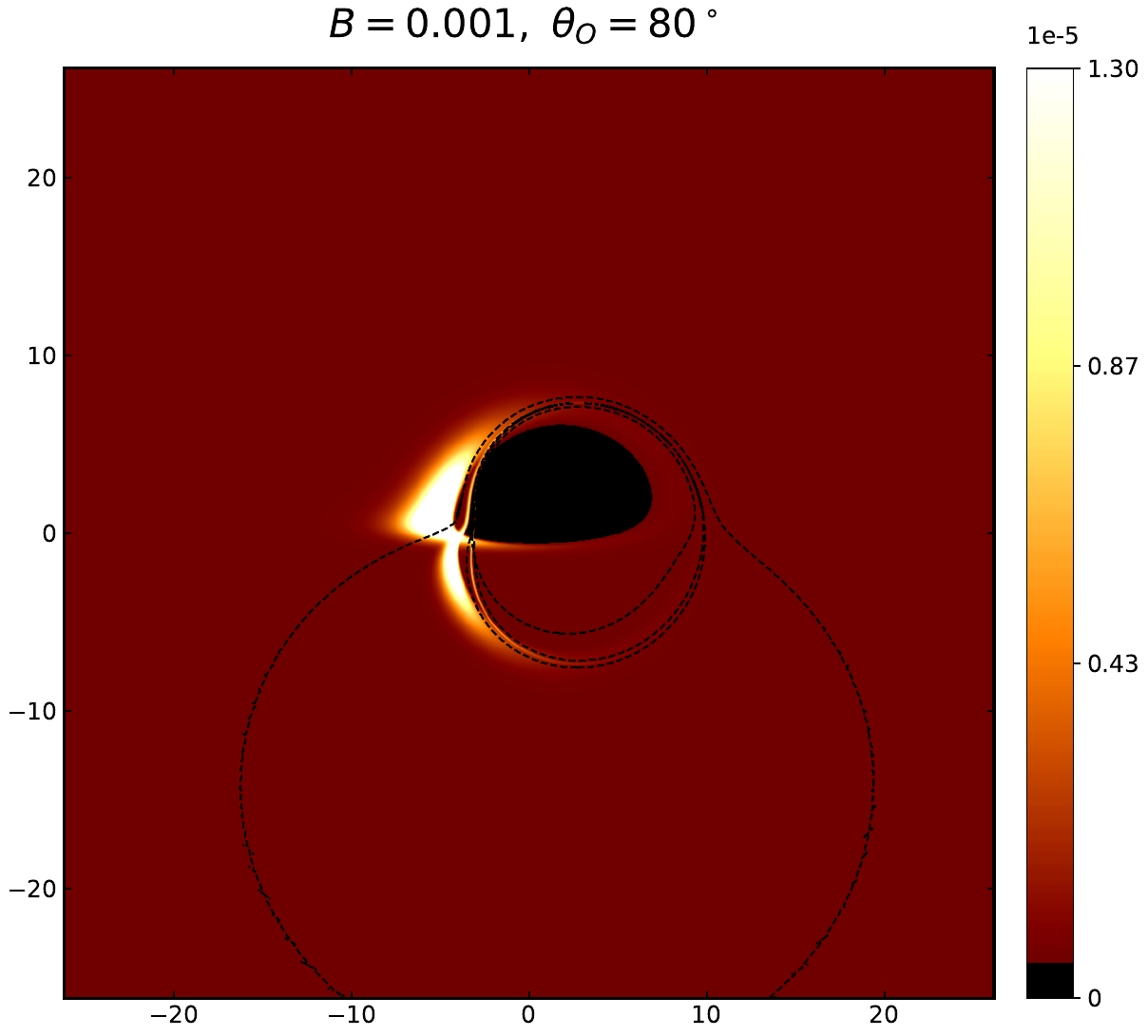} \\
        \includegraphics[width=0.23\textwidth]{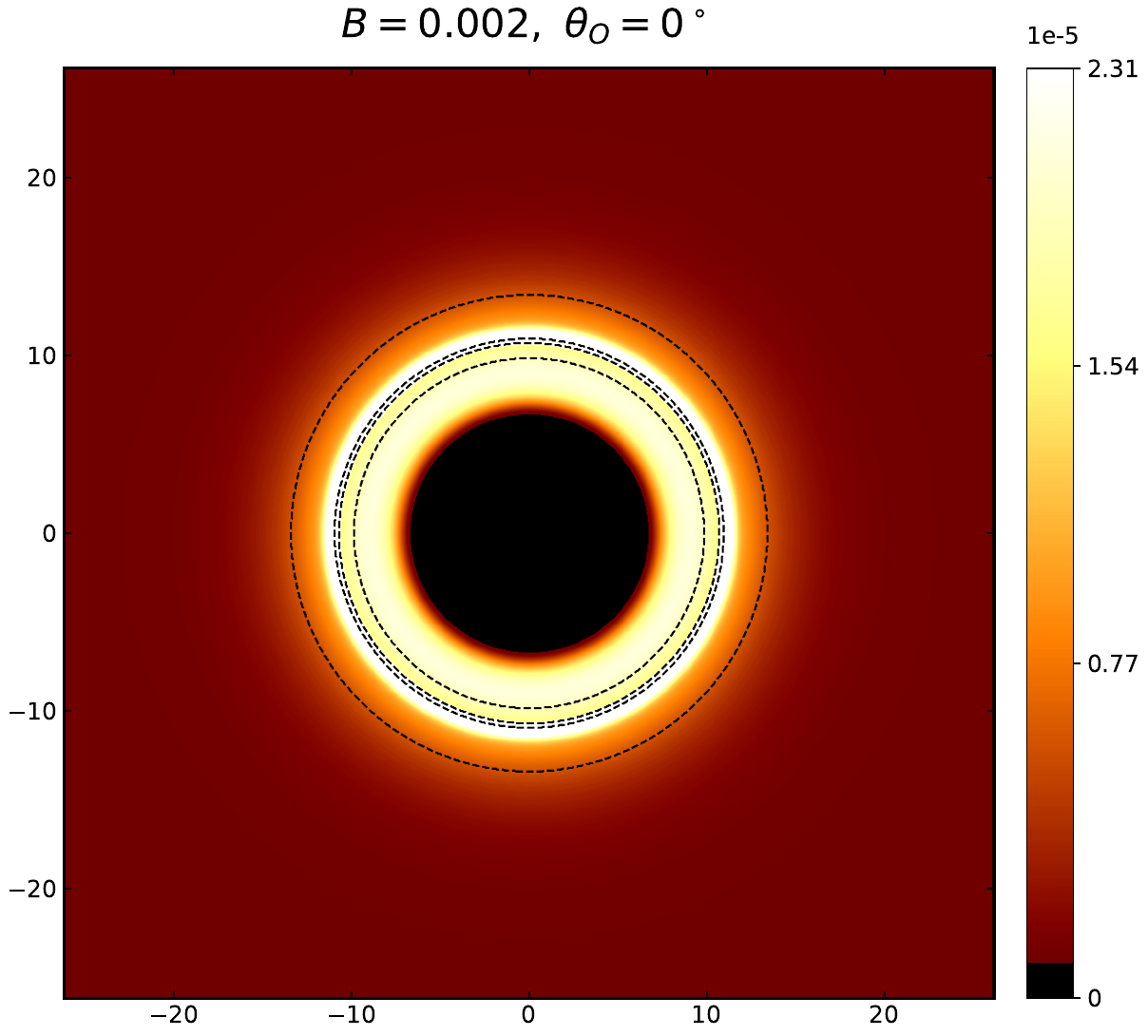} &
        \includegraphics[width=0.23\textwidth]{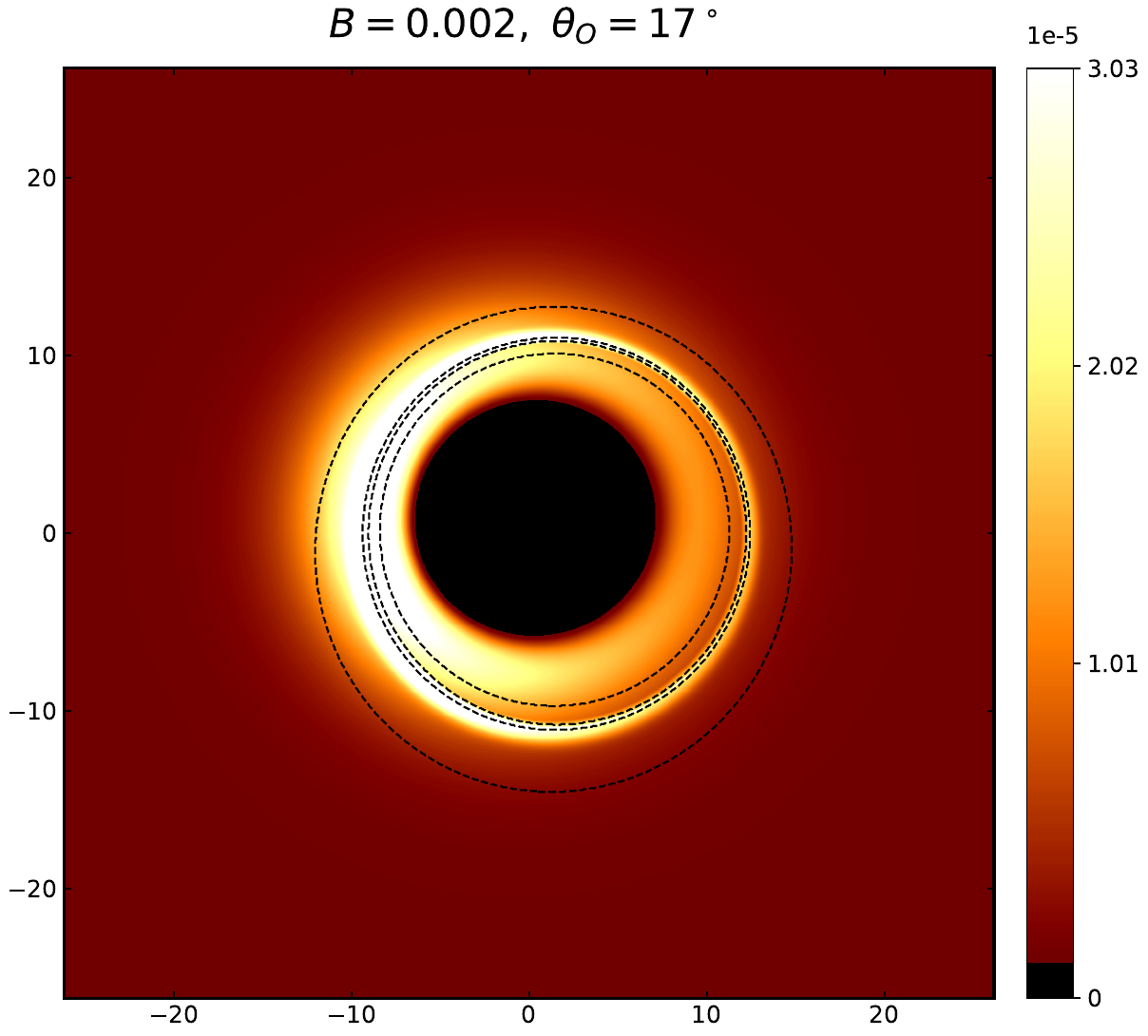} &
        \includegraphics[width=0.23\textwidth]{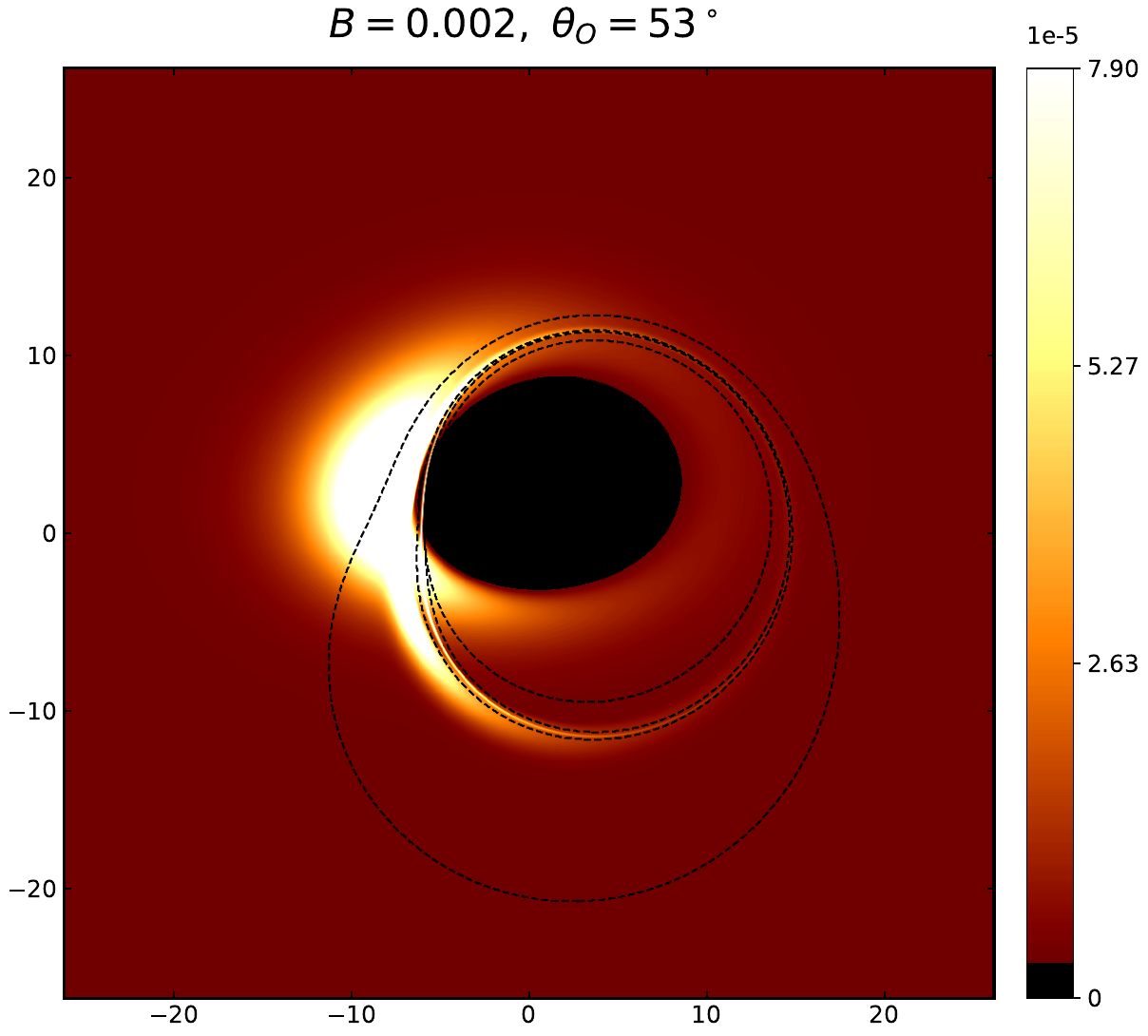} &
        \includegraphics[width=0.23\textwidth]{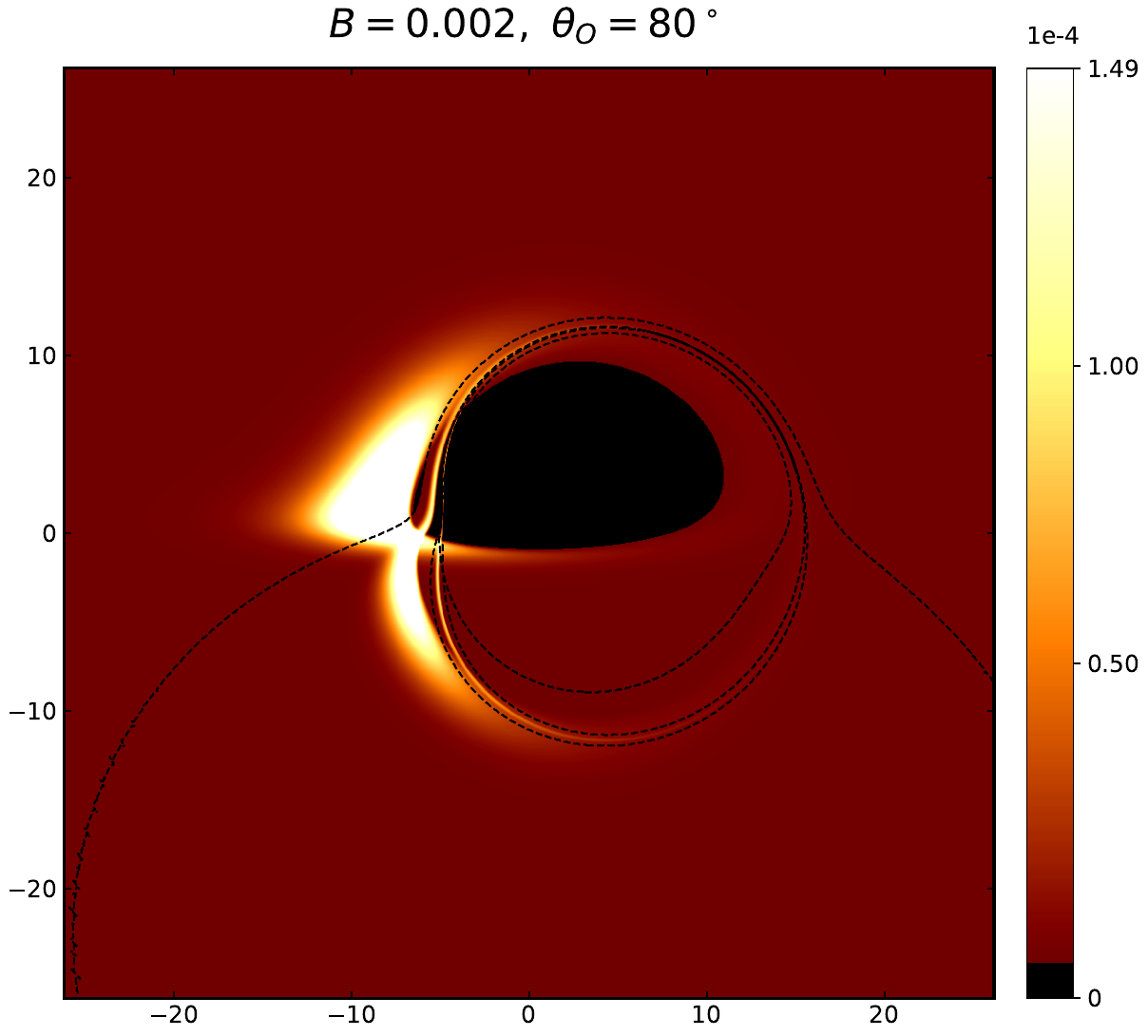} \\
        \includegraphics[width=0.23\textwidth]{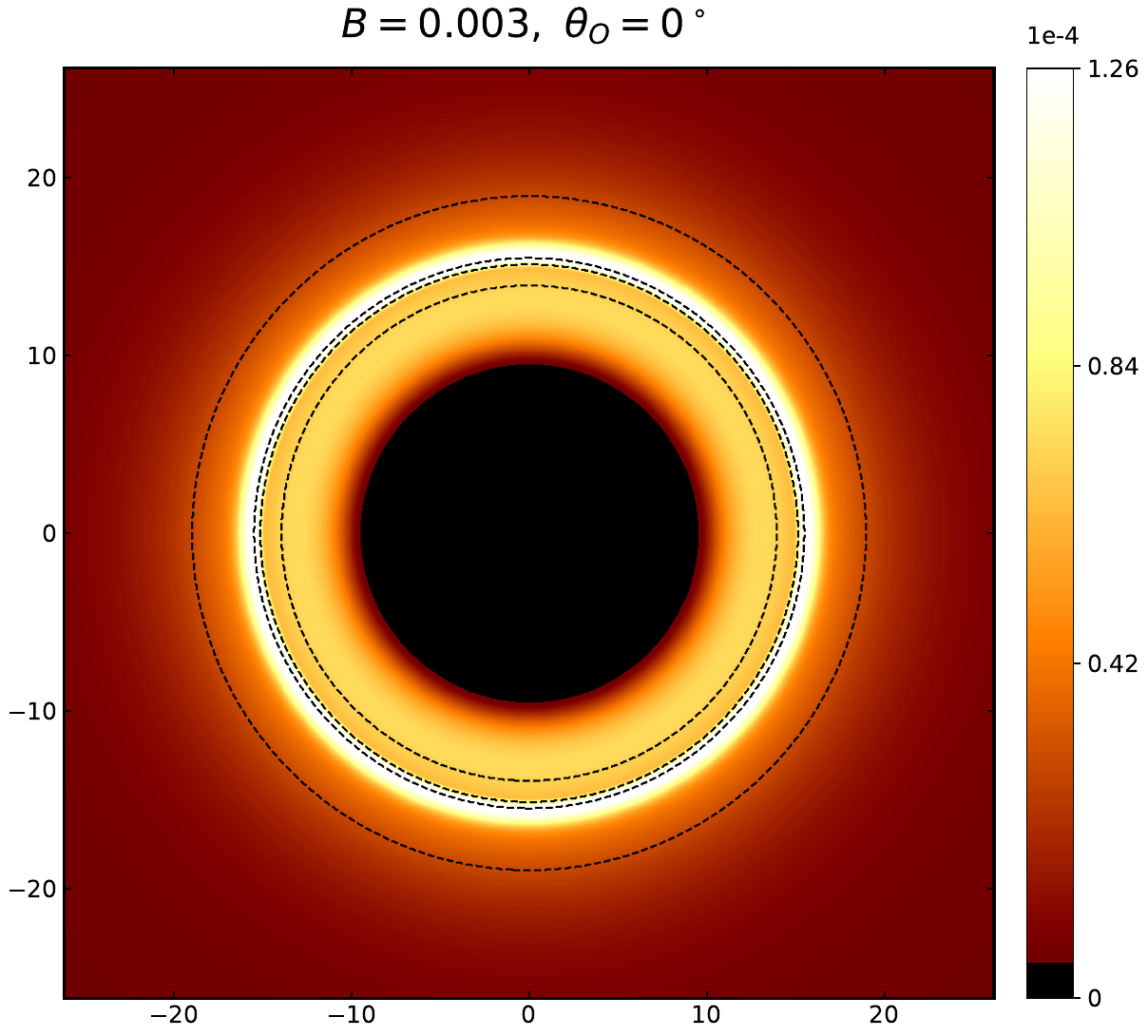} &
        \includegraphics[width=0.23\textwidth]{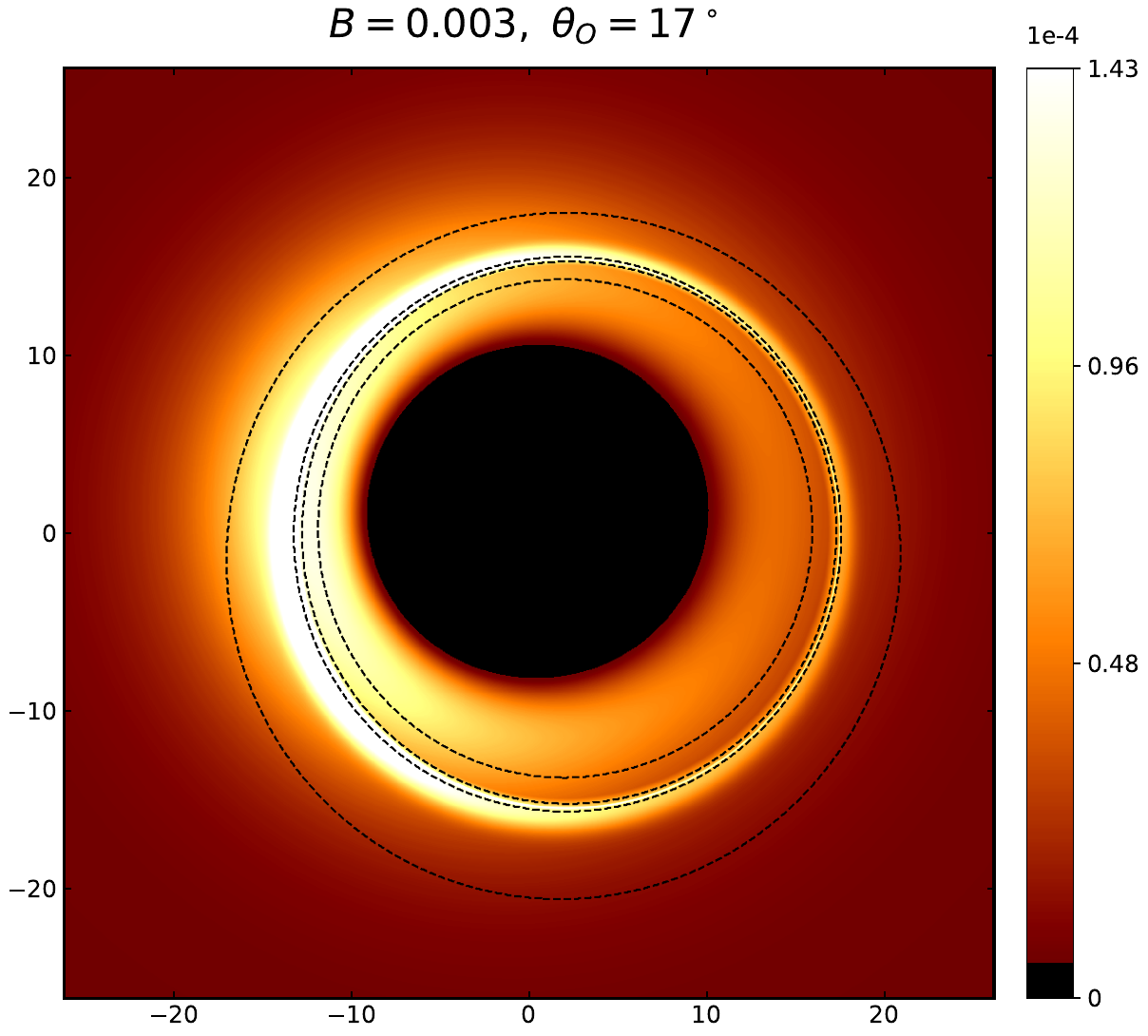} &
        \includegraphics[width=0.23\textwidth]{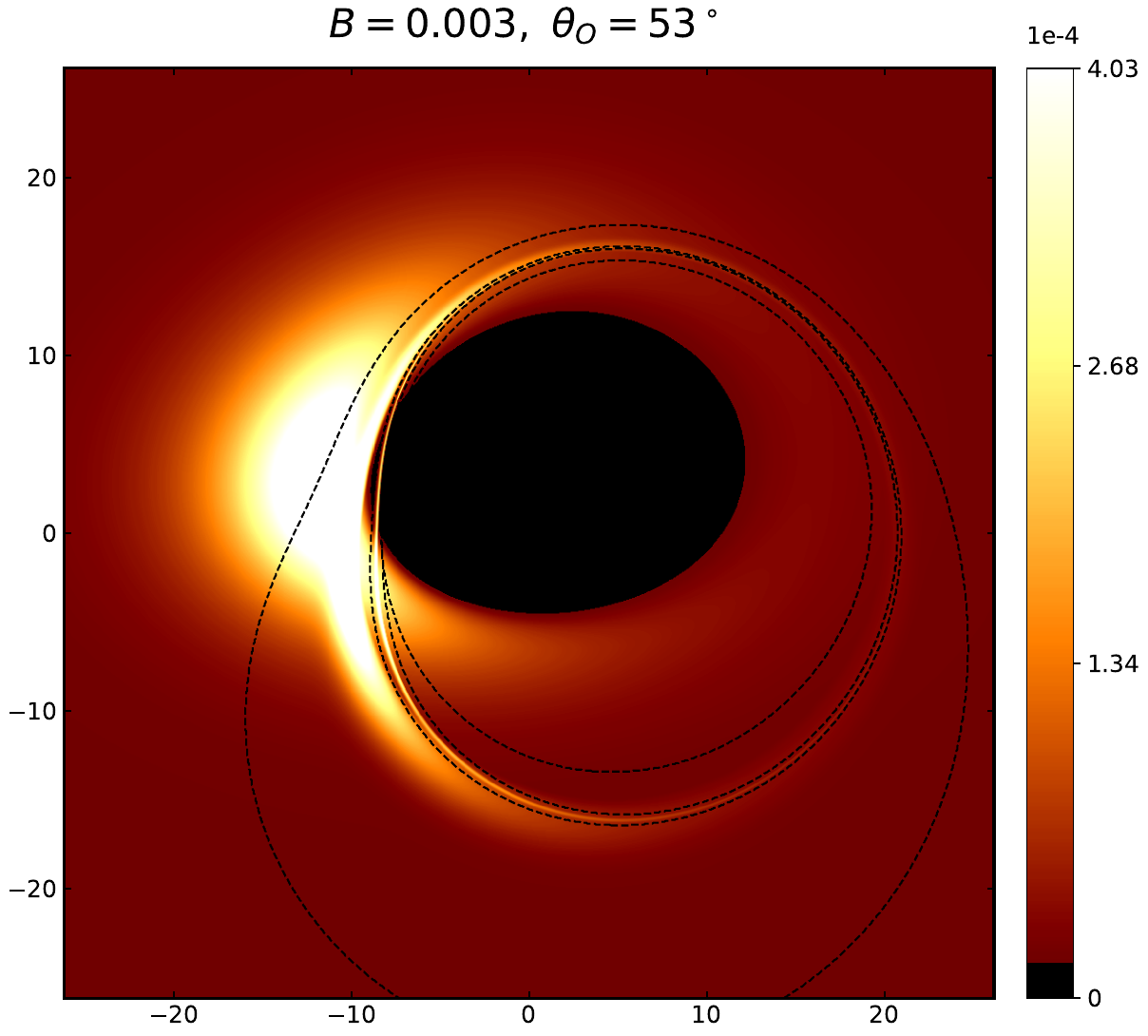} &
        \includegraphics[width=0.23\textwidth]{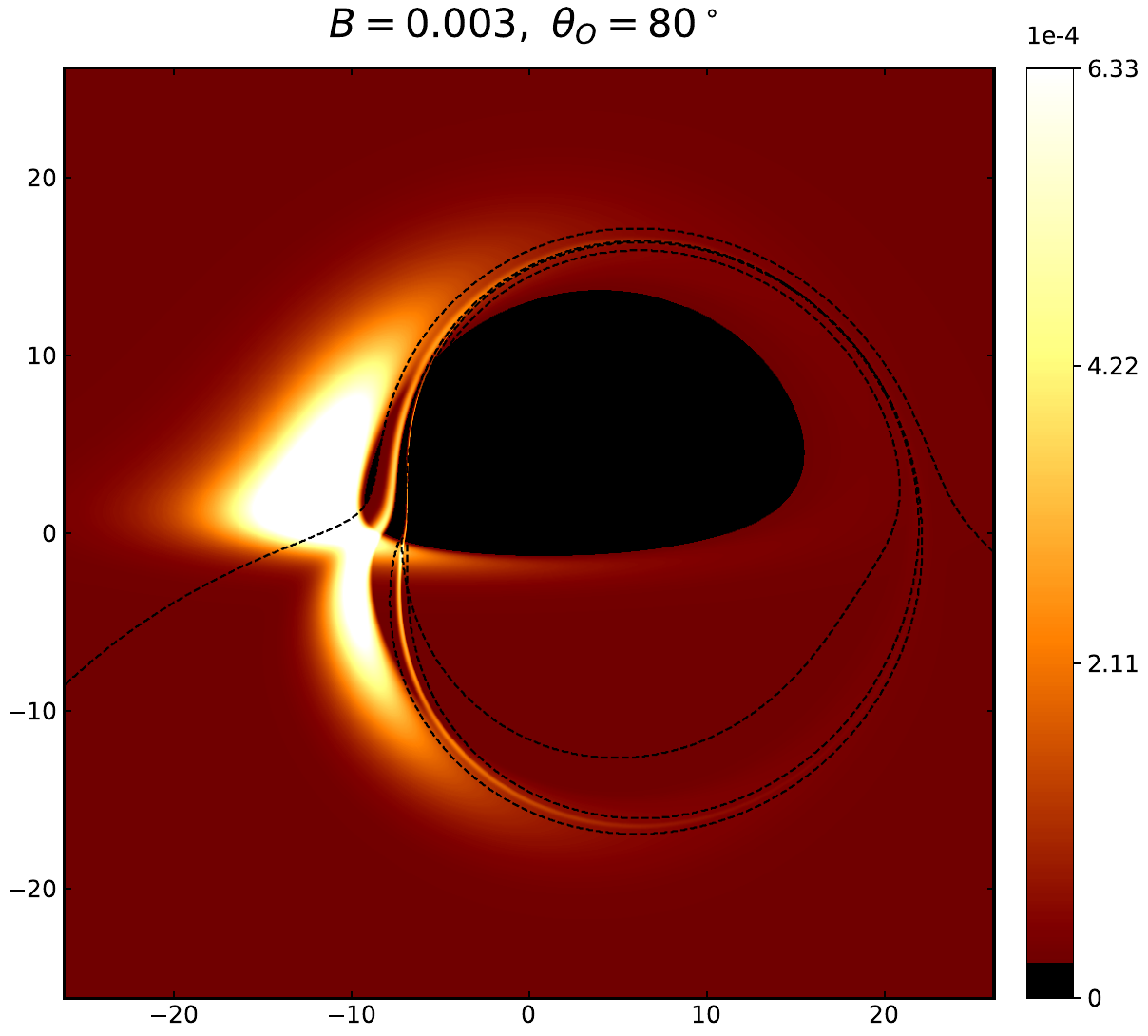} \\
    \end{tabular}
    \caption{Specific-intensity maps of prograde disks with fixed spin $a=0.998$, varying magnetic parameter $B$ from top to bottom and observer inclination $\theta_O$ from left to right. The dashed curves denote the boundaries between the direct image ($n=0$), the lensing-ring contribution ($n=1$), and the higher-order photon-ring subimages ($n\geq 2$). The intensity distribution reflects the combined effects of lensing, redshift, and the magnetic-field-dependent emissivity.}
    \label{fig:flux_varyB}
\end{figure*}

The corresponding specific-intensity maps are shown in Fig.~\ref{fig:flux_varyB}. These images combine the lensing geometry, the redshift factor, and the magnetic-field-dependent emissivity in Eq.~(\ref{eq:final_emissivity}). For the adopted emissivity prescription, the emission is concentrated near the inner part of the disk and then decreases rapidly with radius. Increasing $B$ enhances the effective local magnetic-field factor in the emissivity model and makes the exponential suppression less severe over the displayed radial range. Consequently, the emitting region becomes broader and the overall intensity increases. At large inclination, the Doppler factor further enhances the approaching side of the disk, producing a crescent-like brightness distribution.

\subsubsection{Effect of the Black Hole Spin $a$}

\begin{figure*}[htbp]
    \centering
    \begin{tabular}{cccc}
        \includegraphics[width=0.23\textwidth]{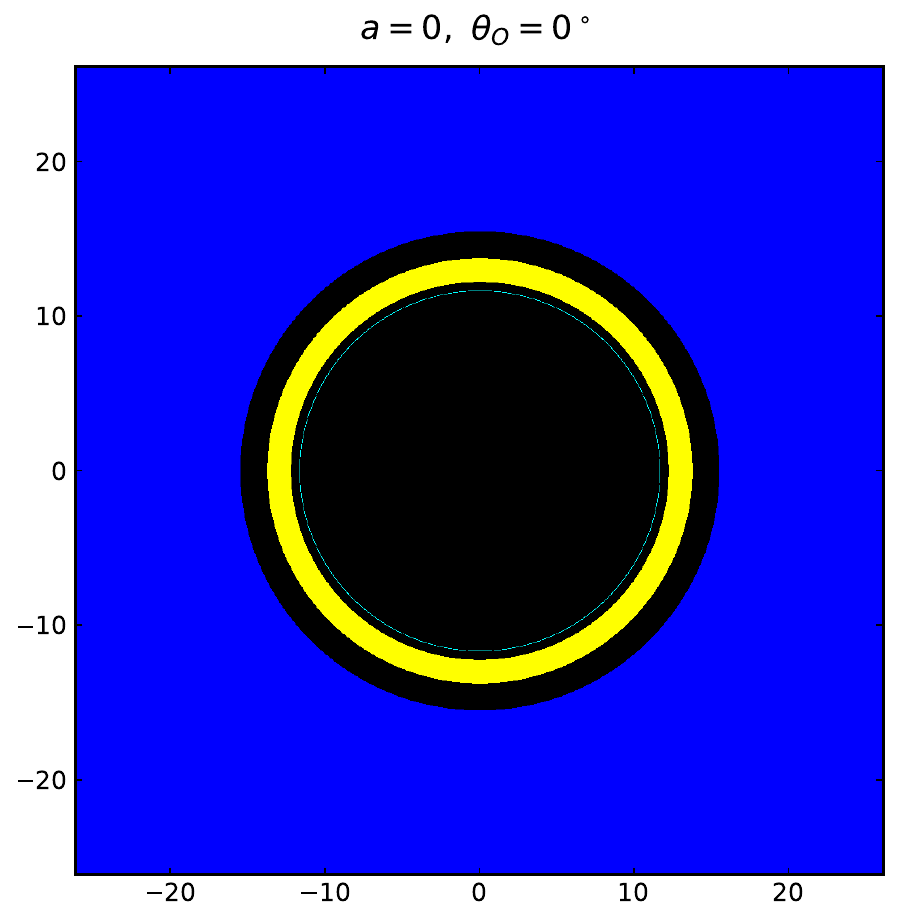} &
        \includegraphics[width=0.23\textwidth]{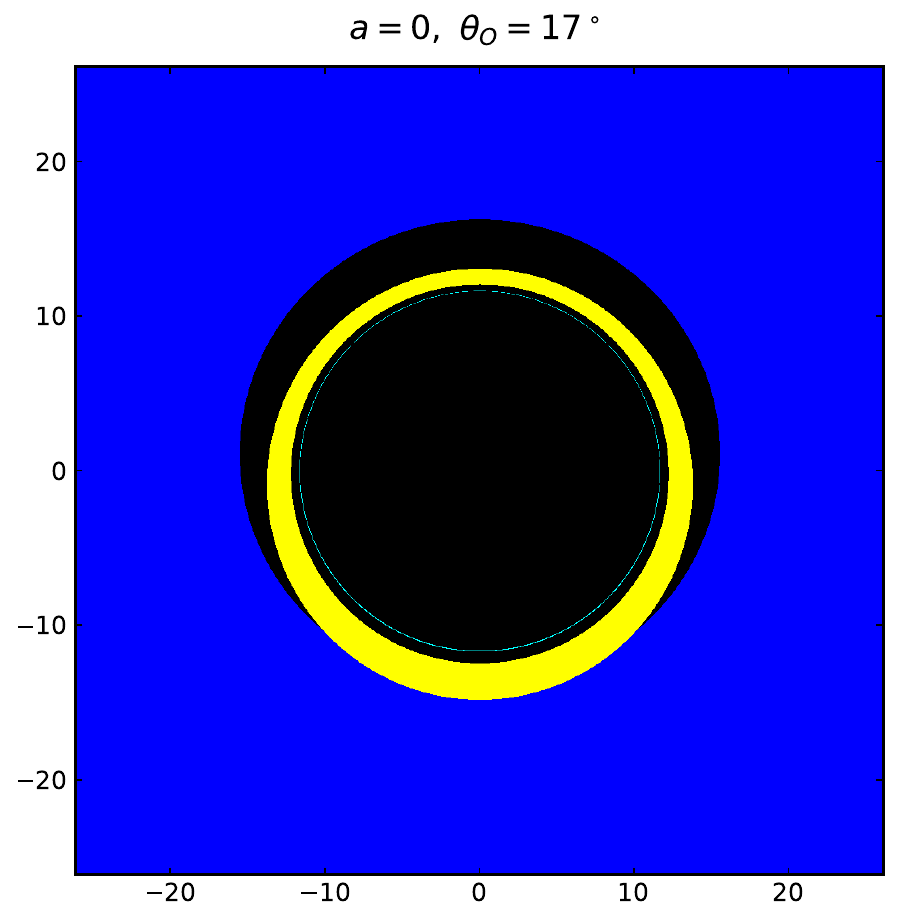} &
        \includegraphics[width=0.23\textwidth]{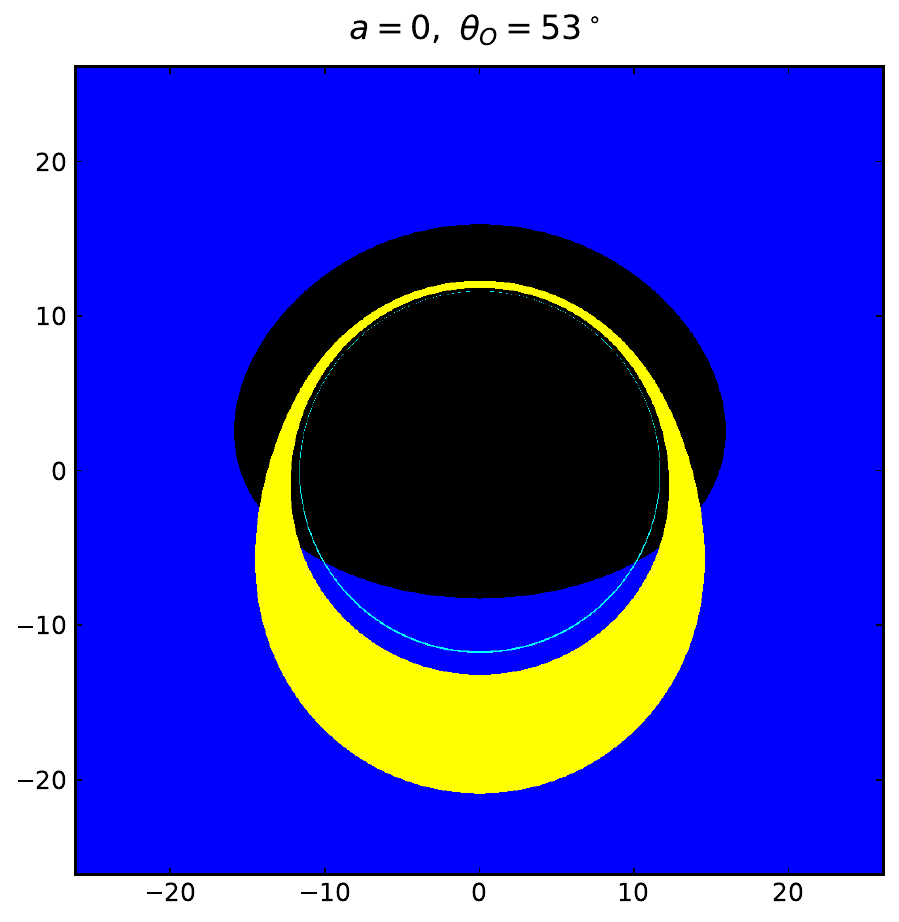} &
        \includegraphics[width=0.23\textwidth]{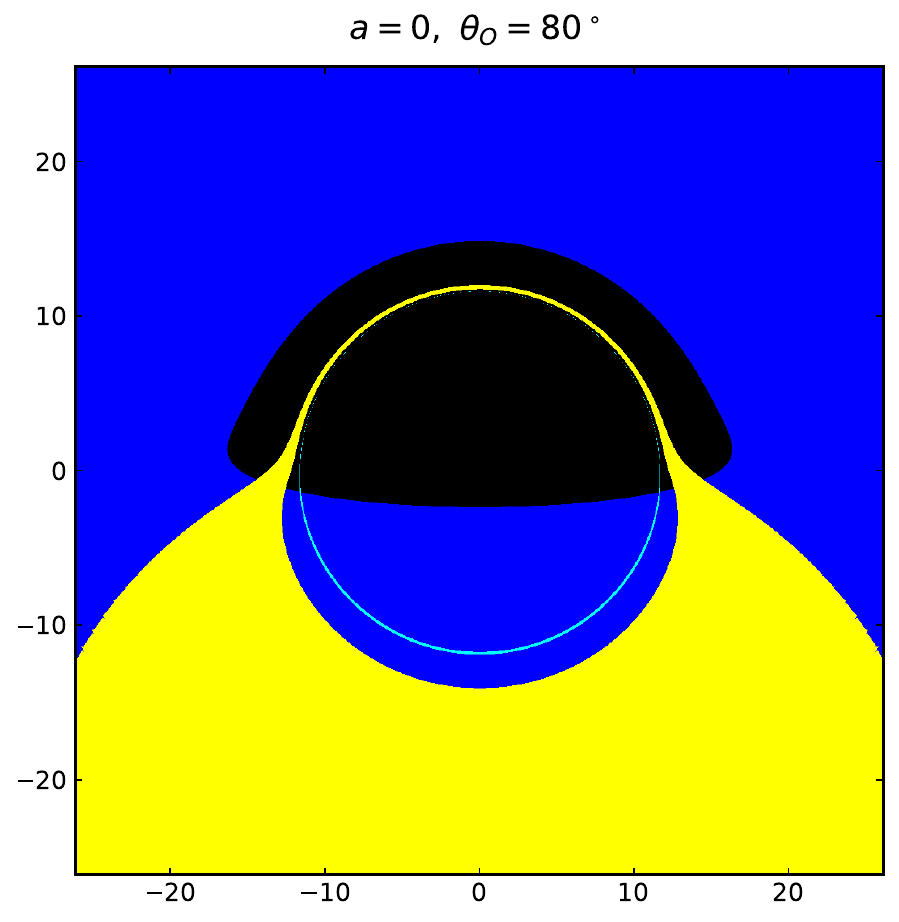} \\
        \includegraphics[width=0.23\textwidth]{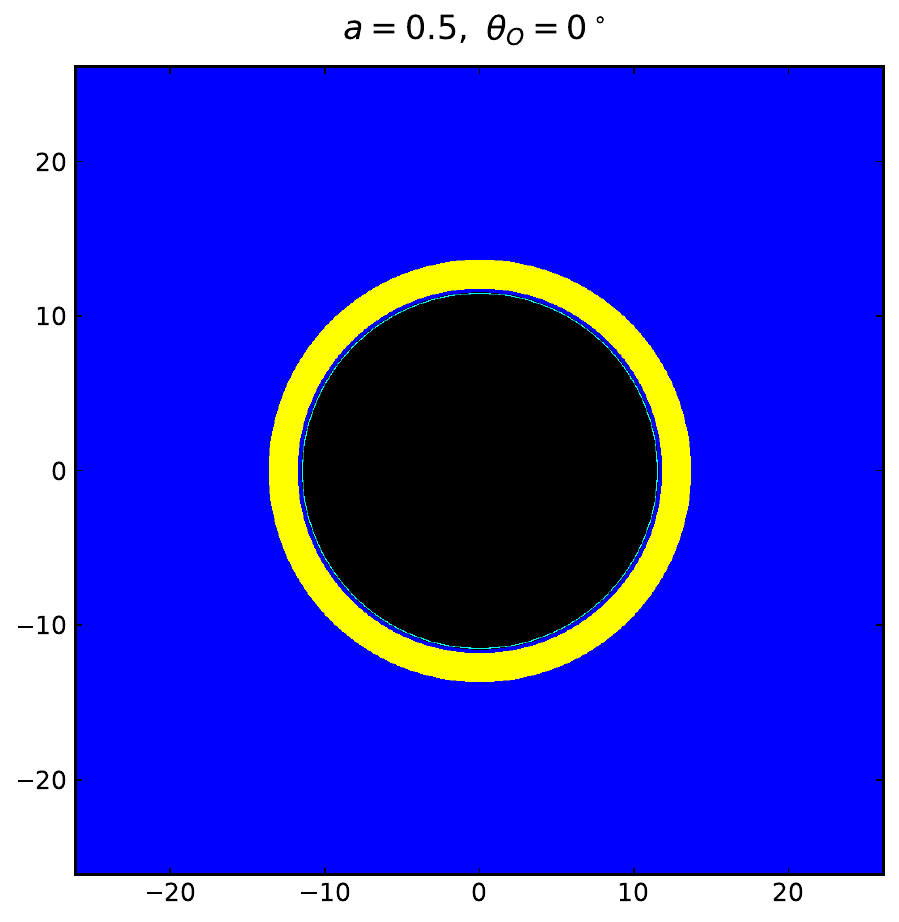} &
        \includegraphics[width=0.23\textwidth]{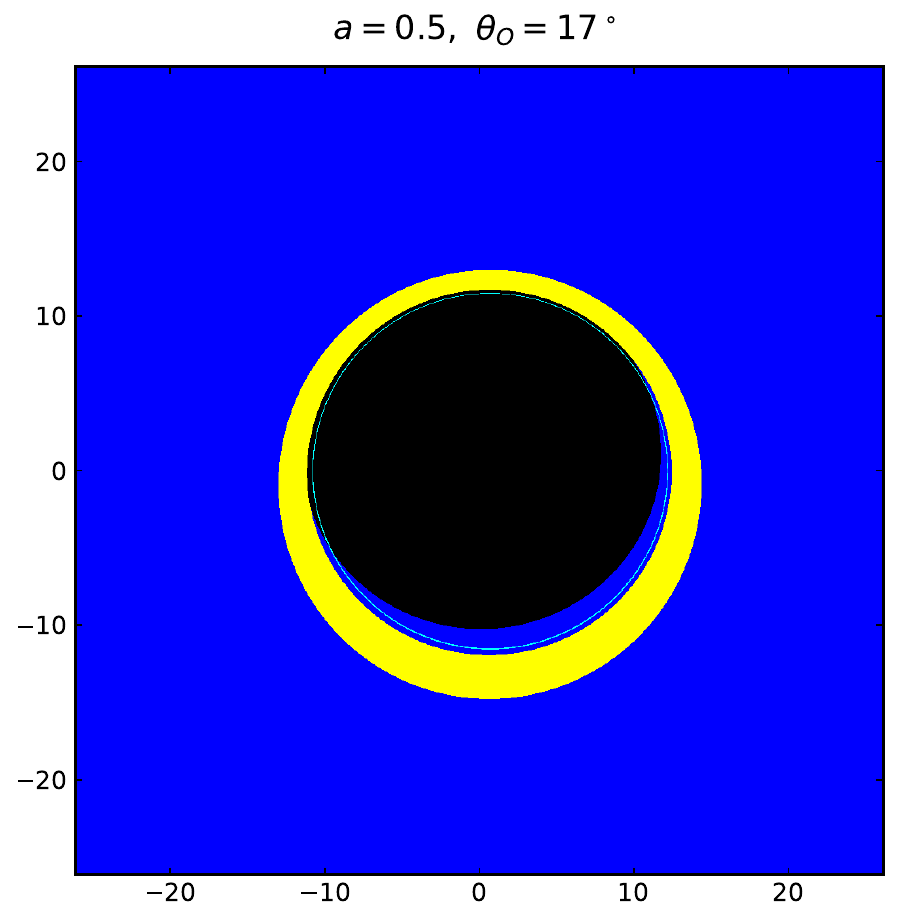} &
        \includegraphics[width=0.23\textwidth]{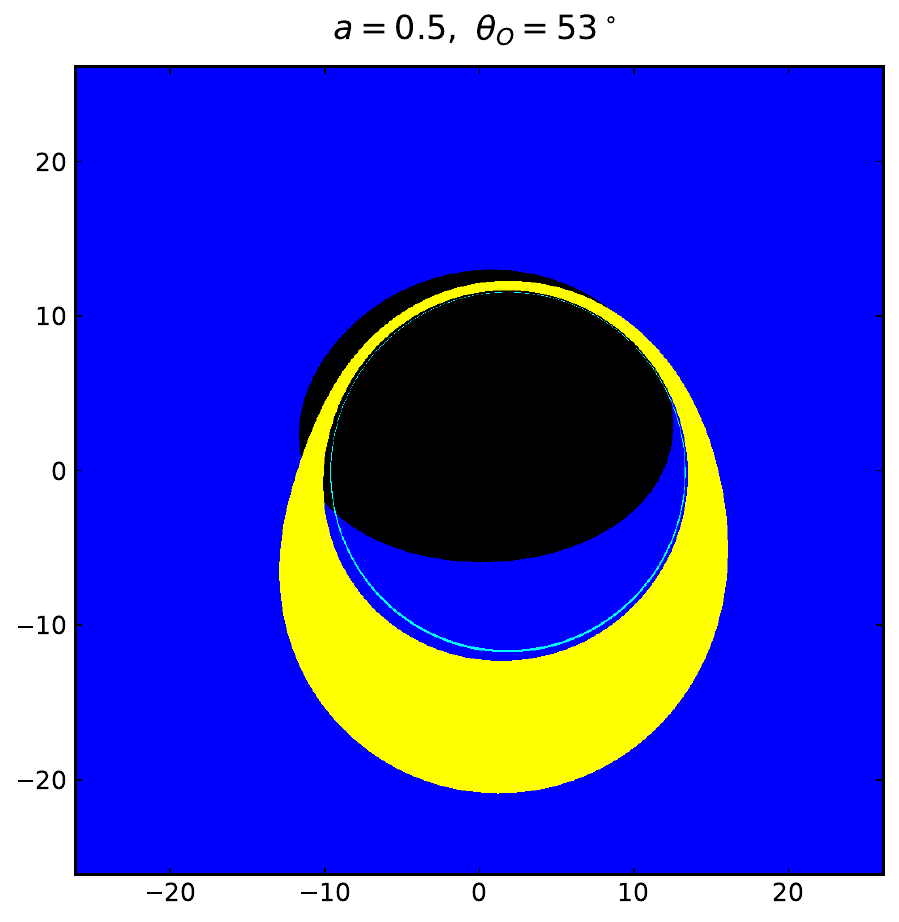} &
        \includegraphics[width=0.23\textwidth]{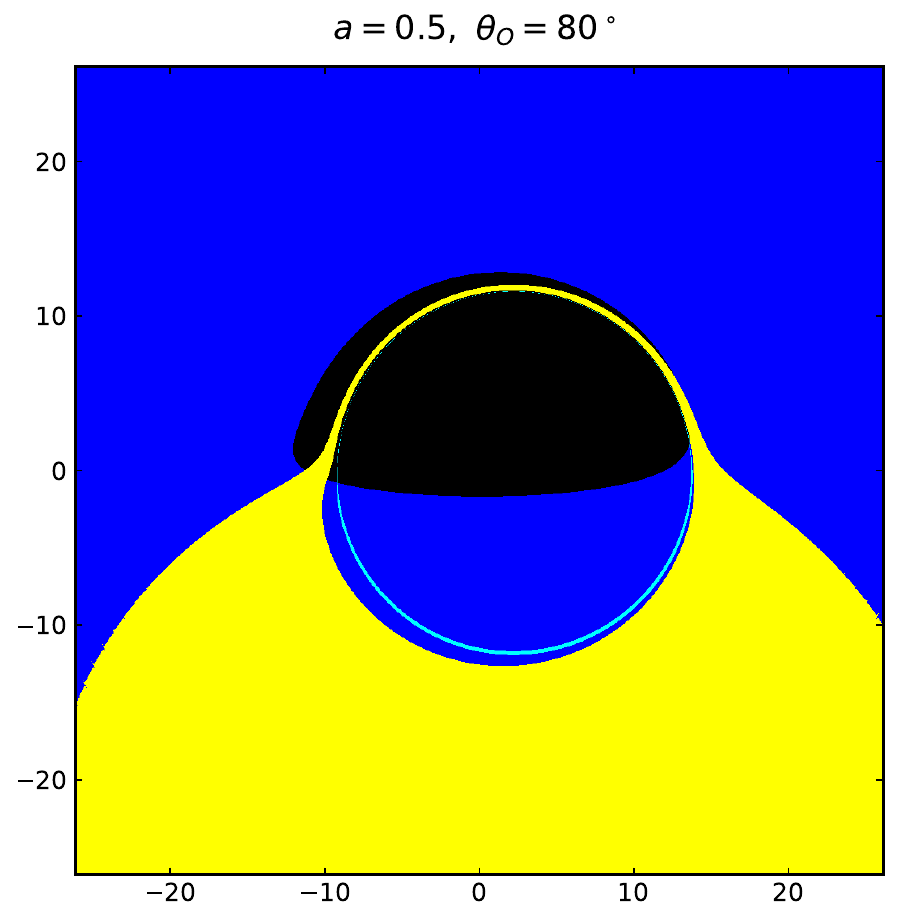} \\
        \includegraphics[width=0.23\textwidth]{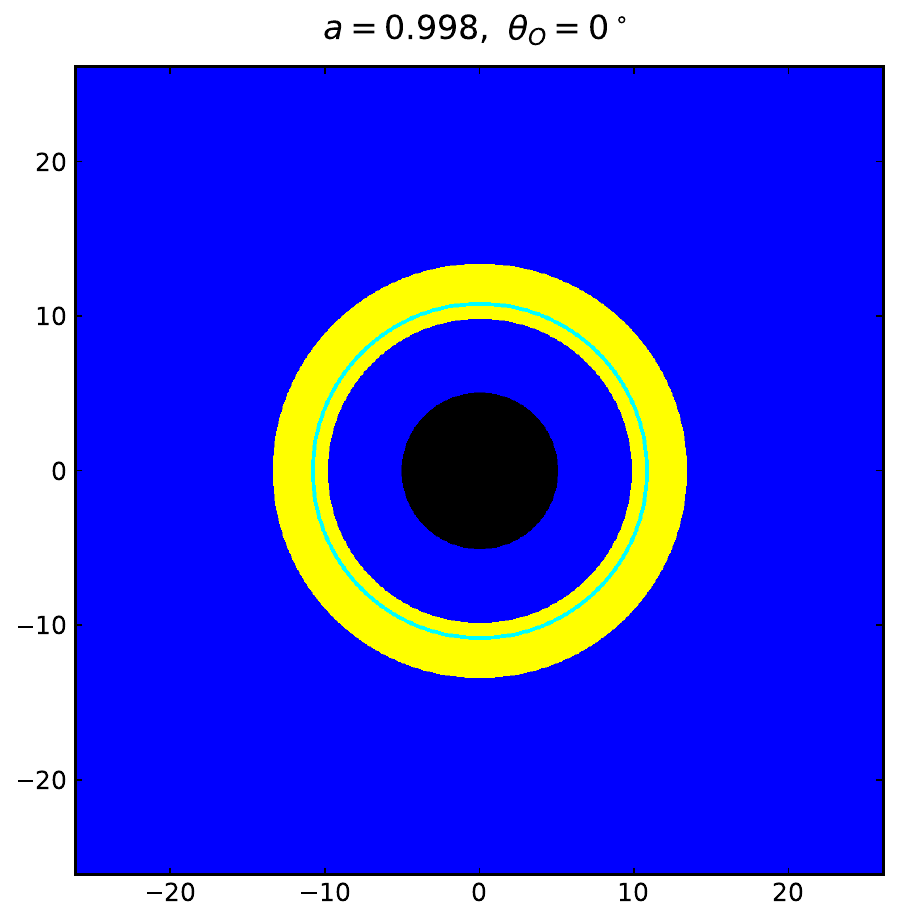} &
        \includegraphics[width=0.23\textwidth]{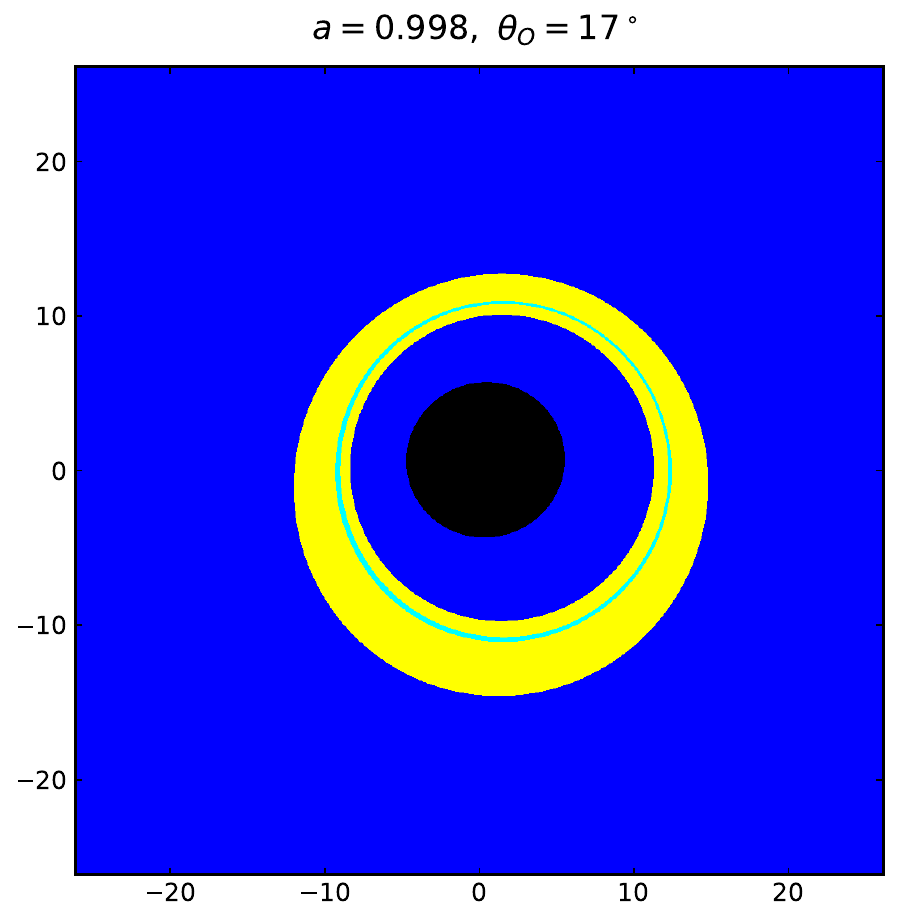} &
        \includegraphics[width=0.23\textwidth]{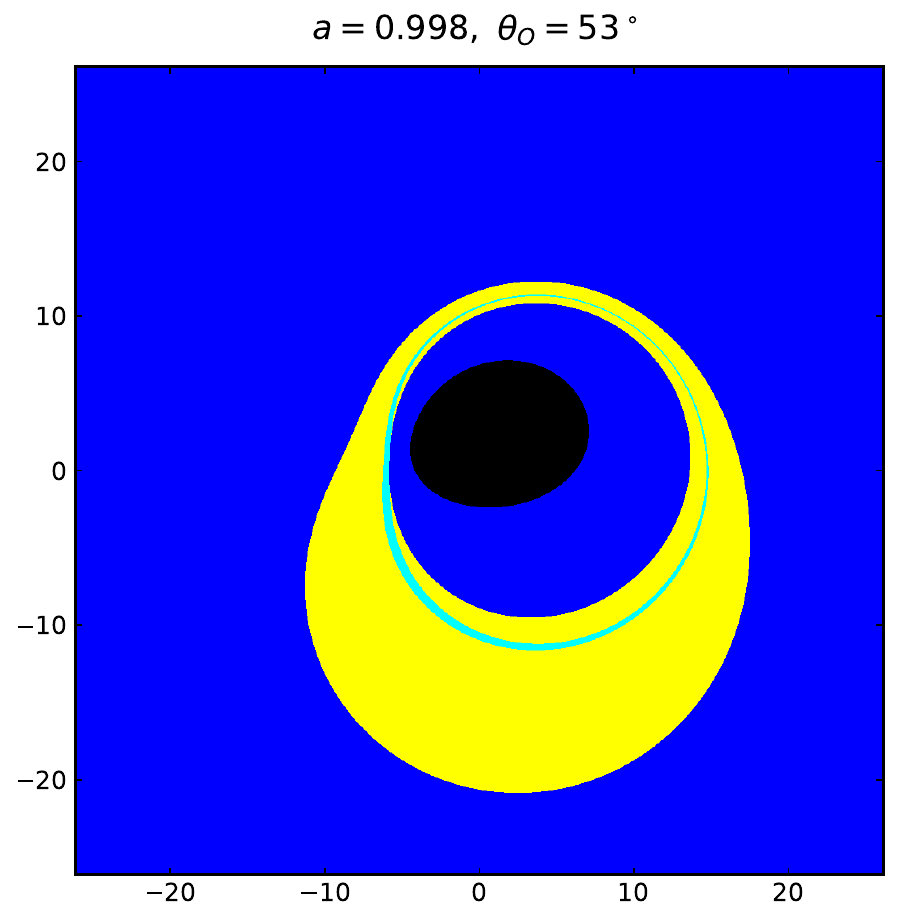} &
        \includegraphics[width=0.23\textwidth]{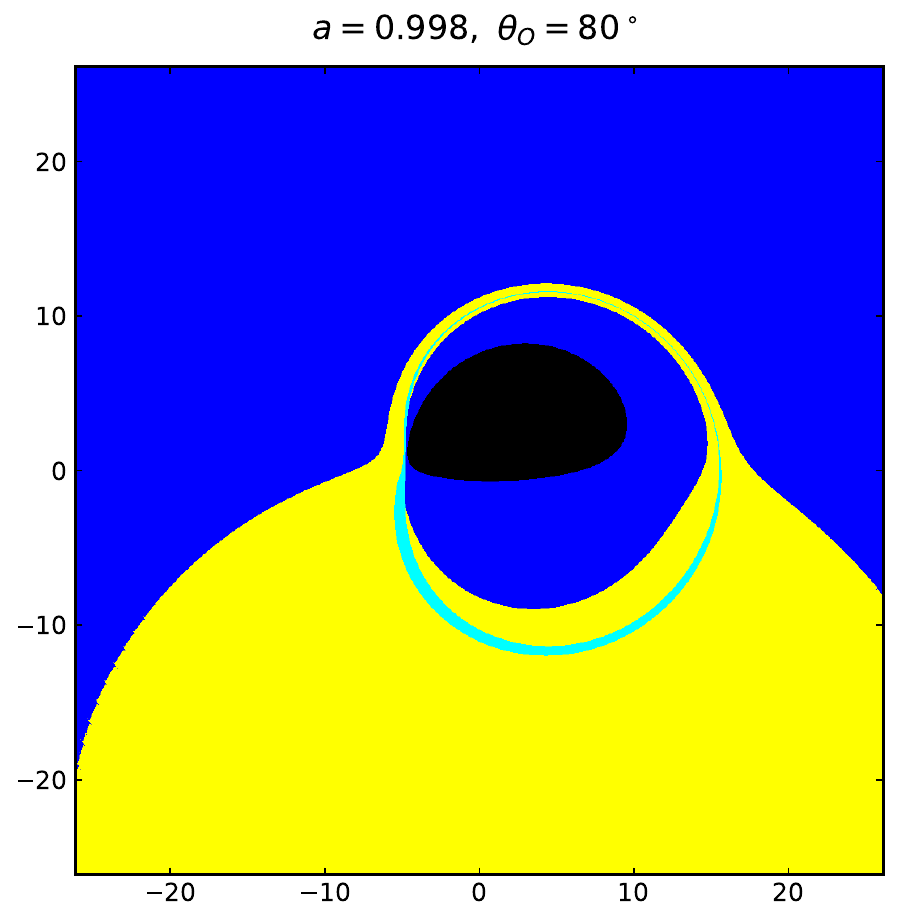} \\
    \end{tabular}
    \caption{Ray-classification maps for prograde disks with fixed $B=0.002$. Rows correspond to $a=0,0.5,0.998$ from top to bottom, and columns correspond to different observer inclinations $\theta_O$. The spin changes both the lensing structure and the effective inner boundary of the emitting disk, thereby modifying the size and shape of the direct, lensing-ring, and higher-order subimage regions.}
    \label{fig:class_varya}
\end{figure*}

\begin{figure*}[htbp]
    \centering
    \begin{tabular}{cccc}
        \includegraphics[width=0.23\textwidth]{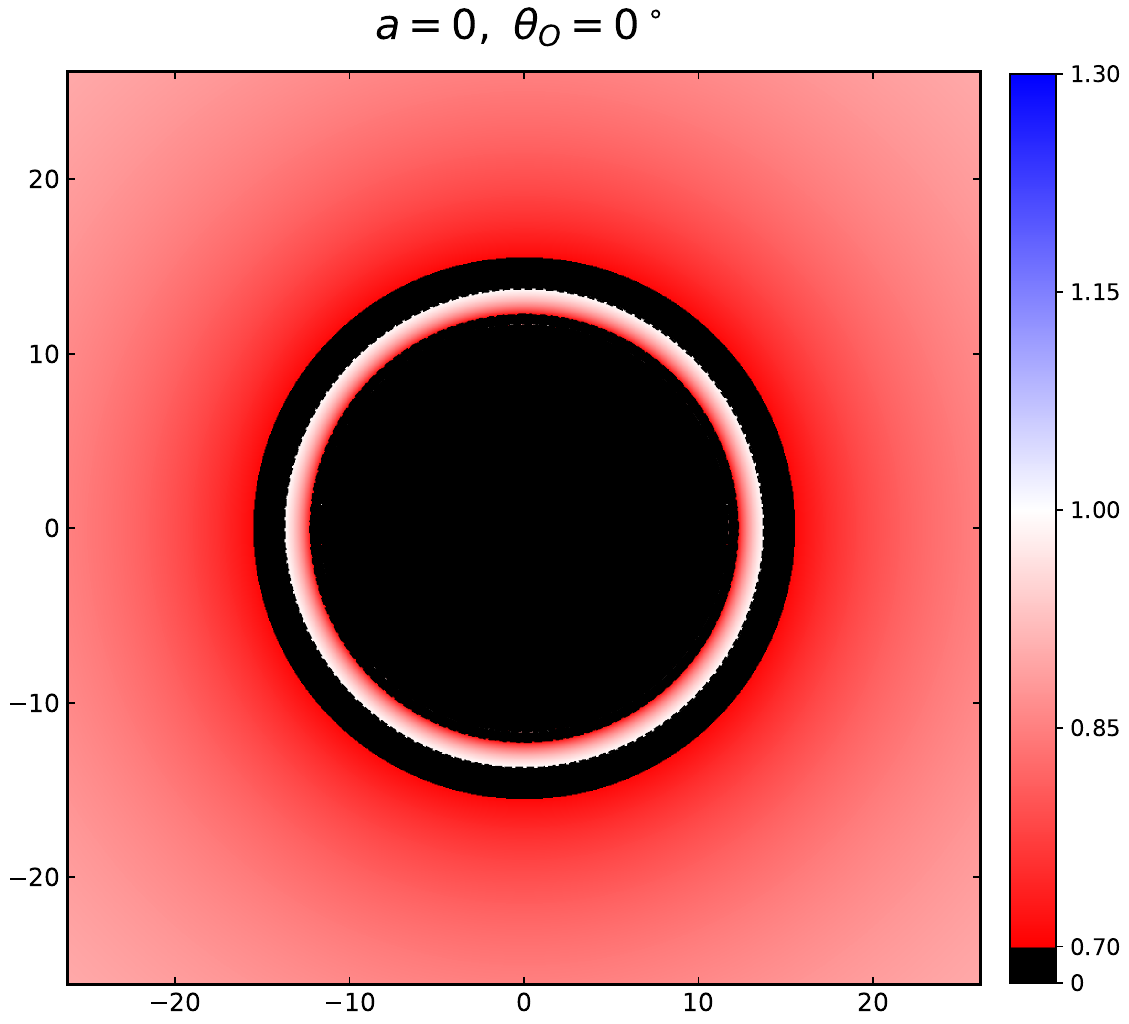} &
        \includegraphics[width=0.23\textwidth]{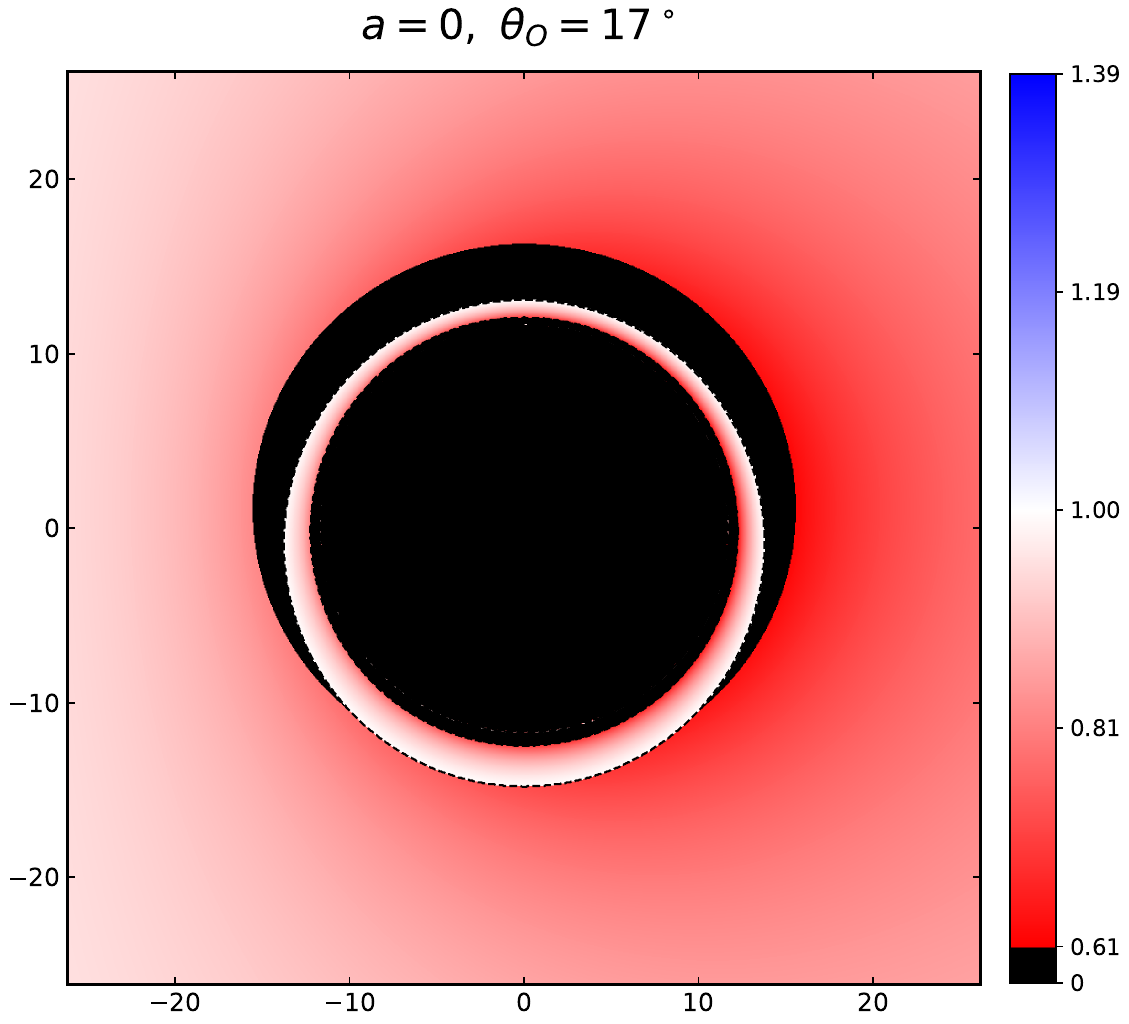} &
        \includegraphics[width=0.23\textwidth]{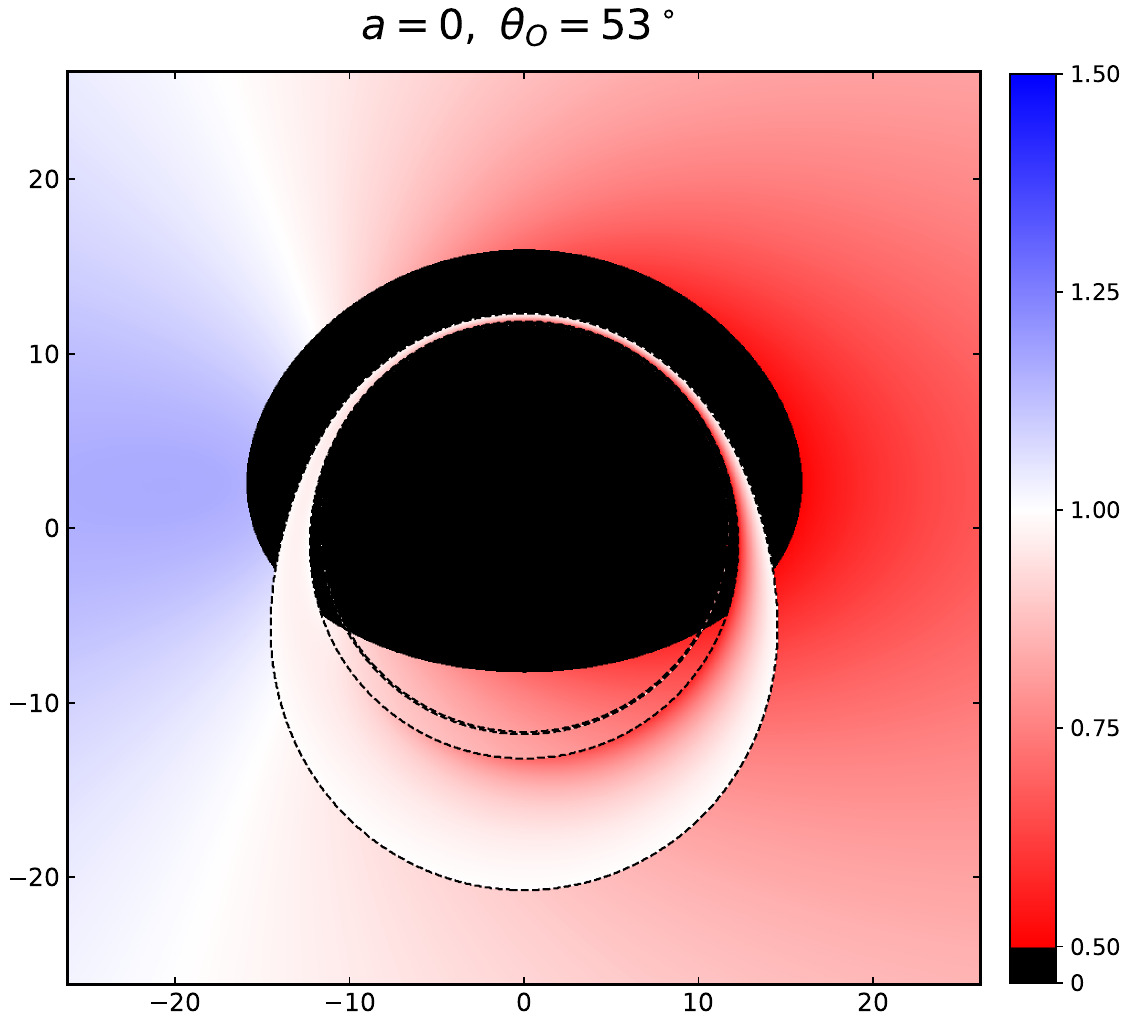} &
        \includegraphics[width=0.23\textwidth]{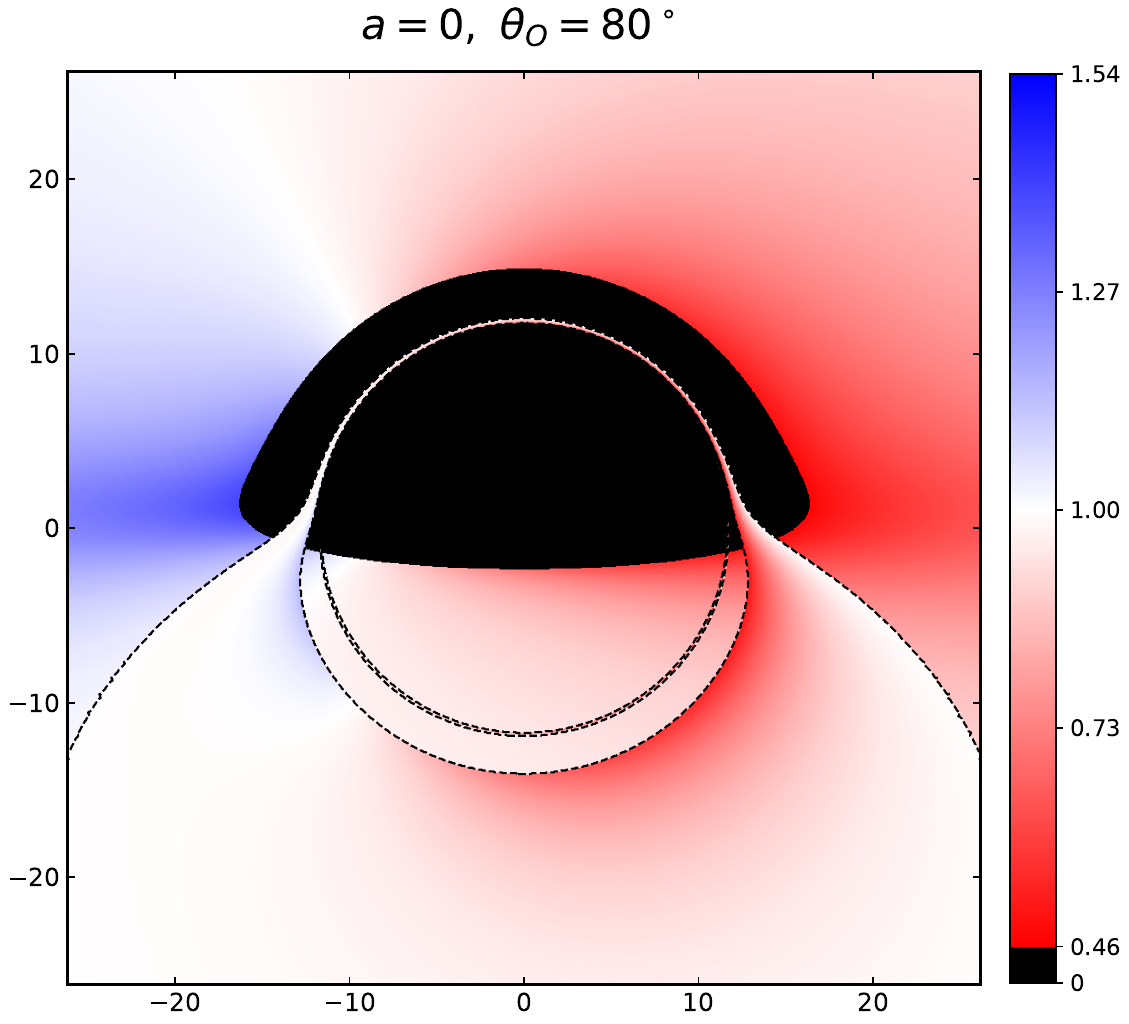} \\
        \includegraphics[width=0.23\textwidth]{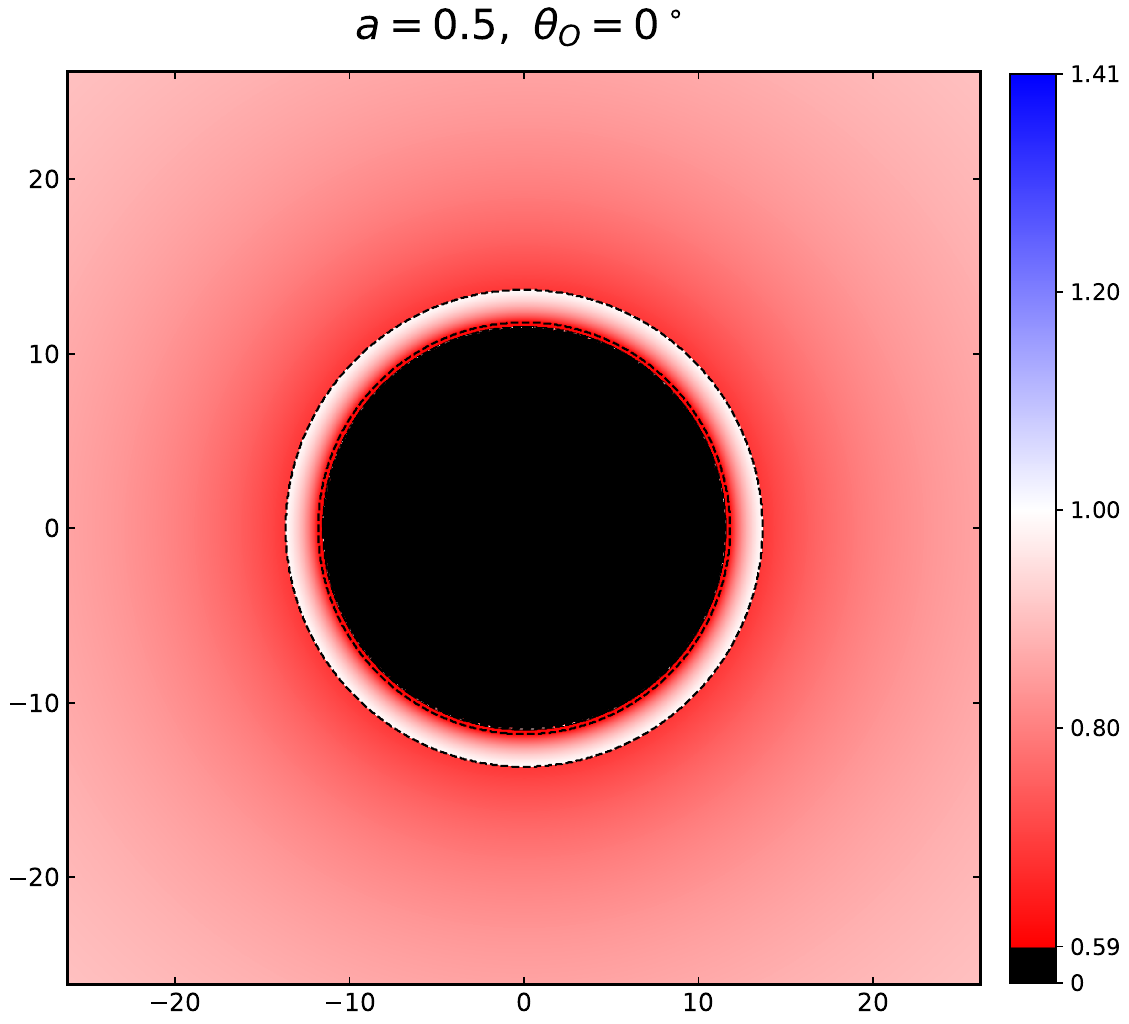} &
        \includegraphics[width=0.23\textwidth]{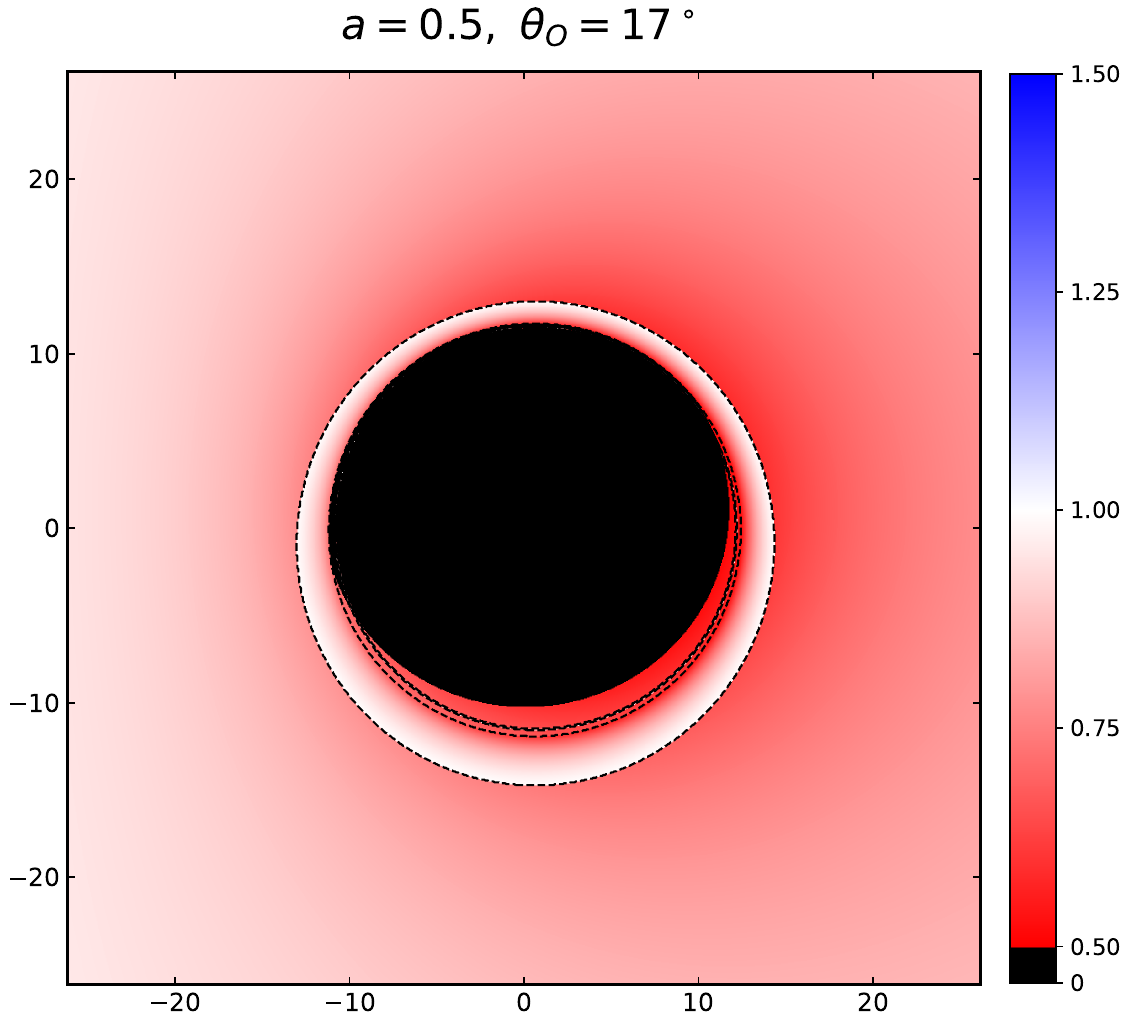} &
        \includegraphics[width=0.23\textwidth]{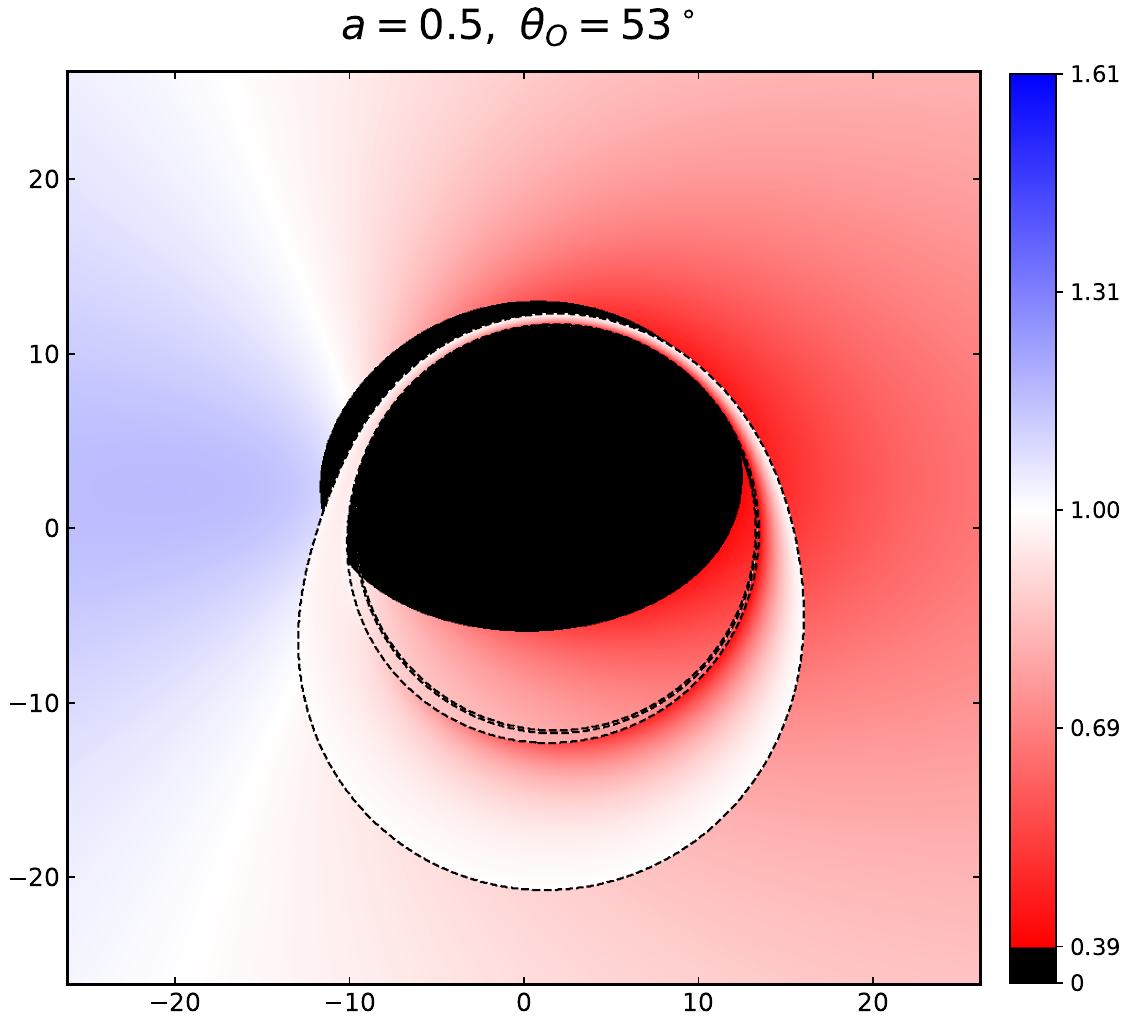} &
        \includegraphics[width=0.23\textwidth]{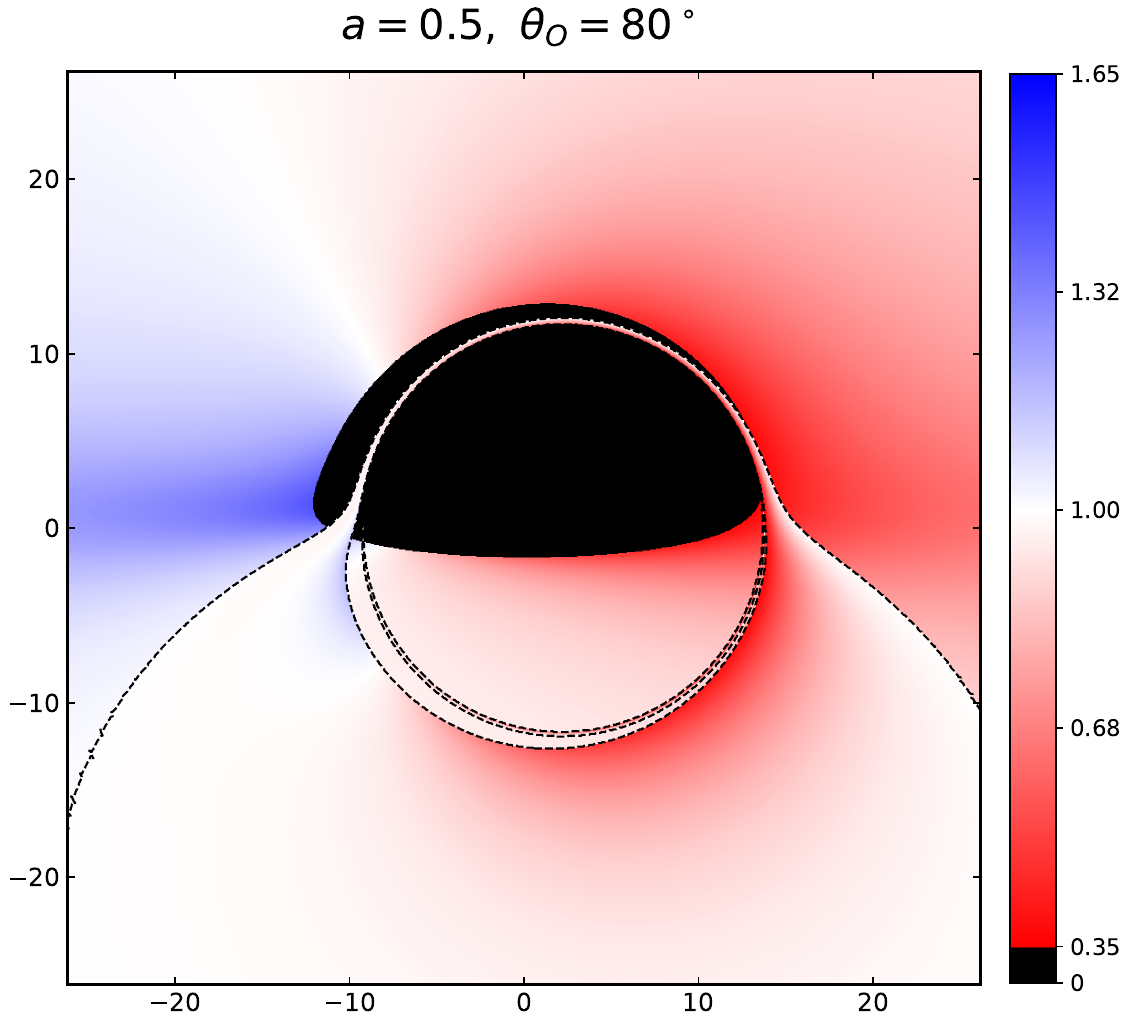} \\
        \includegraphics[width=0.23\textwidth]{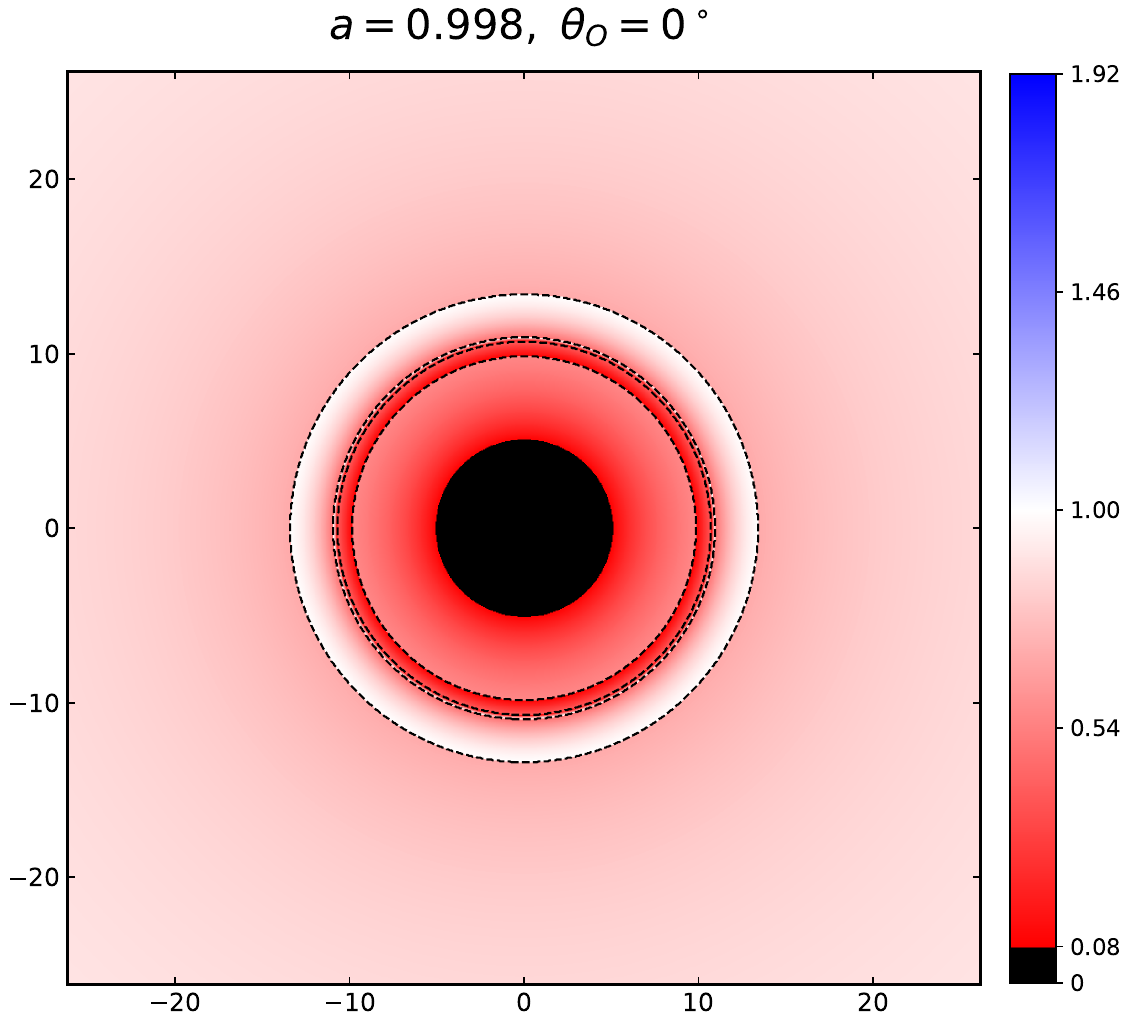} &
        \includegraphics[width=0.23\textwidth]{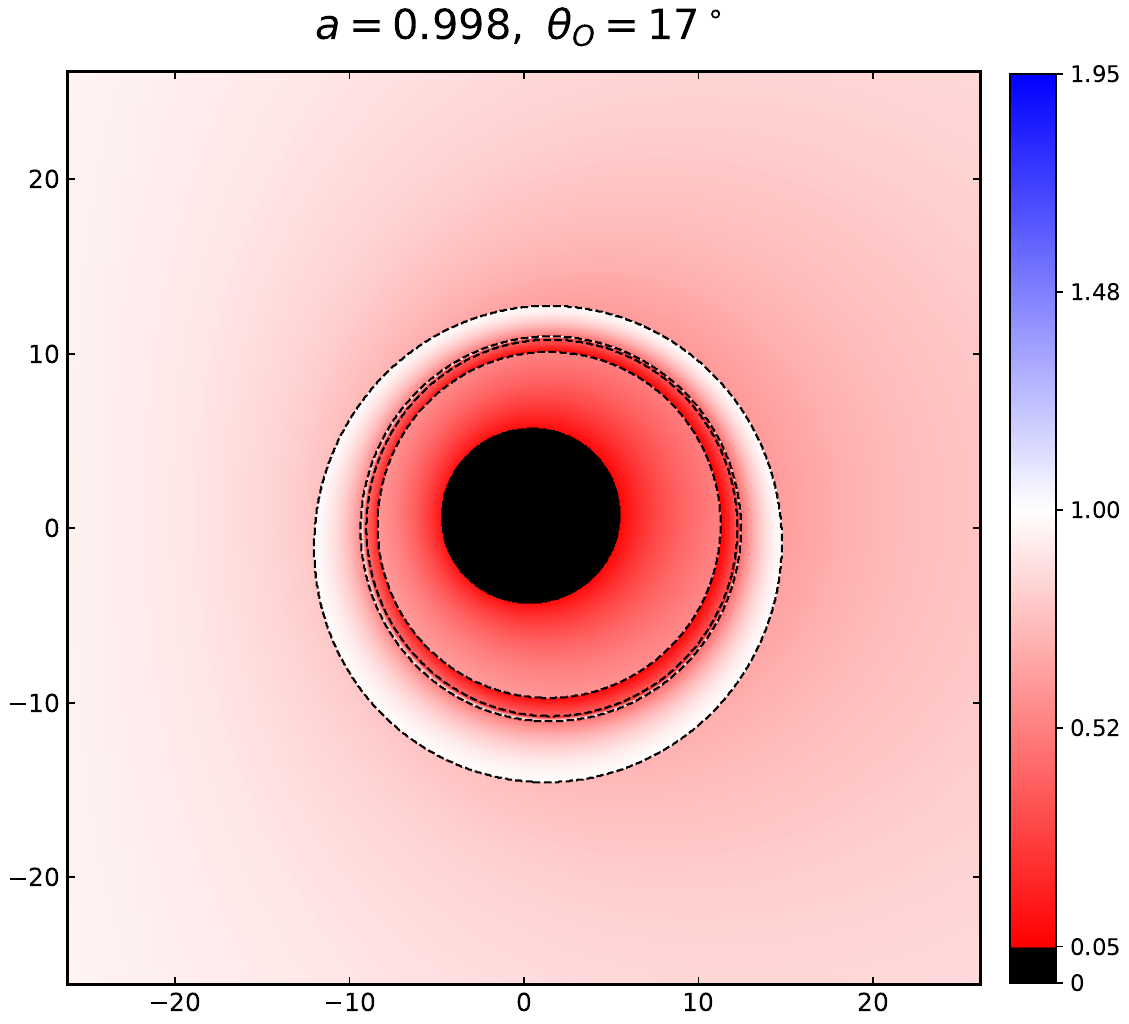} &
        \includegraphics[width=0.23\textwidth]{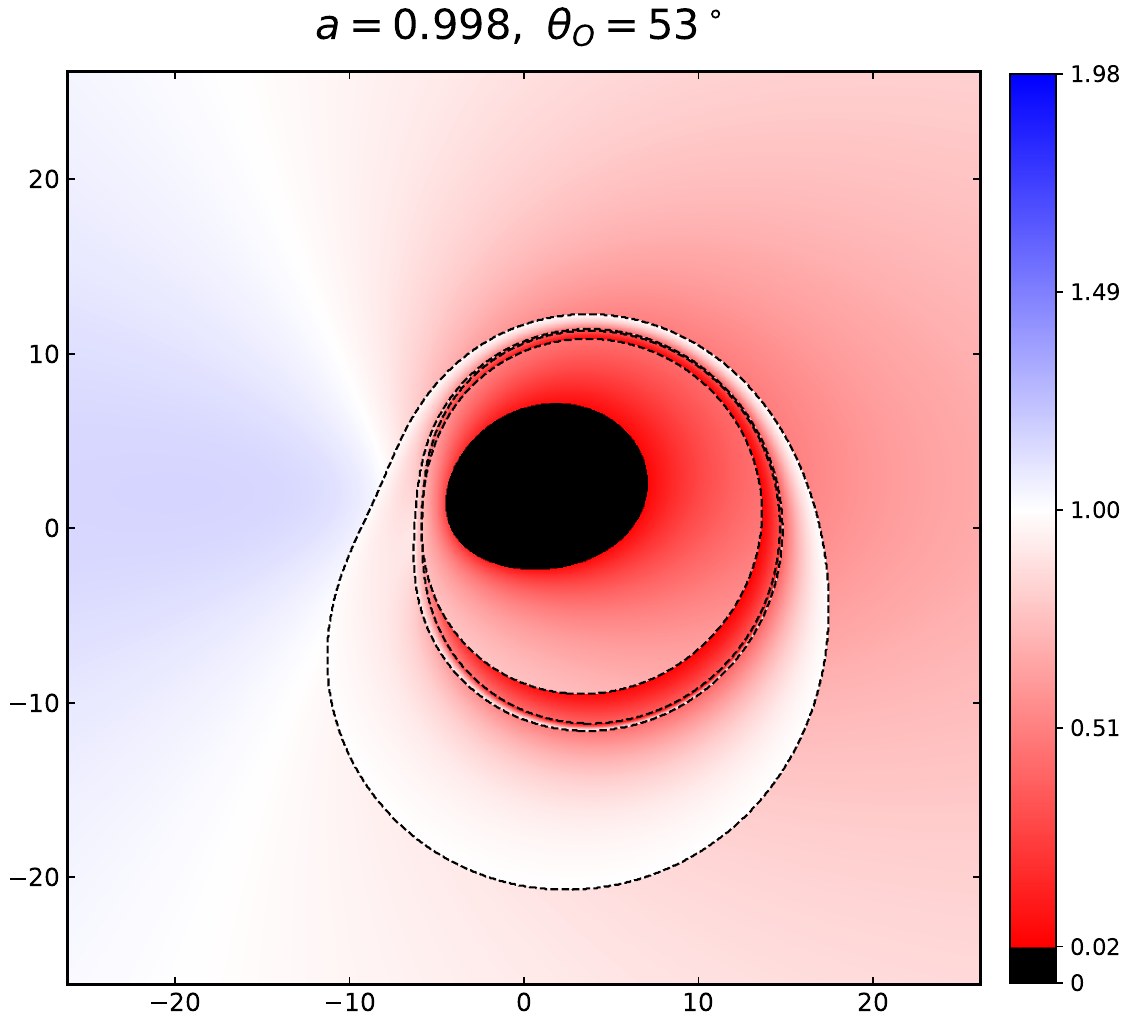} &
        \includegraphics[width=0.23\textwidth]{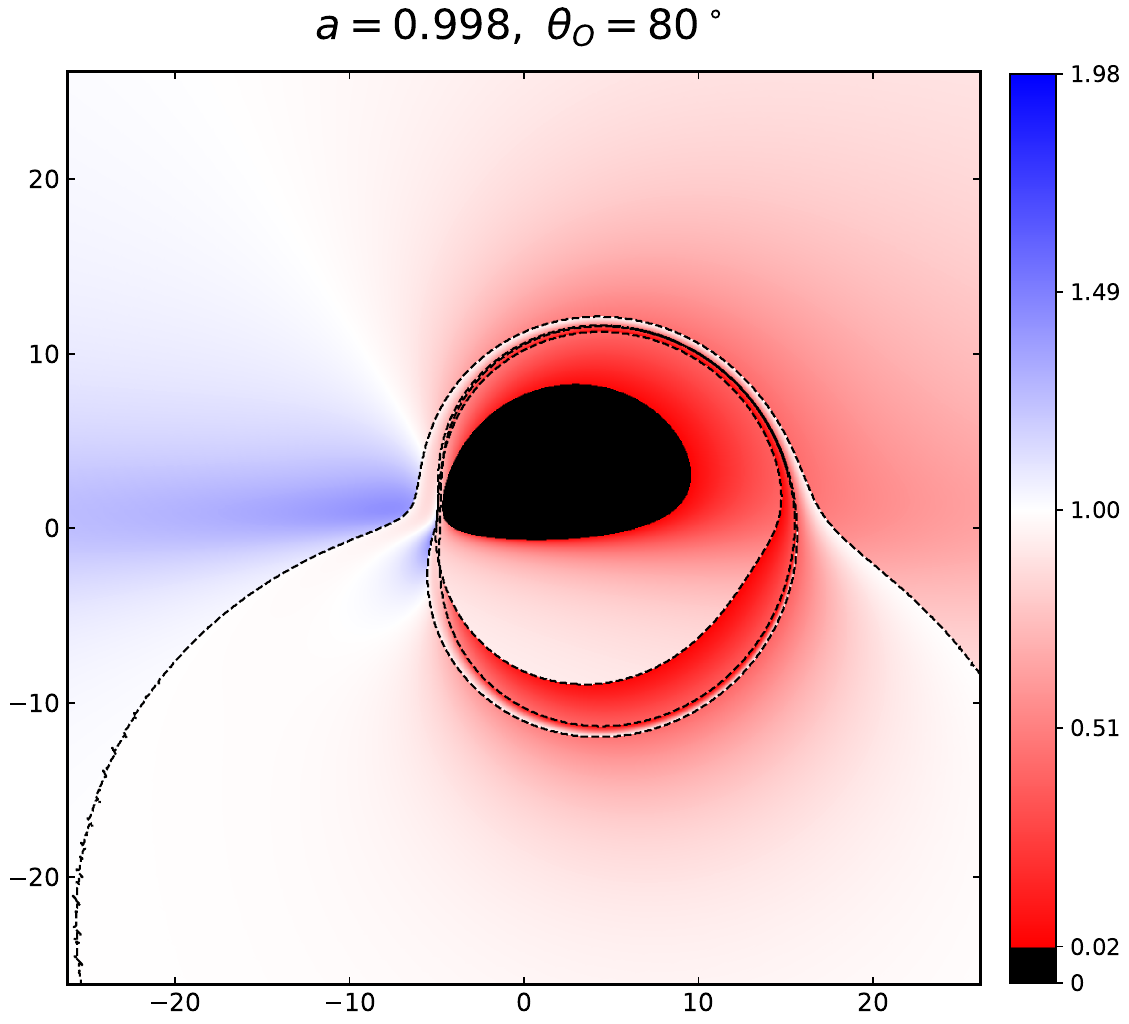} \\
    \end{tabular}
    \caption{Redshift-factor maps for prograde disks with fixed $B=0.002$ and varying spin $a$. The rows correspond to $a=0,0.5,0.998$, and the columns correspond to different inclinations $\theta_O$. Increasing the spin changes the orbital velocity of the emitting gas and enhances the Doppler-induced asymmetry, especially for high-inclination observers.}
    \label{fig:redshift_varya}
\end{figure*}

\begin{figure*}[htbp]
    \centering
    \begin{tabular}{cccc}
        \includegraphics[width=0.23\textwidth]{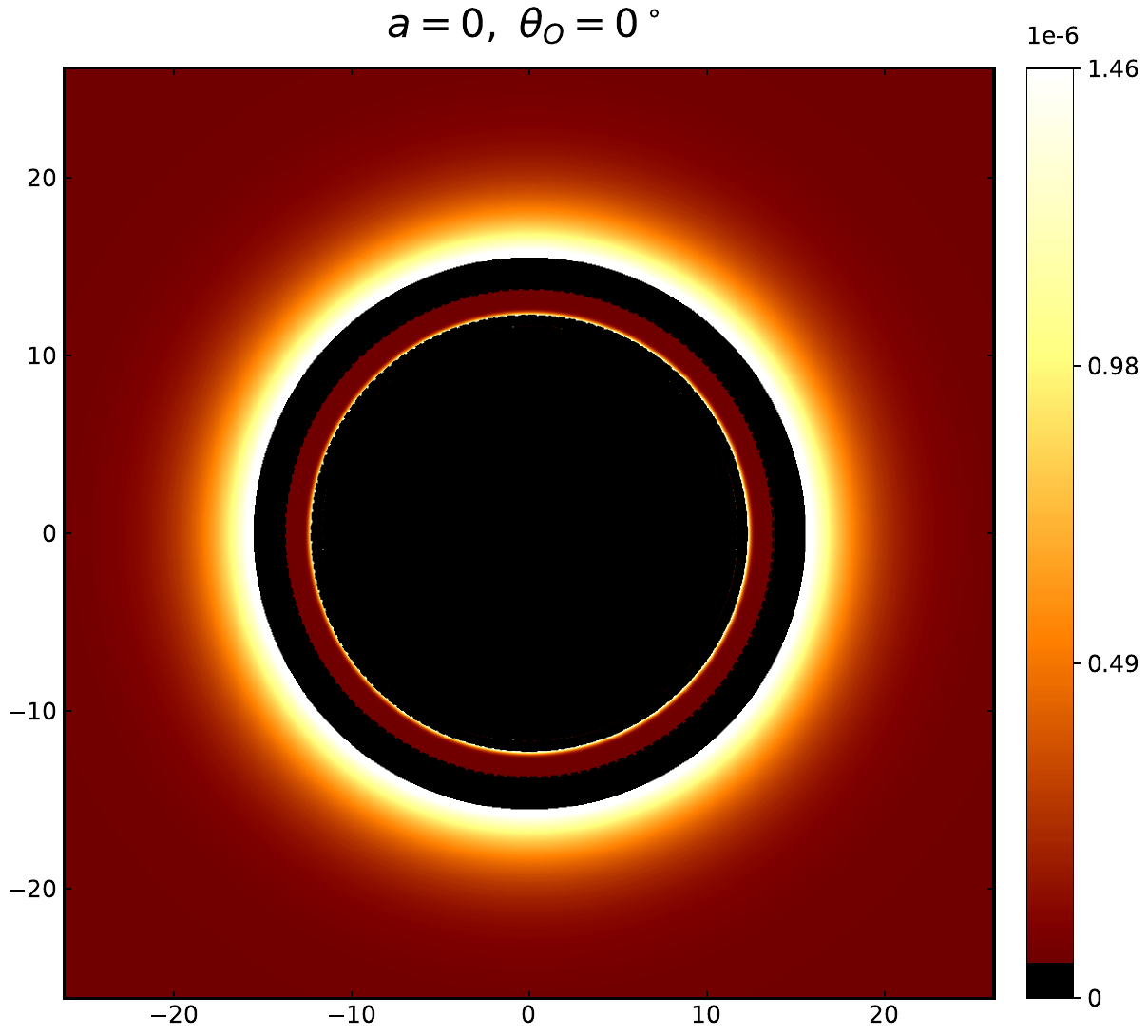} &
        \includegraphics[width=0.23\textwidth]{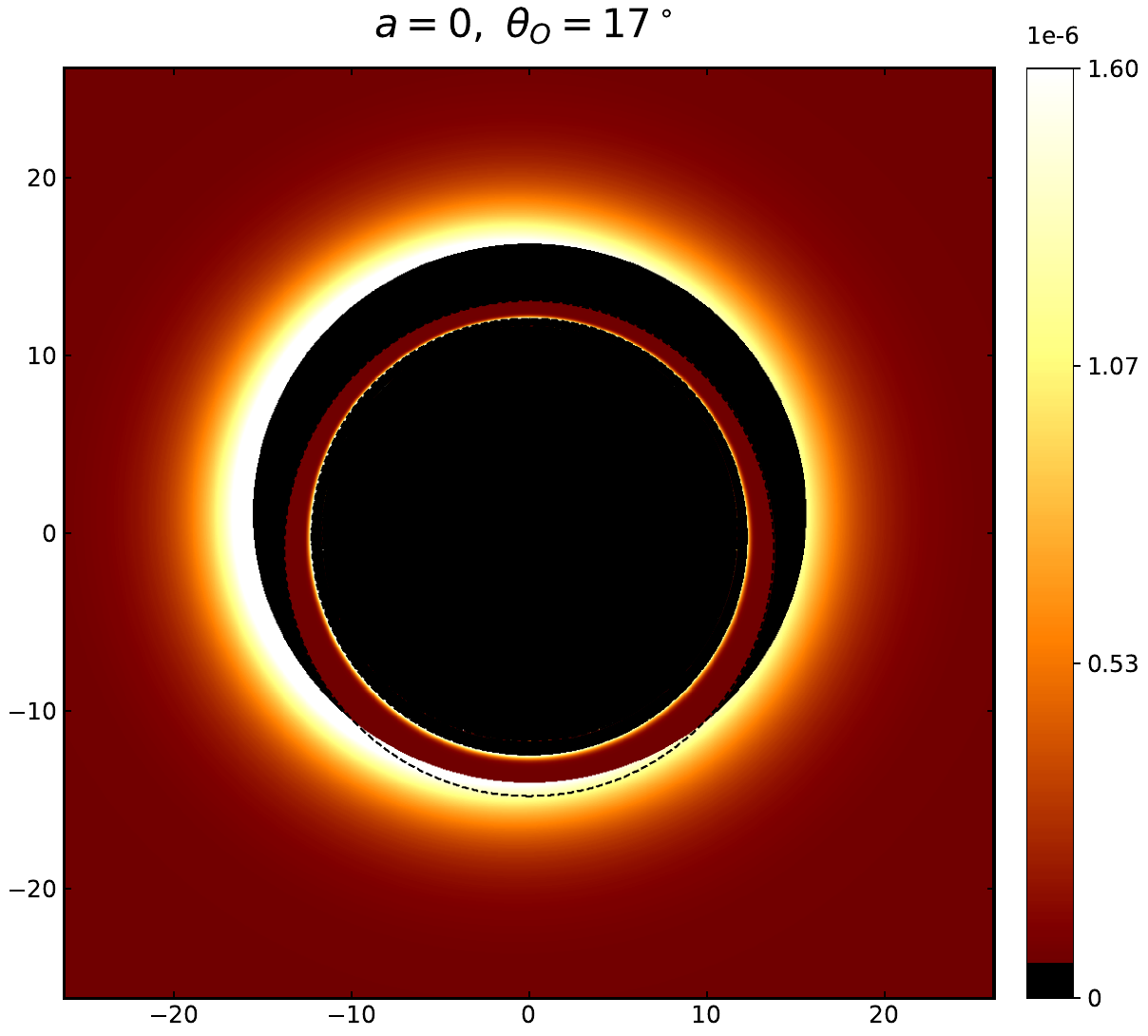} &
        \includegraphics[width=0.23\textwidth]{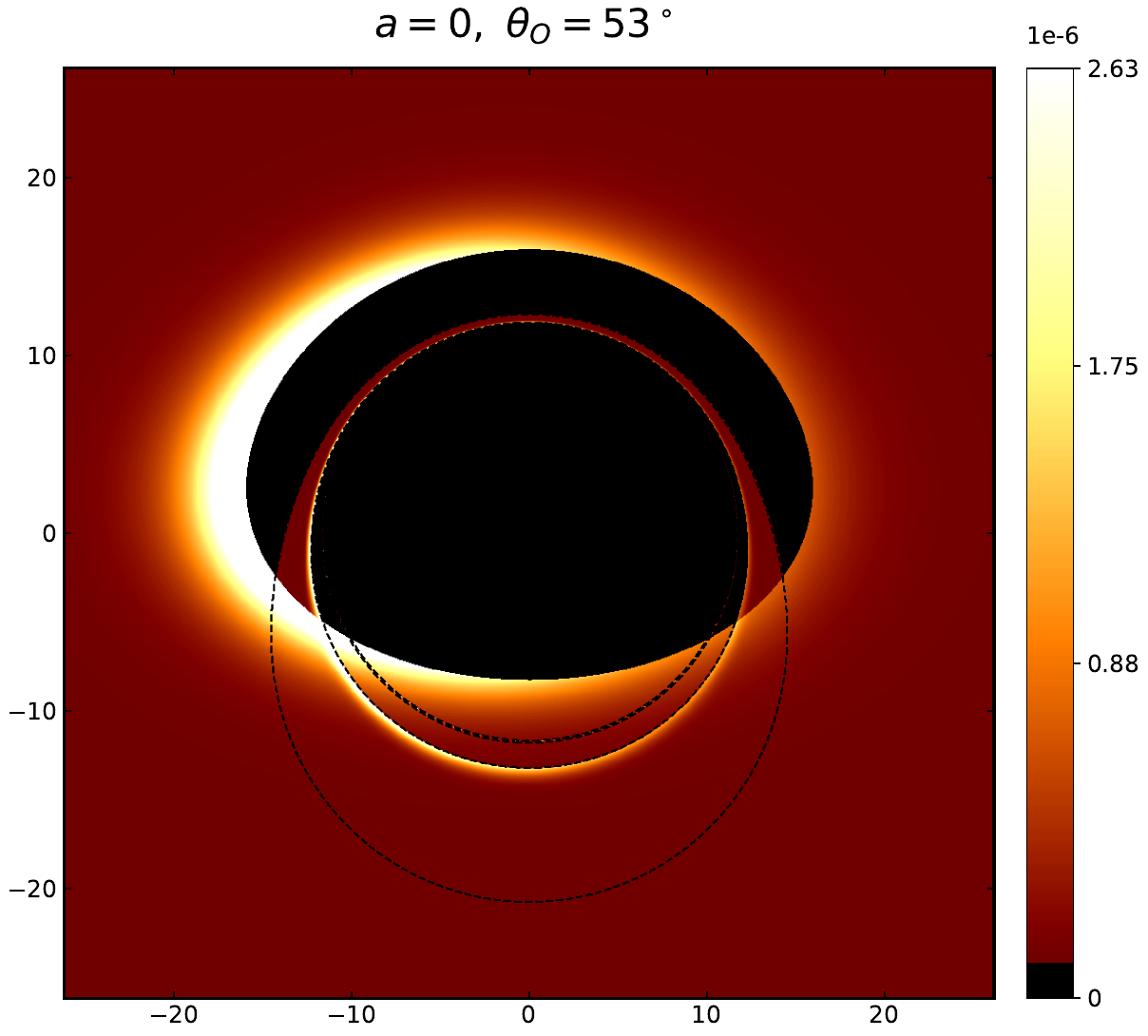} &
        \includegraphics[width=0.23\textwidth]{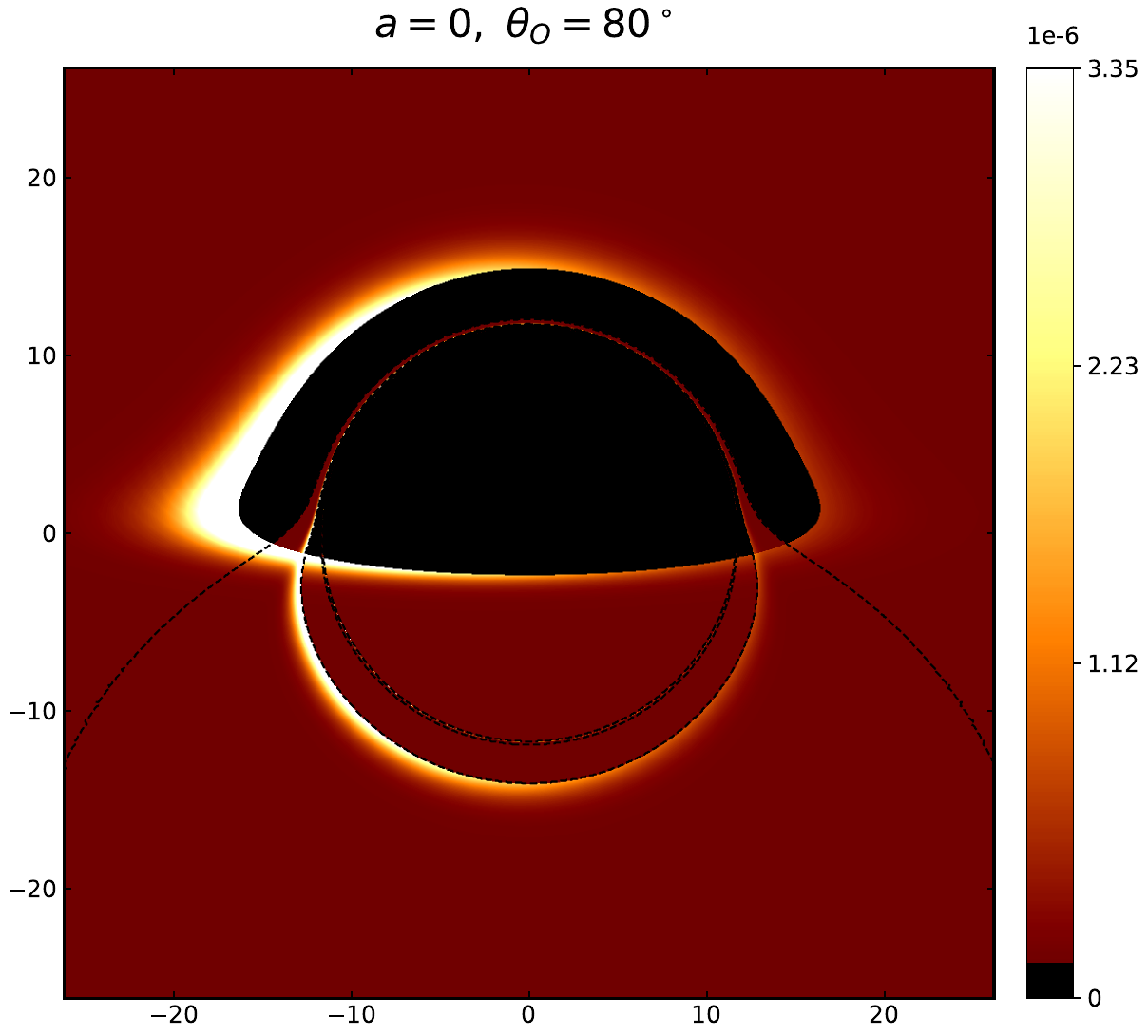} \\
        \includegraphics[width=0.23\textwidth]{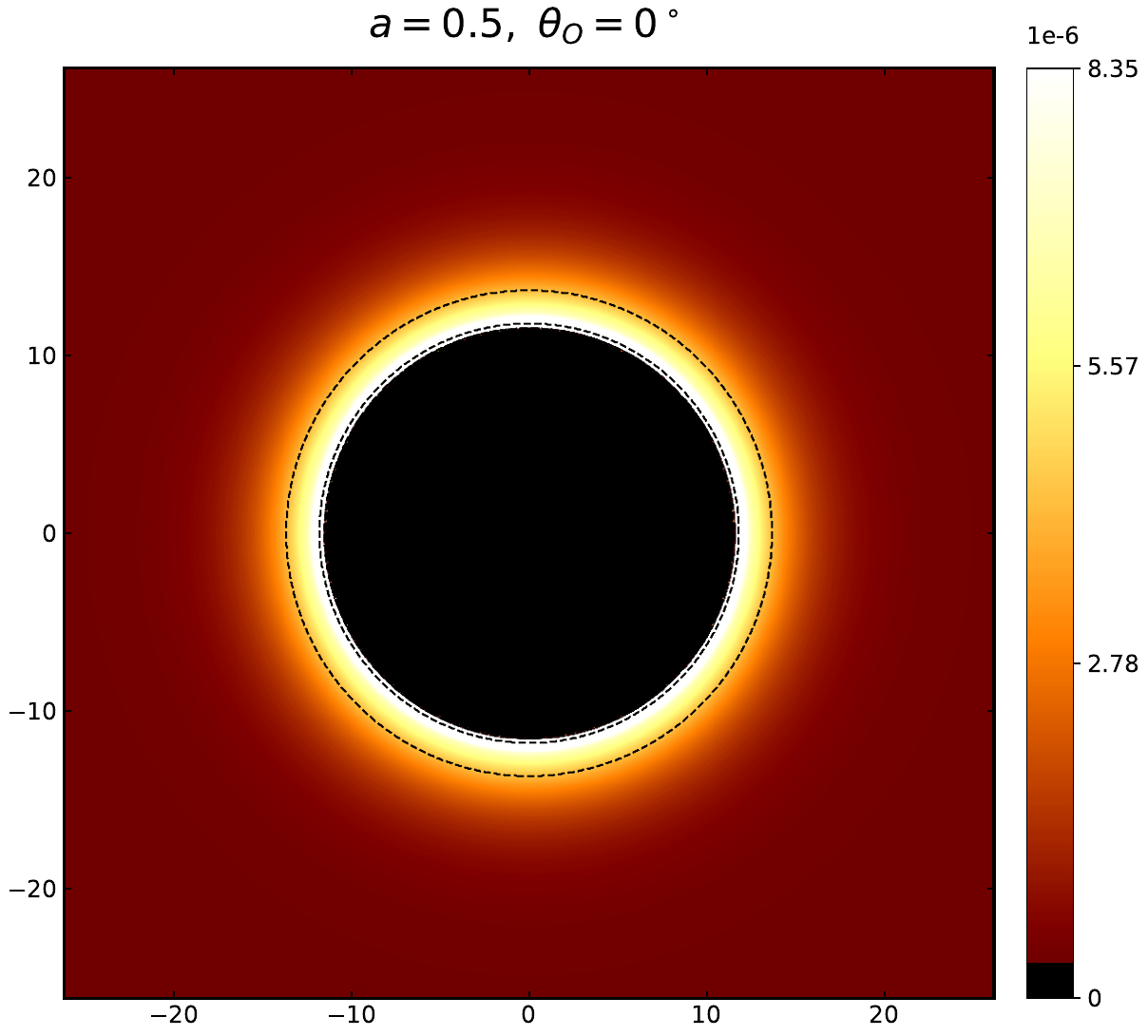} &
        \includegraphics[width=0.23\textwidth]{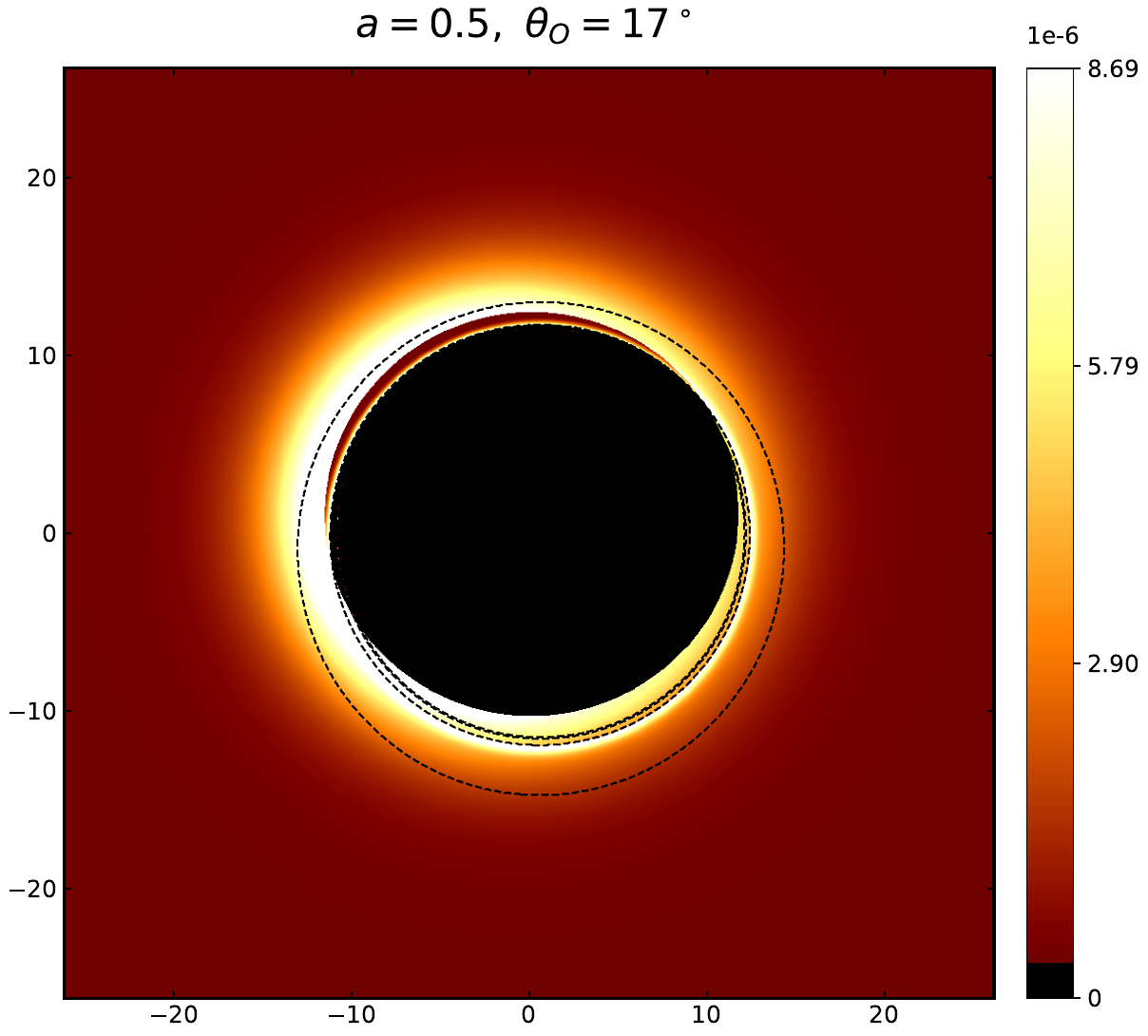} &
        \includegraphics[width=0.23\textwidth]{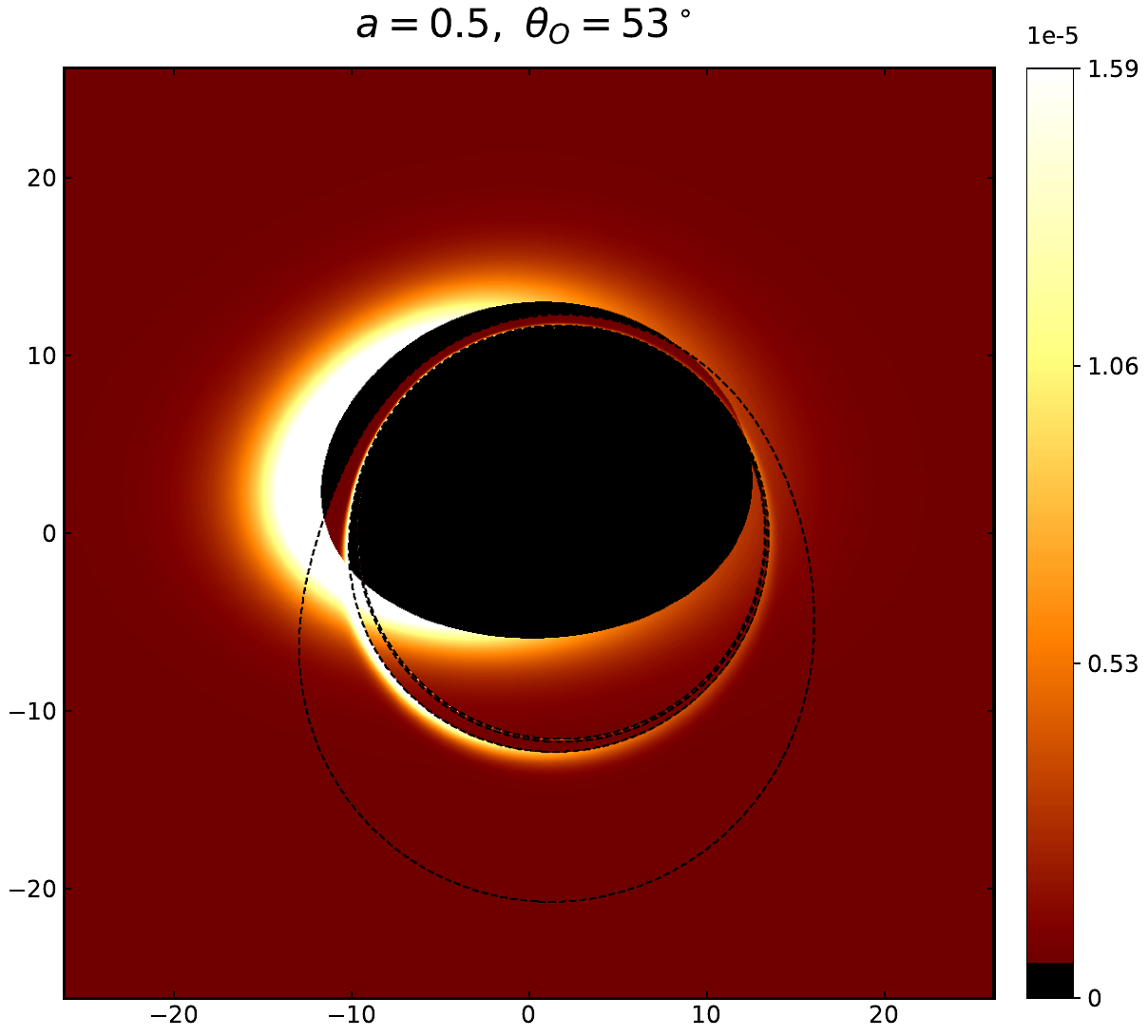} &
        \includegraphics[width=0.23\textwidth]{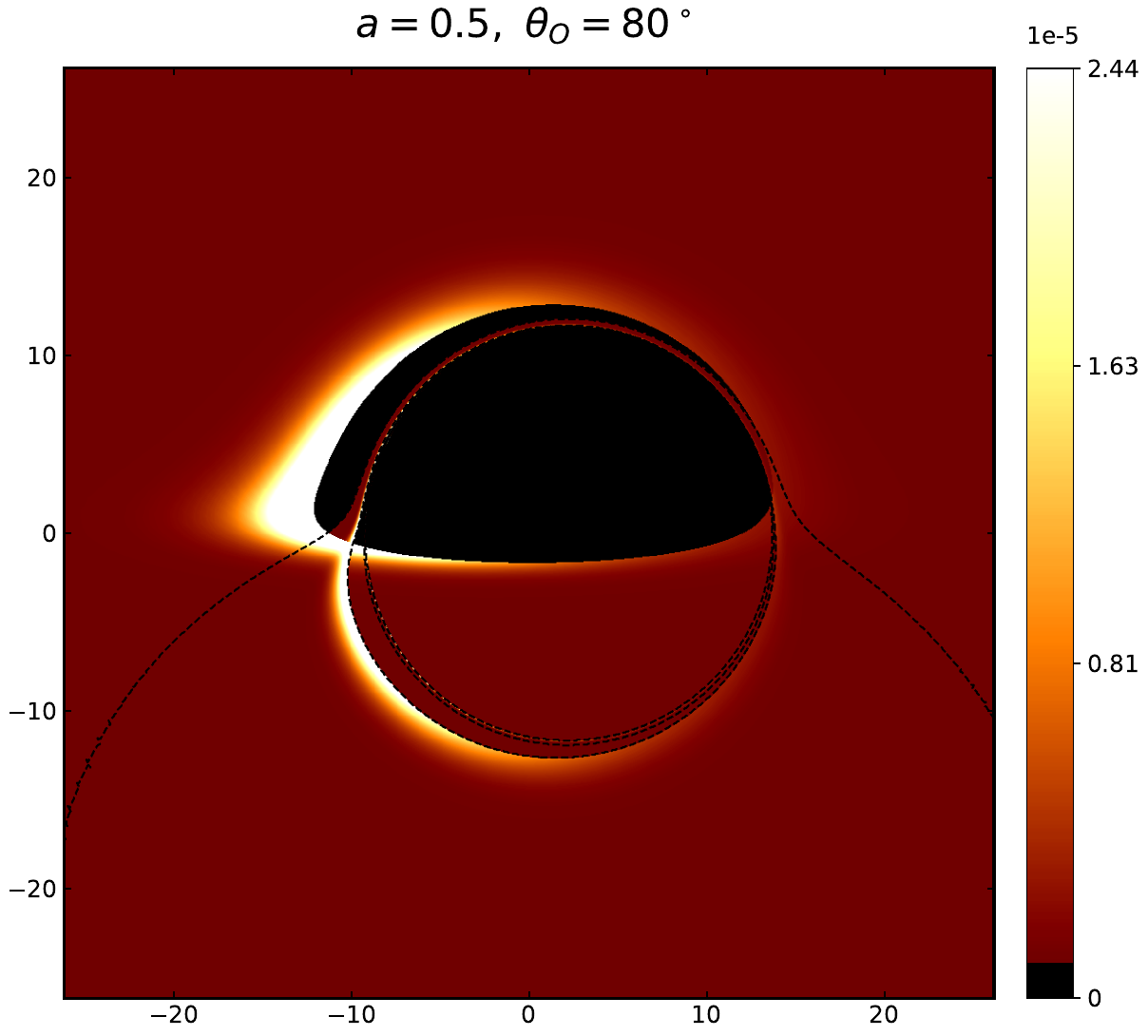} \\
        \includegraphics[width=0.23\textwidth]{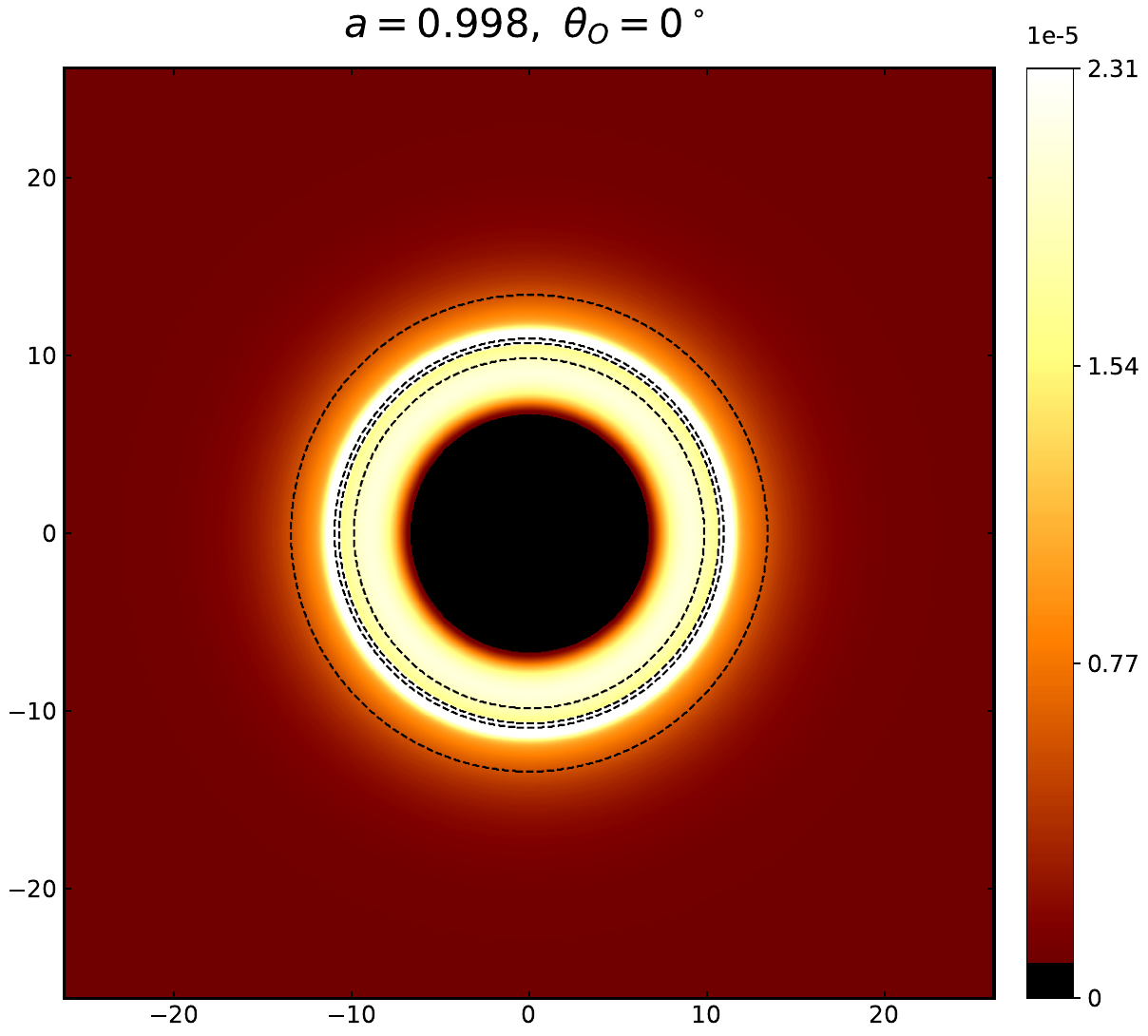} &
        \includegraphics[width=0.23\textwidth]{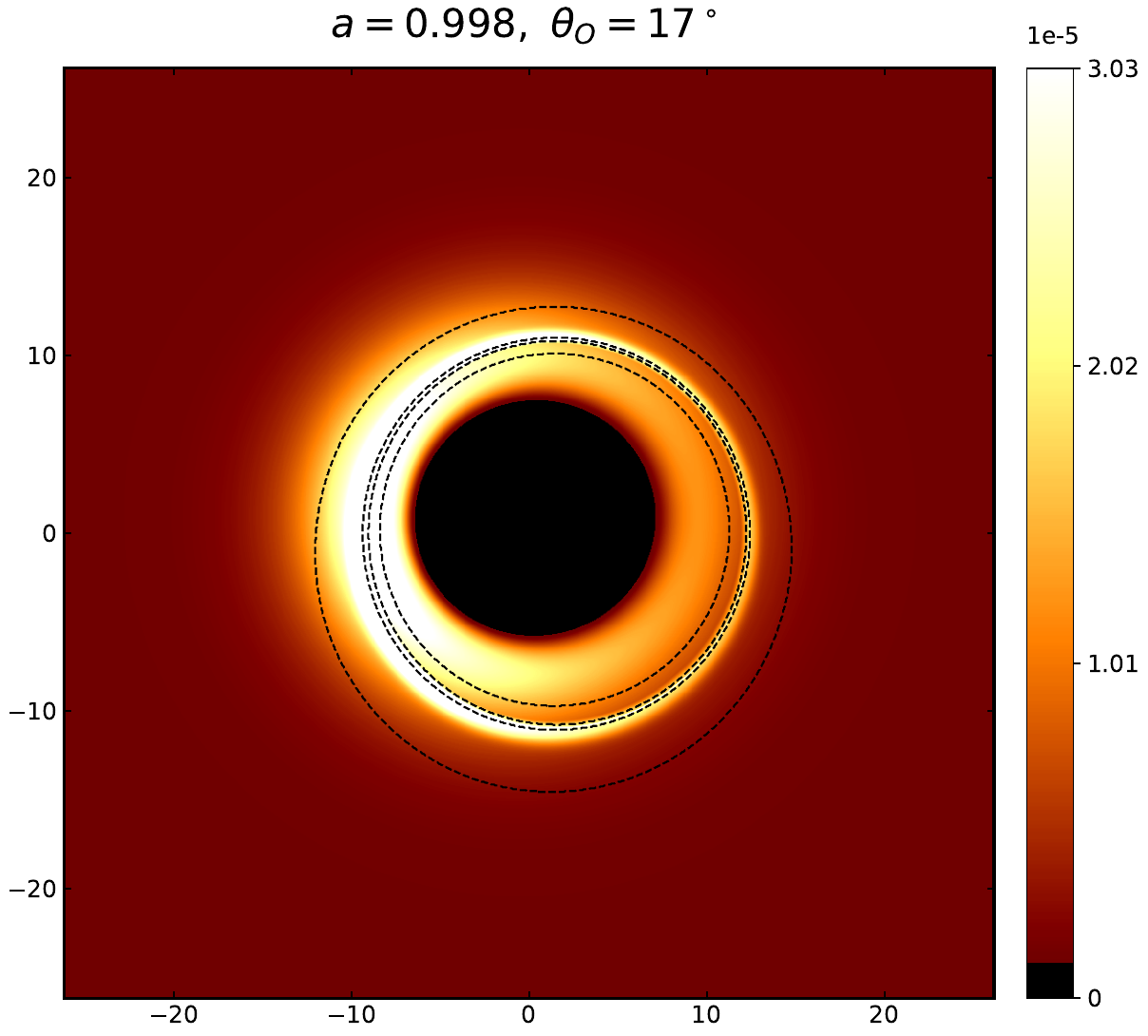} &
        \includegraphics[width=0.23\textwidth]{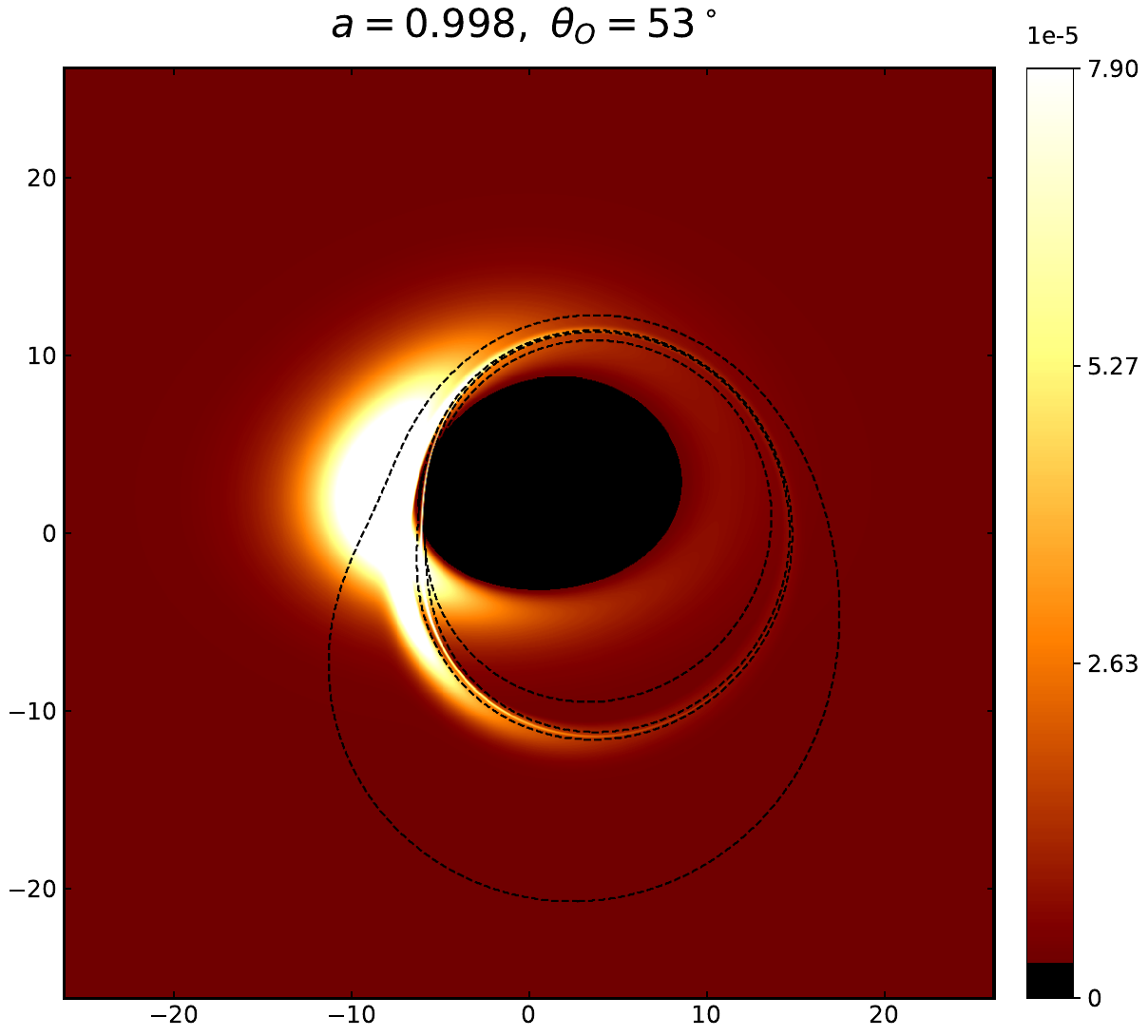} &
        \includegraphics[width=0.23\textwidth]{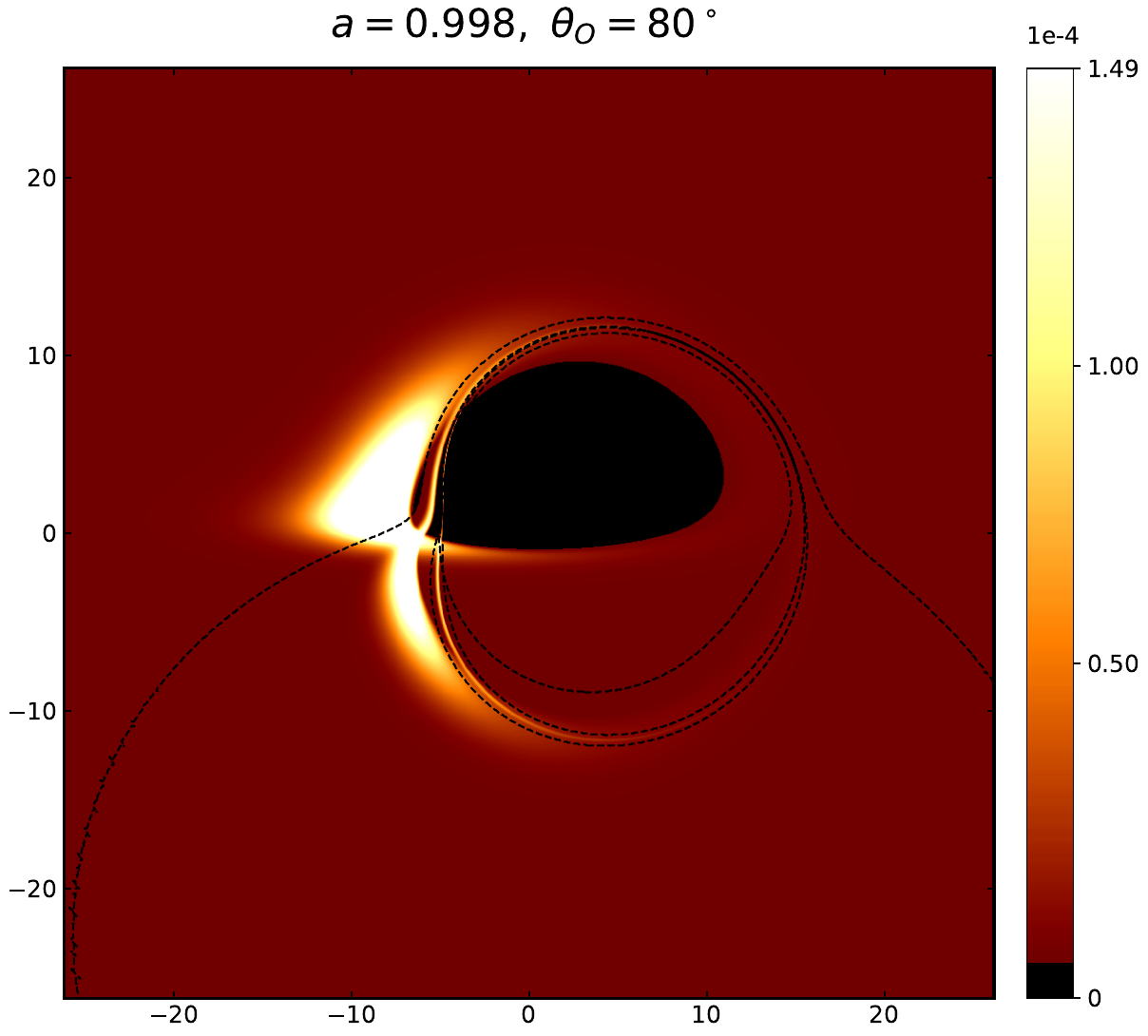} \\
    \end{tabular}
    \caption{Specific-intensity maps of prograde disks with fixed $B=0.002$ and varying spin $a$. The inward shift of the effective inner emitting edge for larger prograde spin reduces the central emission-depleted region and moves the bright emission closer to the image center. At high inclination, the spin also strengthens the Doppler-induced brightness contrast between the approaching and receding sides of the disk.}
    \label{fig:flux_varya}
\end{figure*}

We next examine the effect of the black hole spin. The magnetic parameter is fixed at $B=0.002$, and the spin is varied as
\begin{equation}
    a\in\{0,0.5,0.998\}.
\end{equation}
The corresponding ray-classification maps, redshift-factor maps, and specific-intensity maps are shown in Figs.~\ref{fig:class_varya}, \ref{fig:redshift_varya}, and \ref{fig:flux_varya}, respectively.

Increasing the spin modifies both the photon trajectories and the inner boundary of the emitting region. For the non-rotating case, the ISCO is located relatively far from the horizon. As the spin increases in the prograde direction, the formal ISCO moves inward, allowing the disk emission to originate from smaller radii. For the near-extremal case $a=0.998$, however, the effective inner edge is not necessarily determined by the formal ISCO alone. As discussed in Section~\ref{sec:model}, the magnetically dominated emissivity prescription can cease to be applicable at a radius $r_{\rm md}$ outside the formal ISCO, so that the emitting region is truncated at $r_{\rm in}=\max(r_{\rm ISCO},r_{\rm md})$.

This inward shift of the effective emission boundary has a clear impact on the intensity maps. Compared with the slowly rotating cases, the central emission-depleted region becomes smaller and the bright emission ring is located closer to the image center. At the same time, spin-induced frame dragging changes the orbital velocity of the emitting gas and affects the Doppler factor. As a result, the high-inclination images show a stronger left-right brightness contrast for rapidly rotating configurations. These trends indicate that the spin affects the observed image through both the spacetime lensing structure and the location of the inner emitting edge.

\subsection{One-Dimensional Intensity Profiles}
\label{subsec:1D_profiles}

The two-dimensional images provide a direct visualization of the disk morphology, but the narrow higher-order subimages are more clearly examined through one-dimensional intensity profiles. We therefore extract the specific intensity along the horizontal direction $(\beta=0)$ and the vertical direction $(\alpha=0)$ on the observer's screen. These slices allow us to quantify the positions and relative amplitudes of the direct image, the $n=1$ lensing-ring contribution, and the higher-order $n\geq 2$ photon-ring subimages.

\begin{figure*}[htbp]
    \centering
    \begin{tabular}{ccc}
        \includegraphics[width=0.32\textwidth]{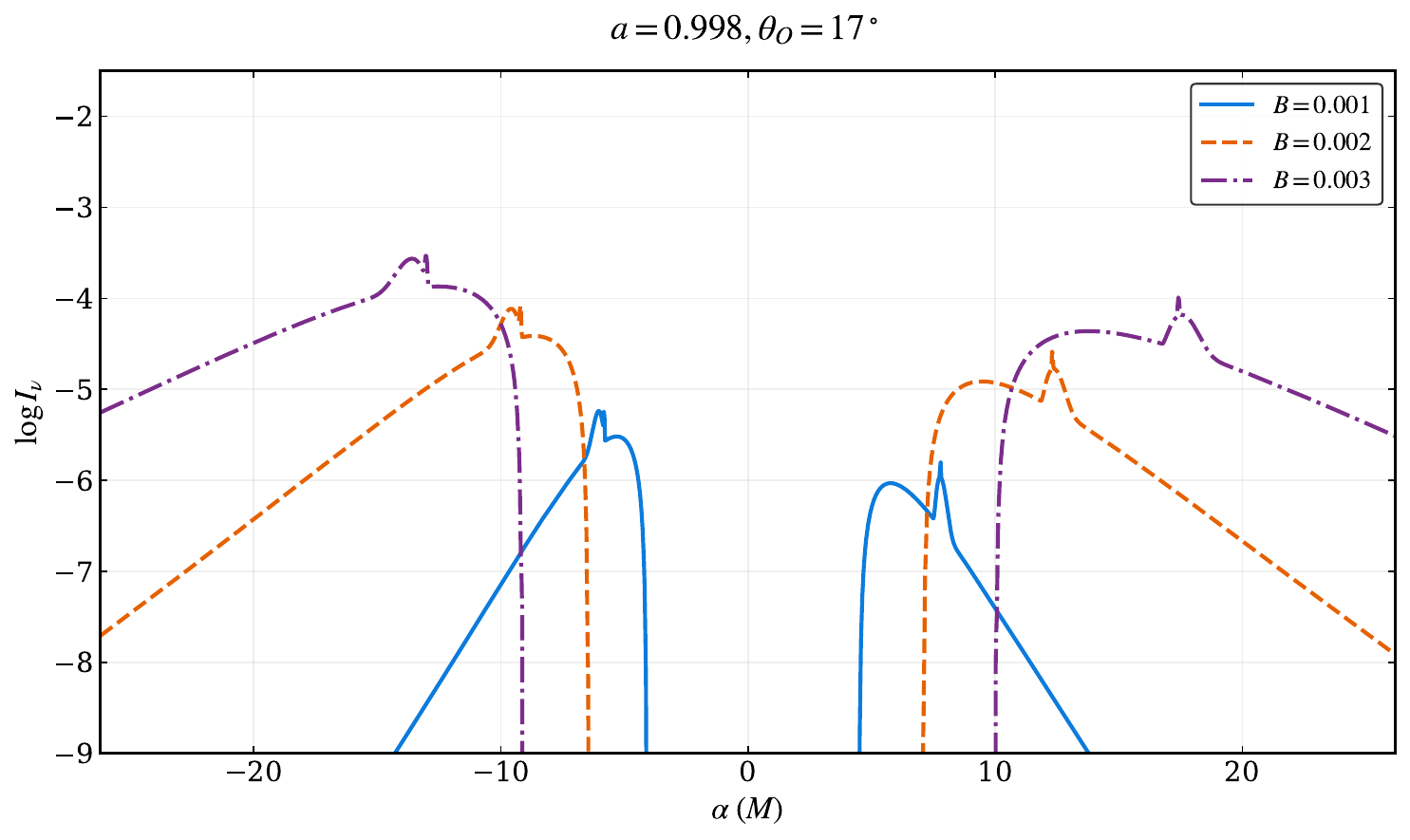} &
        \includegraphics[width=0.32\textwidth]{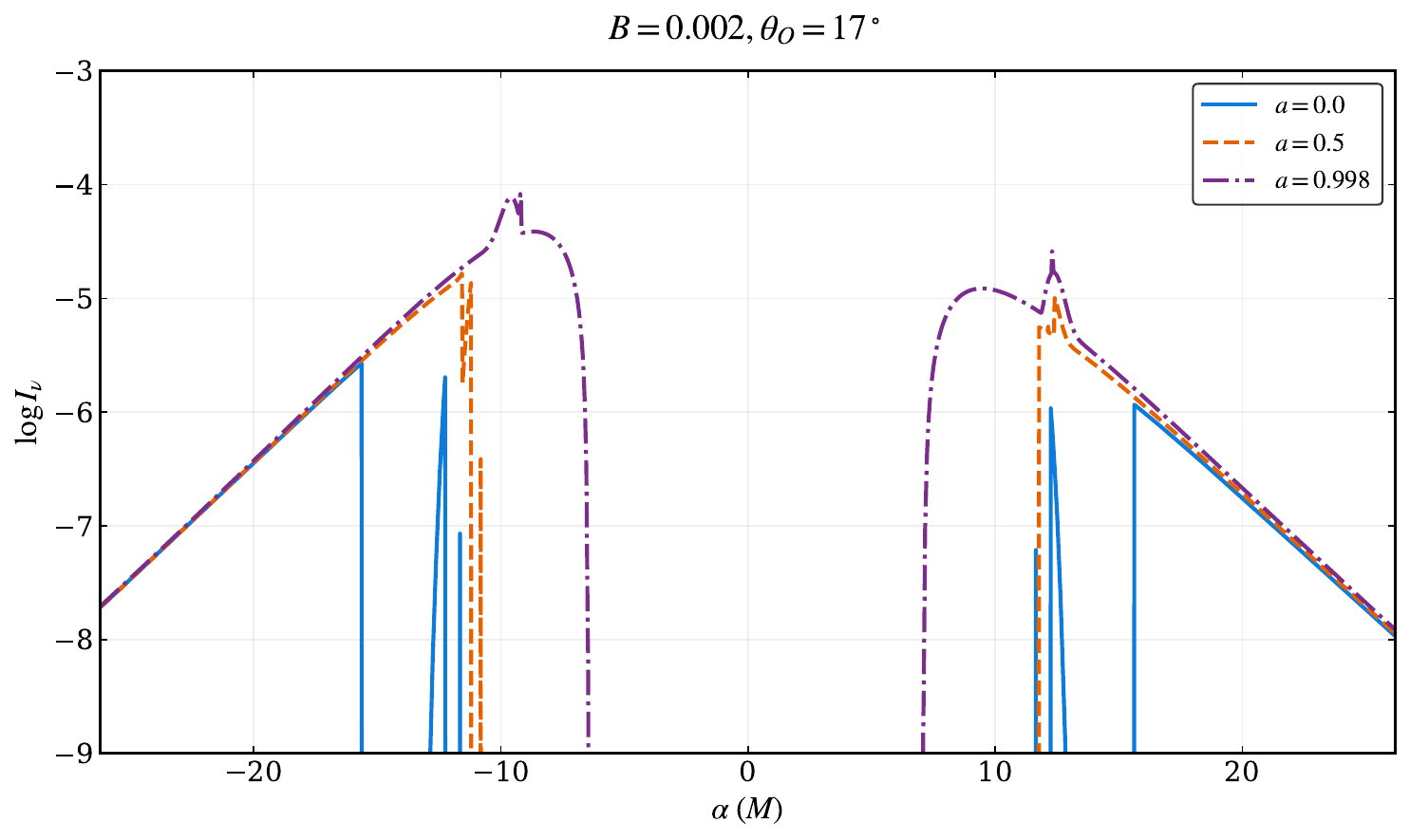} &
        \includegraphics[width=0.32\textwidth]{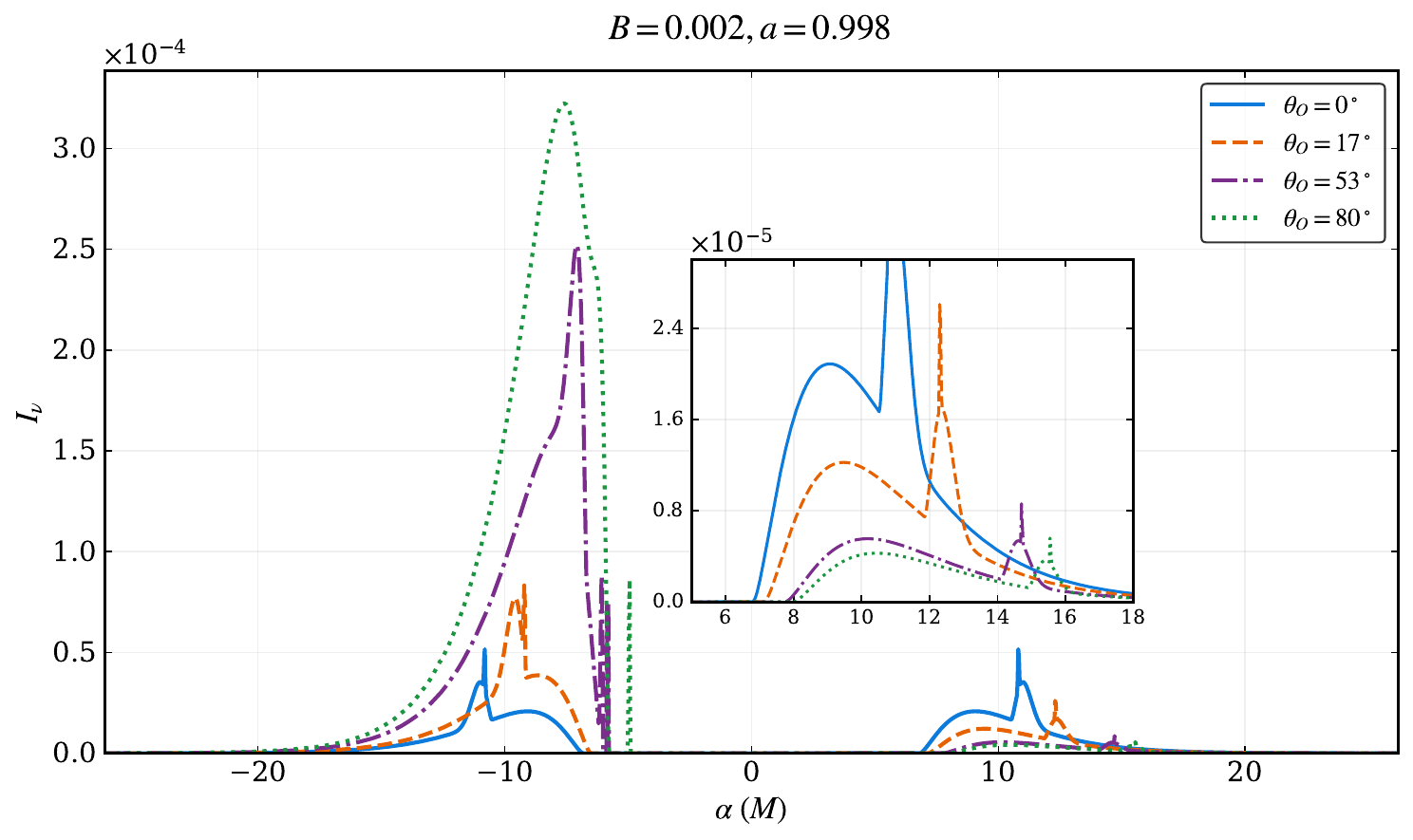} \\
        \includegraphics[width=0.32\textwidth]{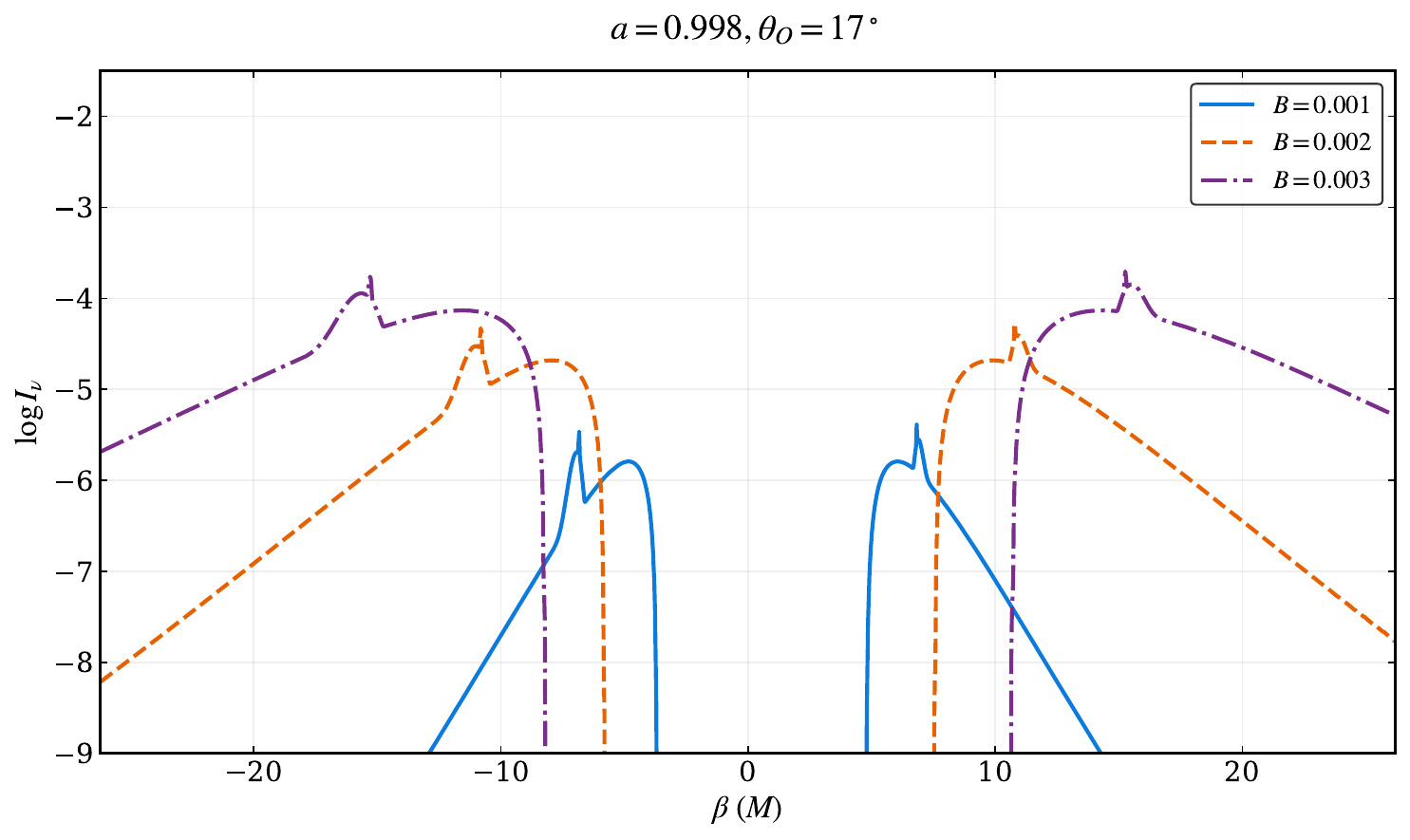} &
        \includegraphics[width=0.32\textwidth]{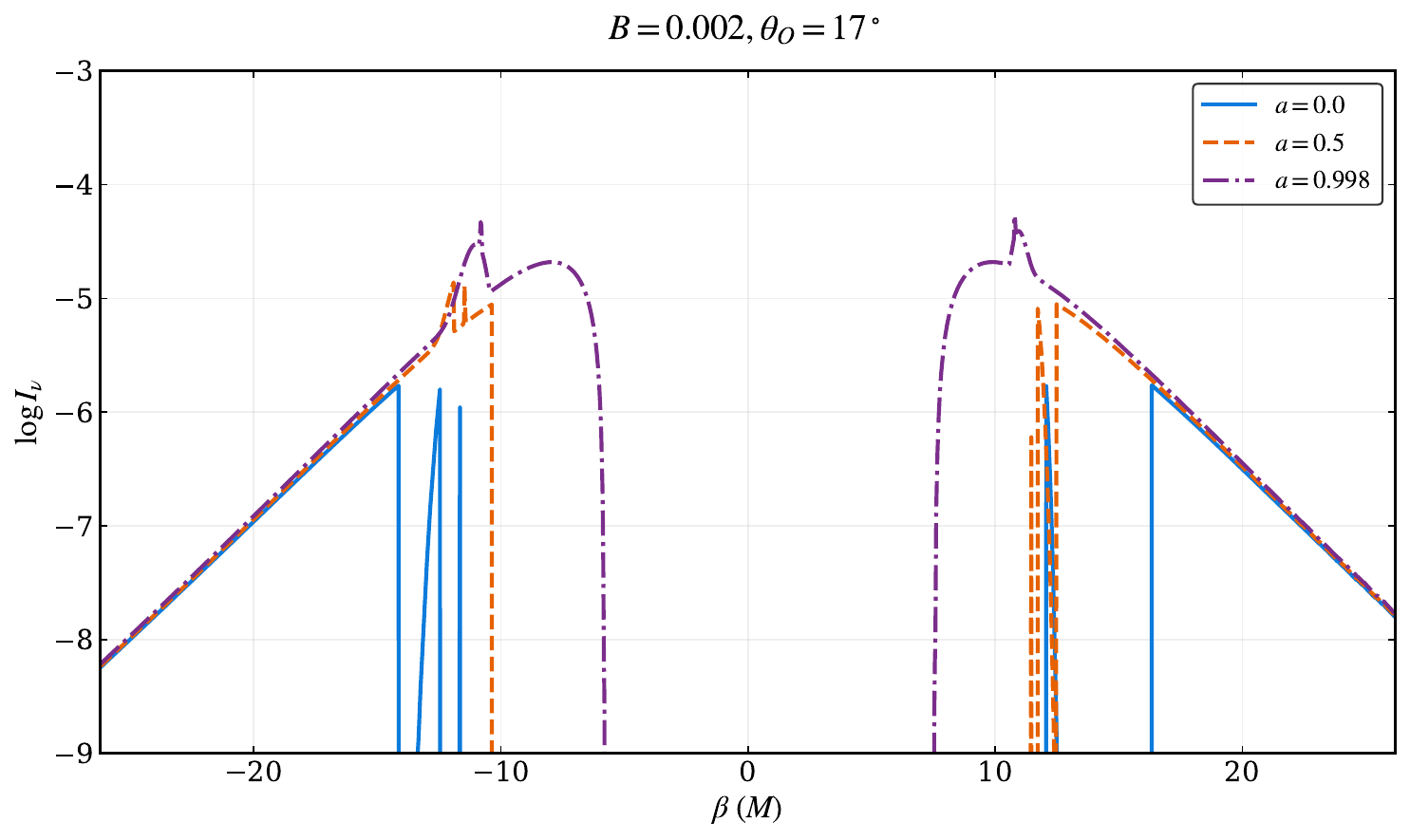} &
        \includegraphics[width=0.32\textwidth]{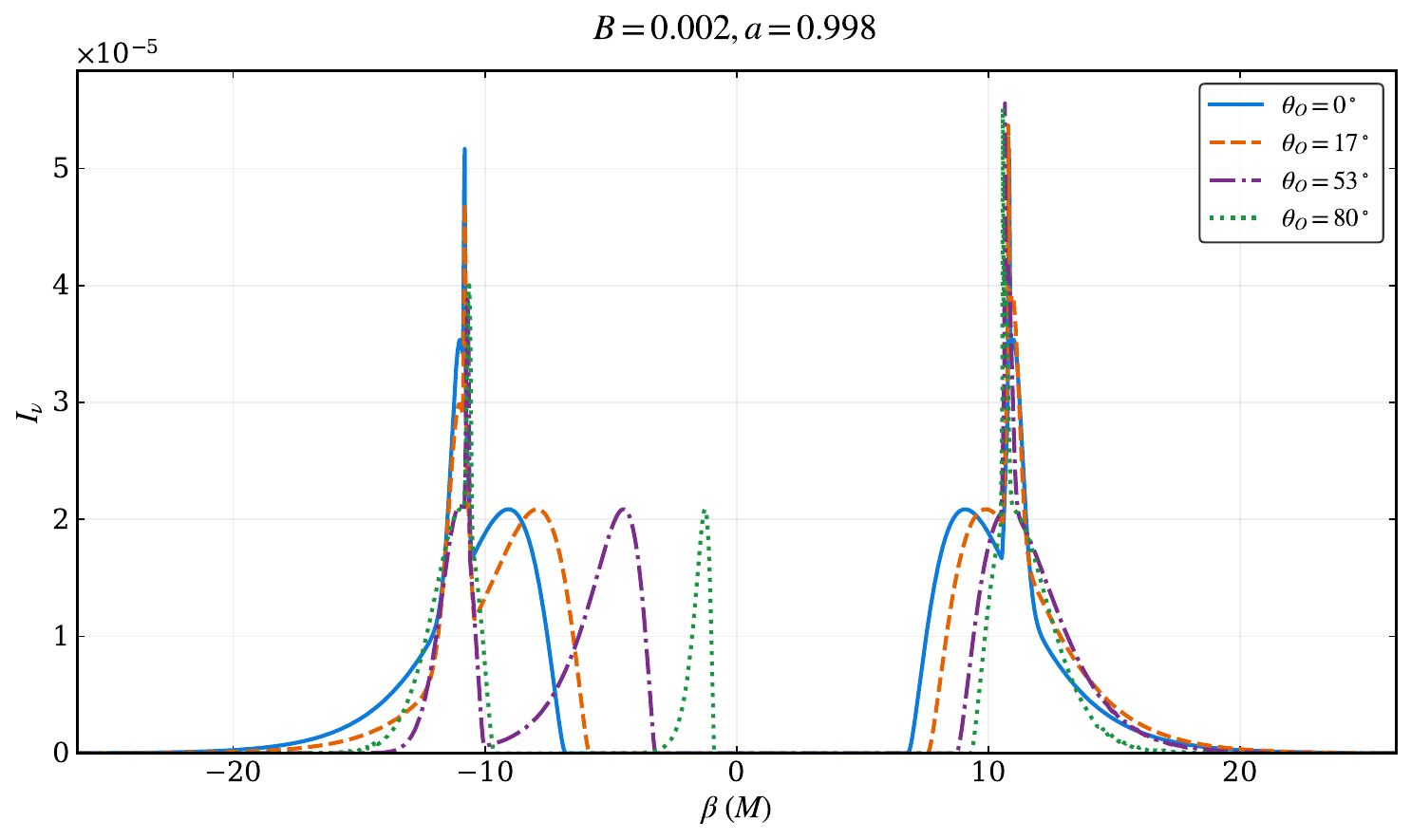} \\
    \end{tabular}
    \caption{One-dimensional specific-intensity profiles along the horizontal direction $(\beta=0$, top row) and vertical direction $(\alpha=0$, bottom row). \textbf{Left column:} Profiles for different magnetic parameters $B$ with fixed $a=0.998$ and $\theta_O=17^\circ$. \textbf{Middle column:} Profiles for different spins $a$ with fixed $B=0.002$ and $\theta_O=17^\circ$. A logarithmic scale is used in the first two columns to show the narrow $n=1$ lensing-ring contribution and the higher-order $n\geq 2$ photon-ring subimages. \textbf{Right column:} Profiles for different observer inclinations $\theta_O$ with fixed $B=0.002$ and $a=0.998$, shown on a linear scale to highlight the Doppler-induced brightness asymmetry.}
    \label{fig:1D_profiles}
\end{figure*}

Figure~\ref{fig:1D_profiles} shows a $2\times 3$ set of intensity profiles. The left column varies the magnetic parameter $B$, the middle column varies the spin $a$, and the right column varies the observer inclination $\theta_O$. In the first two columns, a logarithmic scale is used because the intensity spans several orders of magnitude and because the narrow higher-order subimages would otherwise be difficult to identify. In the right column, a linear scale is used to emphasize the inclination-driven change in the overall brightness asymmetry.

The horizontal profiles in the top row of Fig.~\ref{fig:1D_profiles} sample the approaching and receding sides of the rotating disk. With our screen convention, the negative-$\alpha$ side corresponds to the approaching part of the flow, while the positive-$\alpha$ side corresponds to the receding part. The intensity contrast between the two sides reflects the Doppler factor in the radiative-transfer relation, $I_{\nu_{\rm obs}}\propto g^3$. On the logarithmic scale, the broad envelope is dominated by the direct image, whereas the $n=1$ and $n\geq 2$ components appear as narrow peaks superimposed on this envelope.

The middle column shows that increasing the prograde spin shifts the main emitting region inward. As the formal ISCO moves to smaller radii, and as the effective inner edge is determined by Eq.~(\ref{eq:rin_def}), the central flux depression becomes narrower and the highly deflected subimage peaks move closer to the image center. This trend is consistent with the two-dimensional intensity maps in Fig.~\ref{fig:flux_varya}. The magnetic-dominance cutoff in the near-extremal case prevents the adopted emissivity prescription from being extended arbitrarily inward, but the effective emitting edge still lies much closer to the black hole than in the slowly rotating cases.

The vertical profiles in the bottom row of Fig.~\ref{fig:1D_profiles} probe the upper and lower parts of the image. Along this direction, the line-of-sight component of the orbital velocity is generally smaller than along the horizontal direction, and the Doppler asymmetry is correspondingly weaker. These profiles are therefore useful for isolating the effects of geometric projection and gravitational lensing. The peaks associated with the front and back sides of the disk, together with the narrow higher-order subimage features, can be identified more cleanly in these slices.

The right column of Fig.~\ref{fig:1D_profiles} illustrates the inclination dependence. For a nearly face-on observer, the horizontal and vertical profiles are approximately symmetric. As the inclination increases, the Doppler enhancement of the approaching side becomes more pronounced, leading to a larger difference between the two sides of the horizontal profile. This behavior is consistent with the crescent-like morphology seen in the high-inclination intensity maps.

\subsection{Comparative Study with Retrograde Accretion Disks}
\label{subsec:retrograde}

We finally compare the prograde results with retrograde accretion disks, where the angular momentum of the emitting gas is opposite to the black hole spin. For a rapidly rotating black hole, the retrograde ISCO is located much farther from the horizon than the prograde ISCO. For example, for $a=0.998$ in the parameter range considered here, the inner orbital boundary is shifted from the near-horizon region to radii of order $r_{\rm ISCO}\sim 9M$. This outward displacement strongly affects the spatial extent of the emitting region and, consequently, the observed image.

\begin{figure*}[htbp]
    \centering
    \begin{tabular}{ccc}
        % Row 1: B=0.001
        \includegraphics[width=0.232\textwidth]{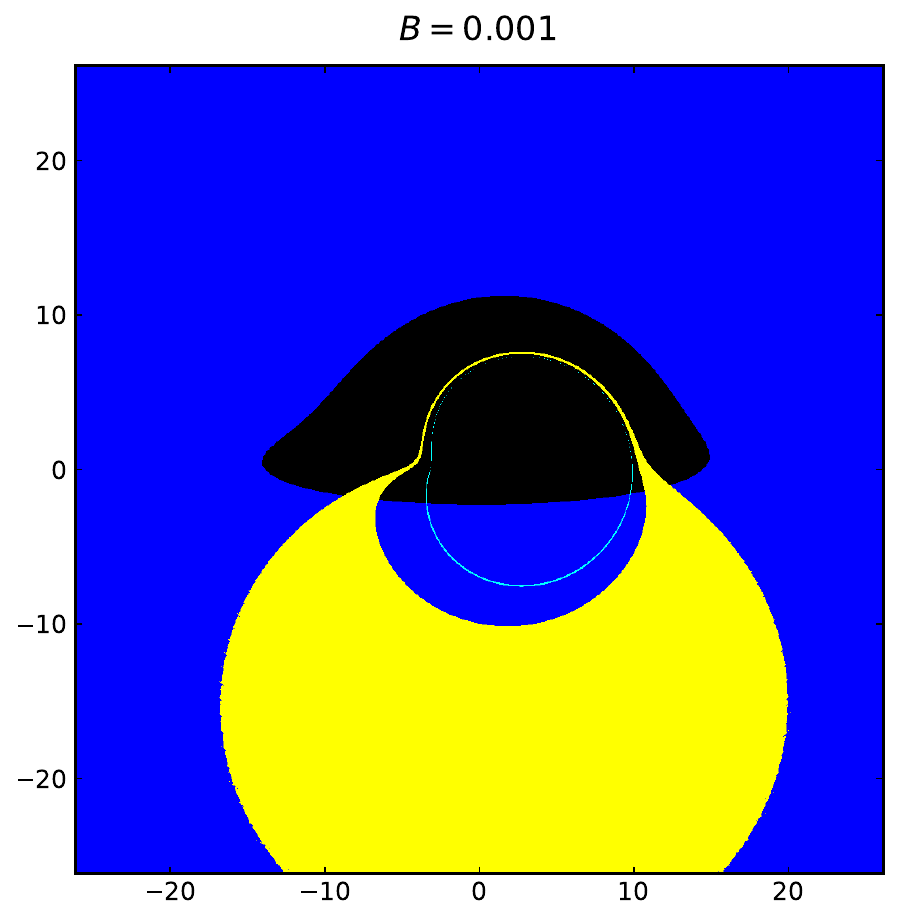} &
        \includegraphics[width=0.26\textwidth]{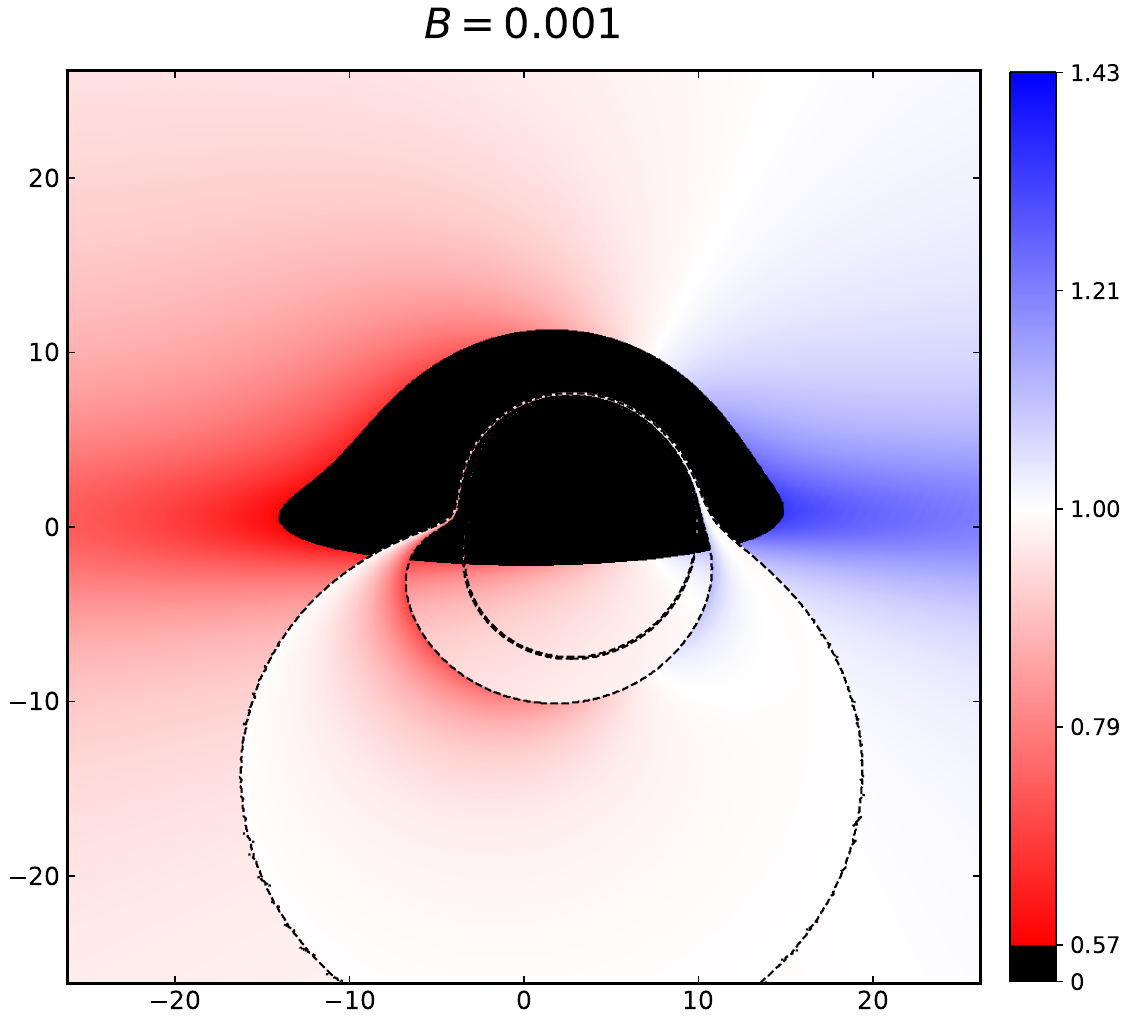} &
        \includegraphics[width=0.26\textwidth]{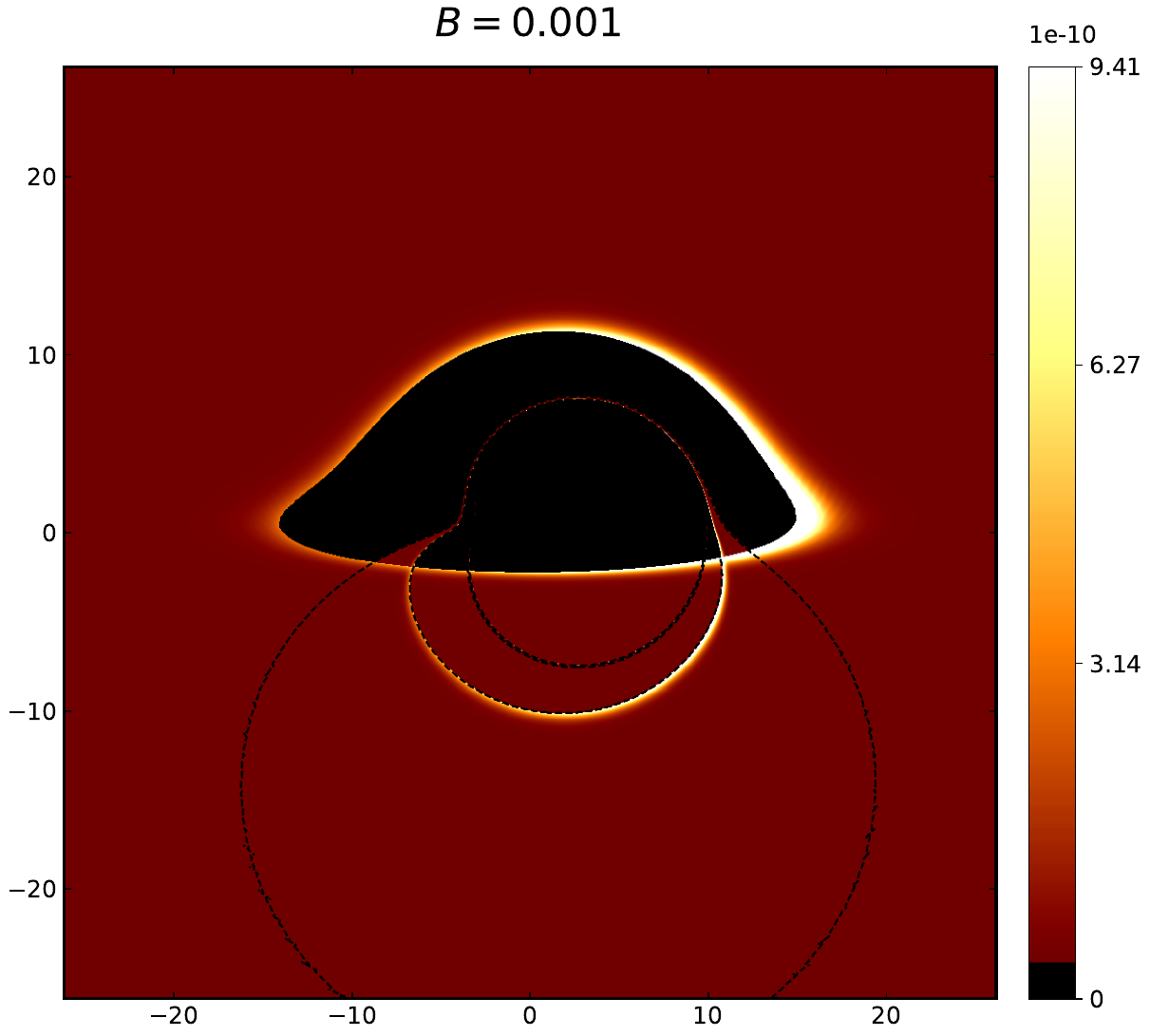} \\
        % Row 2: B=0.002
        \includegraphics[width=0.232\textwidth]{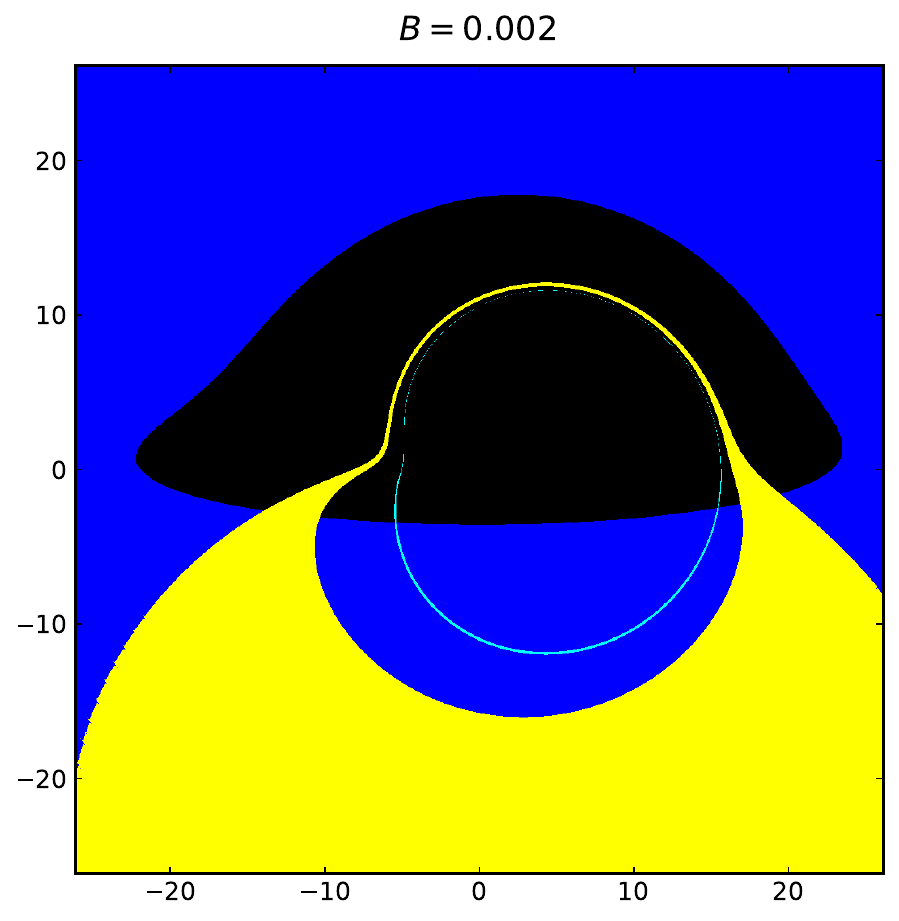} &
        \includegraphics[width=0.26\textwidth]{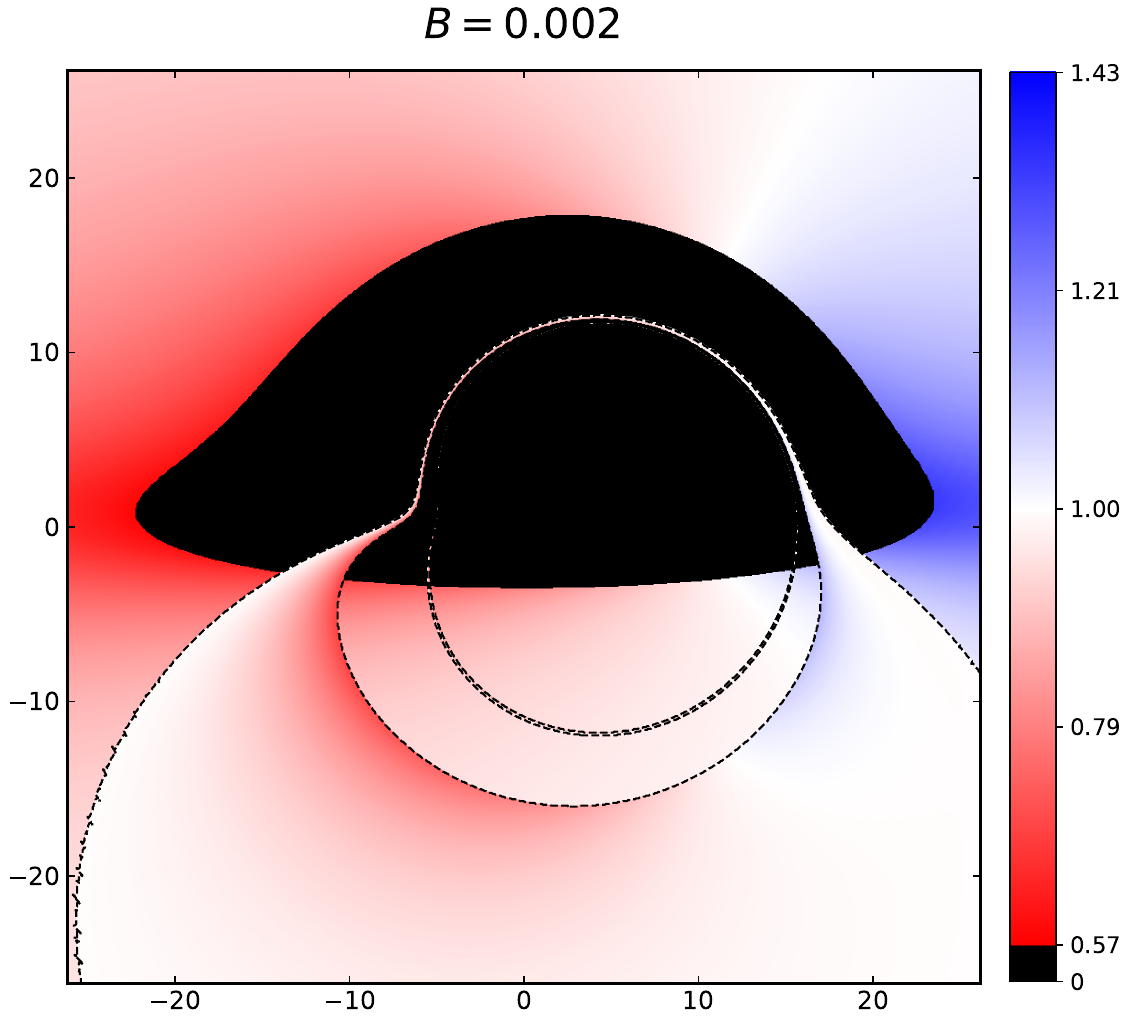} &
        \includegraphics[width=0.26\textwidth]{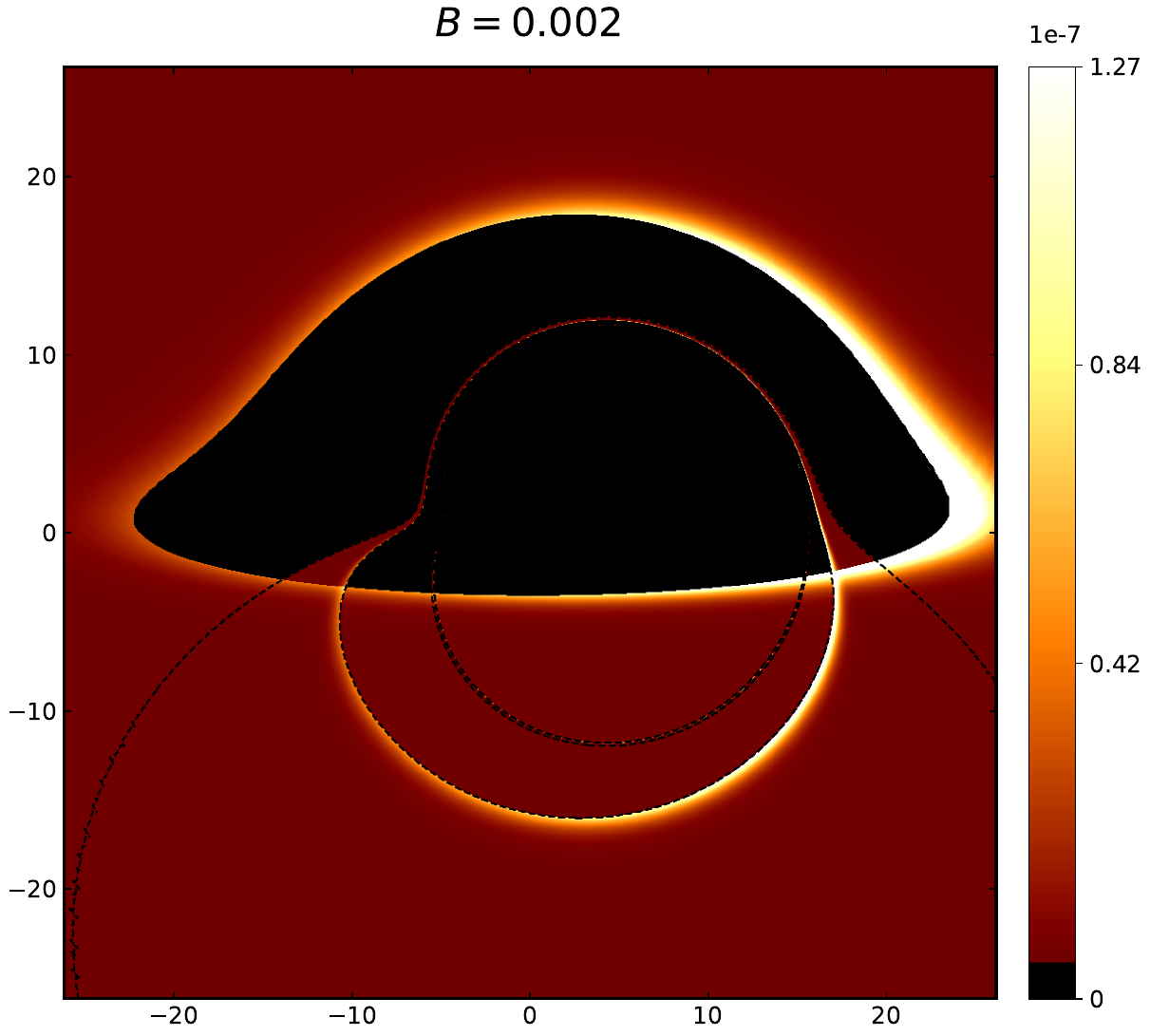} \\
        % Row 3: B=0.003
        \includegraphics[width=0.232\textwidth]{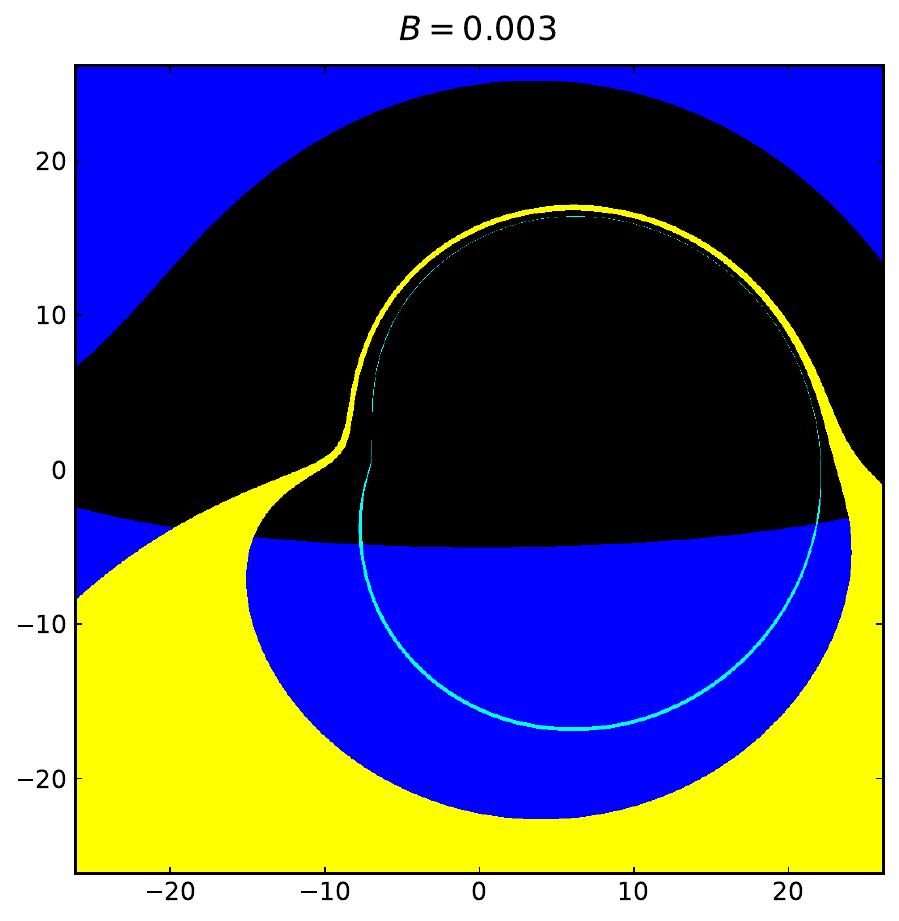} &
        \includegraphics[width=0.26\textwidth]{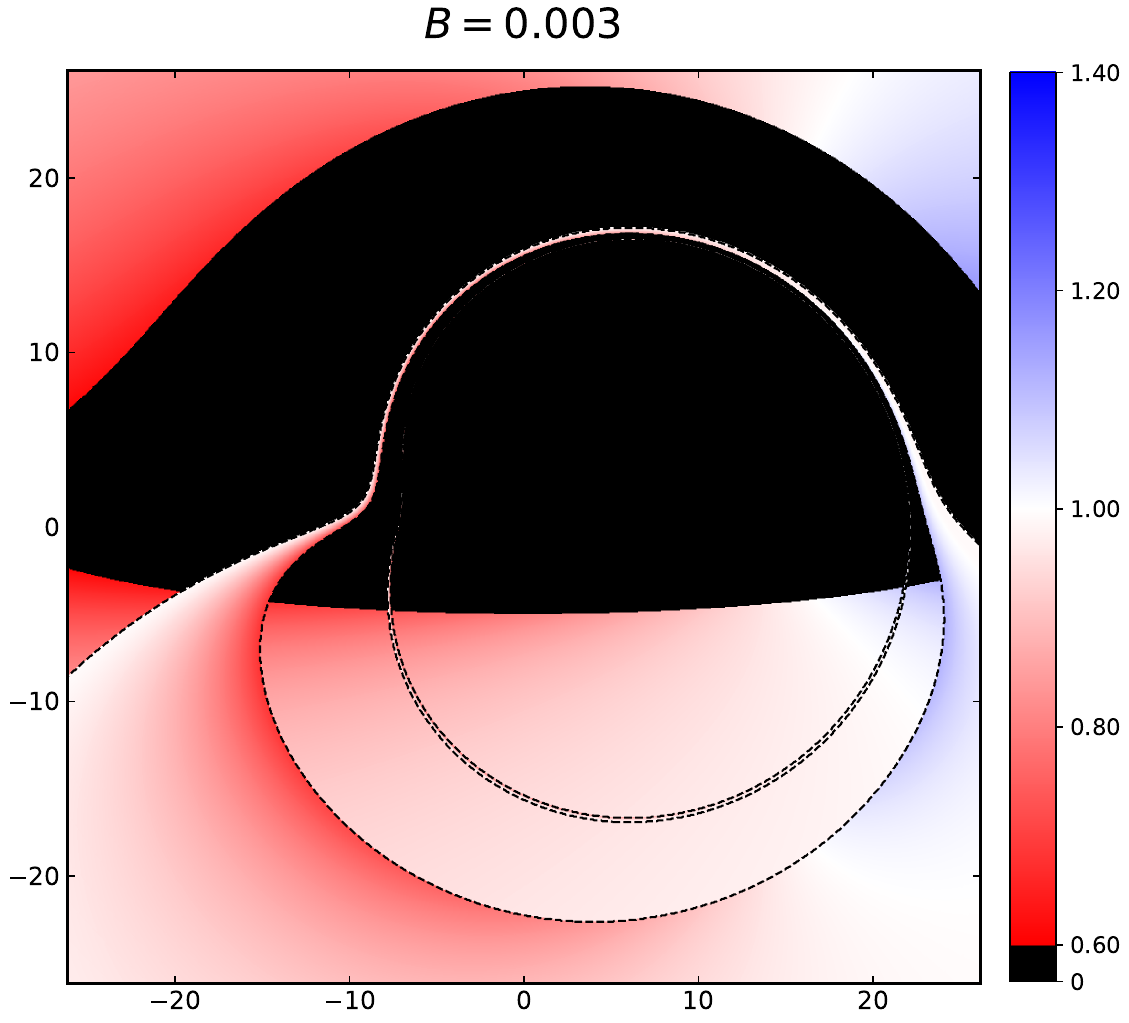} &
        \includegraphics[width=0.26\textwidth]{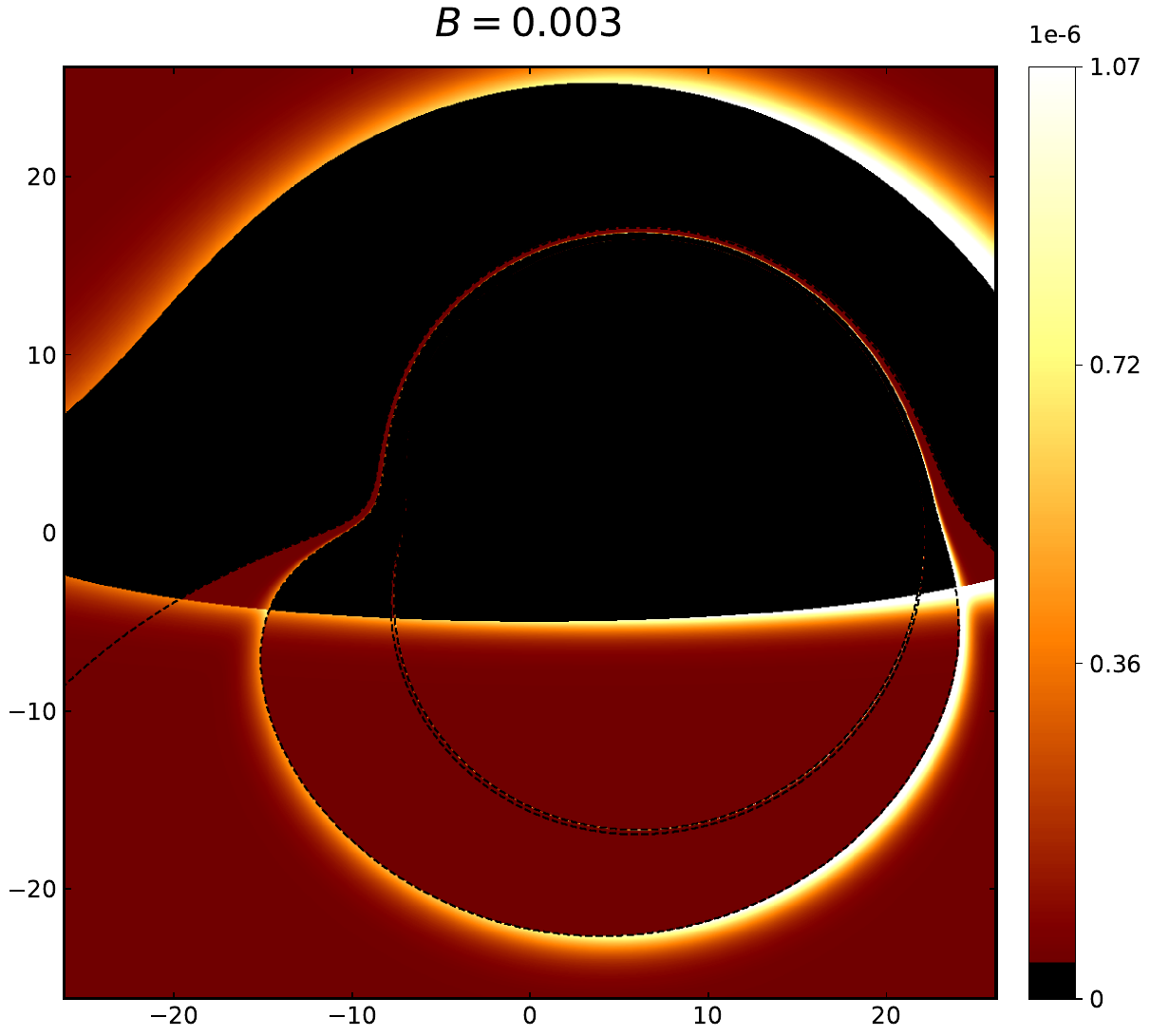} \\
    \end{tabular}
    \caption{Optical appearance of retrograde disks with fixed spin $a=0.998$ and inclination $\theta_O=80^\circ$, for different magnetic parameters $B$ from top to bottom. \textbf{Left column:} Ray-classification maps. \textbf{Middle column:} Redshift-factor maps. \textbf{Right column:} Specific-intensity maps. The outward displacement of the retrograde ISCO produces a wide emission-depleted central region and separates the direct emission from the higher-order lensed components.}
    \label{fig:retro_2D_maps}
\end{figure*}

To illustrate this effect, we fix $a=0.998$ and $\theta_O=80^\circ$, and compute retrograde disk images for
\begin{equation}
    B\in\{0.001,0.002,0.003\}.
\end{equation}
The corresponding ray-classification maps, redshift-factor maps, and specific-intensity maps are shown in Fig.~\ref{fig:retro_2D_maps}.

The ray-classification maps in the left column of Fig.~\ref{fig:retro_2D_maps} show a much wider central region without direct disk emission than in the prograde case. This occurs because the retrograde ISCO lies at larger radii, so the direct image can only originate from the outer disk. The central dark region in these maps should therefore be interpreted as an emission-depleted region determined by the adopted disk inner boundary, not simply as an enlargement of the geometrical black hole shadow. The higher-order lensed components remain tied to the strong-lensing structure of the spacetime and can appear inside this dark region.

The redshift maps in the middle column show that the left-right Doppler asymmetry persists for retrograde disks, but its magnitude is reduced compared with the near-extremal prograde case. This is expected because the emitting gas is located at larger radii, where the orbital velocity is lower and the gravitational redshift is weaker. The intensity maps in the right column combine this redshift structure with the magnetic-field-dependent emissivity. The bright direct emission is displaced to larger impact parameters, leaving a broad central flux depression.

\begin{figure*}[htbp]
    \centering
    \begin{tabular}{cc}
        \includegraphics[width=0.48\textwidth]{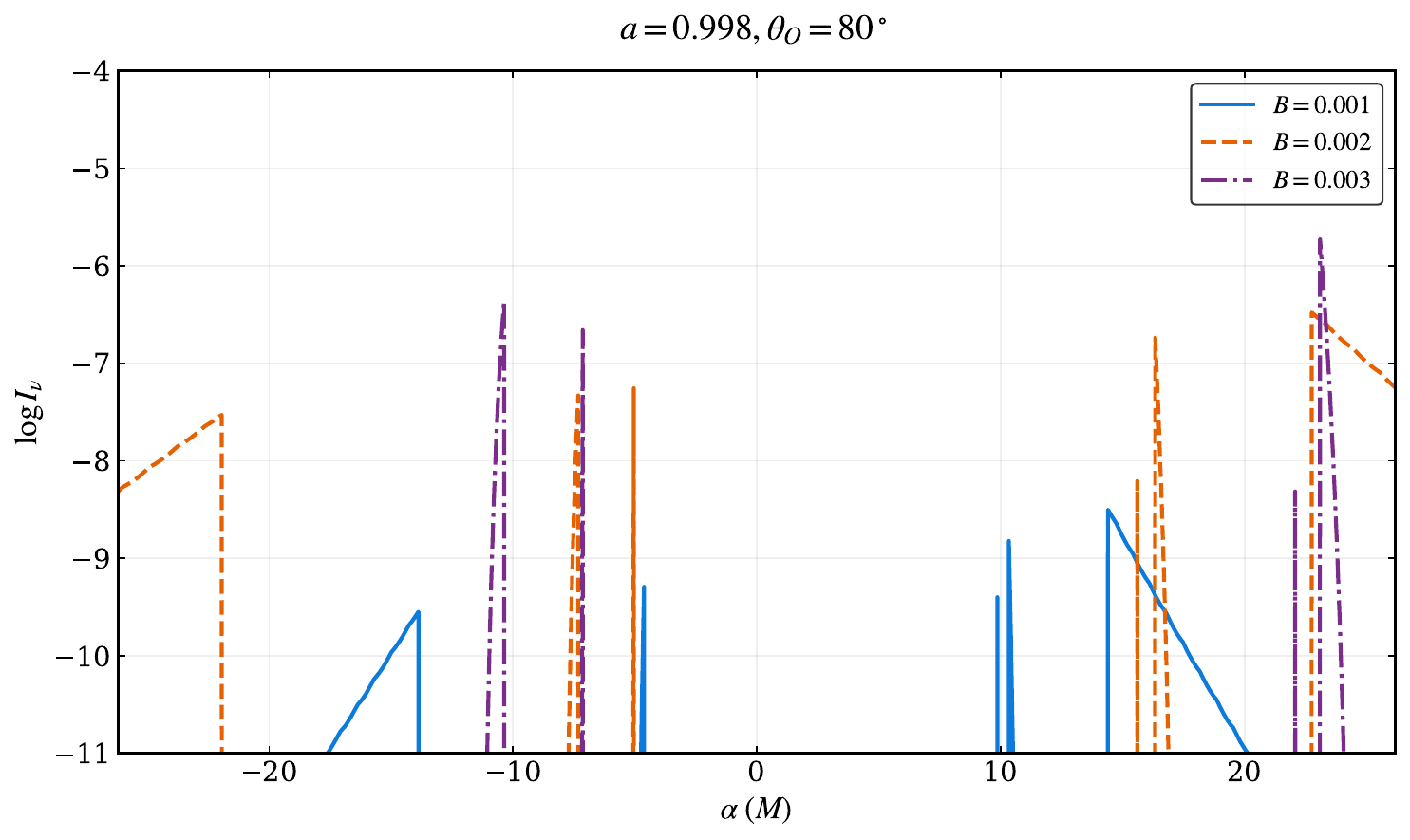} &
        \includegraphics[width=0.48\textwidth]{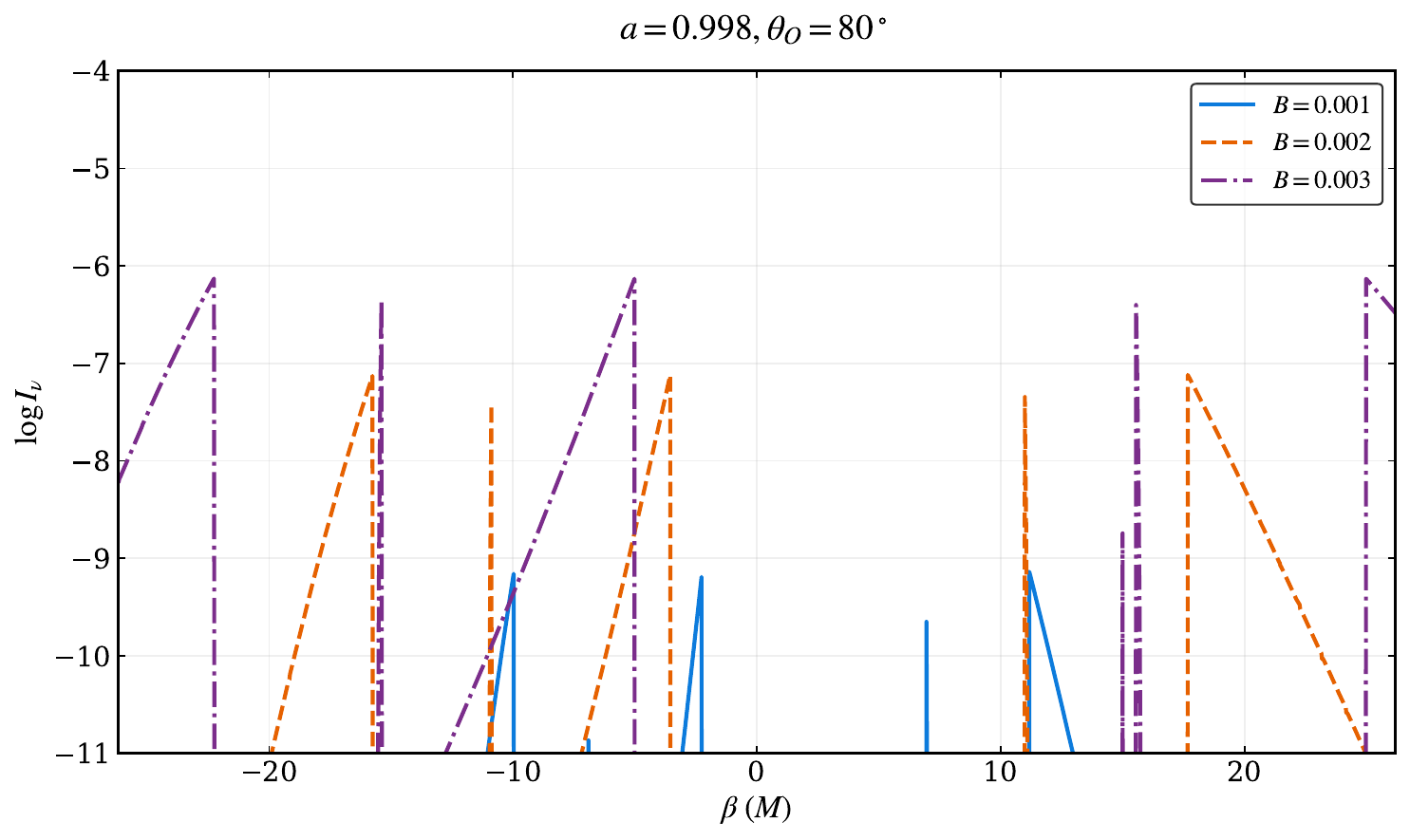} \\
    \end{tabular}
    \caption{One-dimensional specific-intensity profiles for retrograde disks along the horizontal direction $(\beta=0$, left) and vertical direction $(\alpha=0$, right), with $a=0.998$, $\theta_O=80^\circ$, and varying $B$. The logarithmic scale highlights the narrow lensed features. Because the retrograde ISCO is located at larger radii, the primary emission peaks are shifted to larger impact parameters, making the $n=1$ and $n\geq 2$ subimages more clearly separated from the direct emission.}
    \label{fig:retro_1D_profiles}
\end{figure*}

To quantify this behavior, Fig.~\ref{fig:retro_1D_profiles} shows the horizontal and vertical intensity profiles for the same retrograde configurations. In contrast to the prograde case, the primary emission peaks occur at larger impact parameters, roughly $|\alpha|,|\beta|\gtrsim 10M$ for the cases displayed. The central region between these primary peaks is therefore much wider. Within this region, the $n=1$ lensing-ring contribution and the higher-order $n\geq 2$ photon-ring subimages appear as narrow peaks that are less contaminated by the direct emission.

This comparison shows that the macroscopic image morphology is controlled not only by the critical curve of the background spacetime, but also by the location of the disk inner edge. The critical photon region determines where highly deflected trajectories accumulate, whereas the ISCO and the model-dependent emissivity cutoff determine which parts of these trajectories are actually illuminated. In this sense, prograde and retrograde disks around the same Kerr-BR black hole can produce markedly different brightness distributions even when the underlying spacetime parameters are fixed.

\section{Conclusion}
\label{sec:conclusion}

In this paper, we have investigated the optical appearance and one-dimensional (1D) intensity profiles of a Kerr-Bertotti-Robinson (Kerr-BR) black hole surrounded by a geometrically and optically thin accretion disk. Instead of adopting a purely phenomenological power-law emissivity, we employed a magnetically driven synchrotron emissivity proxy in which the local emission depends on the electromagnetic environment of the Kerr-BR spacetime. Using a backward ray-tracing framework, we examined how the black hole spin $a$, the magnetic parameter $B$, and the observer inclination $\theta_O$ affect the ray-classification maps, redshift distributions, and specific-intensity images. Our main results can be summarized as follows.

First, the inner boundary of the emitting region is controlled by both the orbital structure and the validity range of the adopted emissivity prescription. In the standard thin-disk picture, the ISCO provides the baseline inner edge of the disk. In the Kerr-BR spacetime, the ISCO position depends on both the spin parameter $a$ and the magnetic parameter $B$. For rapidly rotating prograde configurations, the formal ISCO can move into a region where the magnetically dominated approximation used to construct the synchrotron emissivity proxy is no longer applicable. In our model, this introduces an additional inner cutoff, so that the effective emitting edge is determined by $r_{\rm in}=\max(r_{\rm ISCO},r_{\rm md})$. This cutoff should be interpreted as the validity boundary of the adopted magnetic-field-dependent emissivity model, rather than as a proof that all physical radiation must vanish inside this radius. It can nevertheless have a visible impact on the central brightness depression and on the relative contribution of higher-order lensed emission.

Second, the two-dimensional images show how the ray topology, redshift distribution, and emissivity profile combine to determine the observed morphology. The ray-classification maps separate the direct image ($n=0$), the $n=1$ lensing-ring contribution, and the higher-order $n\geq 2$ photon-ring subimages. The redshift maps show the expected Doppler-induced left-right asymmetry at high inclination, while the final intensity maps demonstrate how this asymmetry is further weighted by the magnetic-field-dependent emissivity. Varying the magnetic parameter $B$ changes both the spacetime lensing properties and the emissivity profile, leading to visible changes in the size of the emission-depleted central region and in the width and brightness of the lensed components.

Third, the 1D intensity profiles provide a useful quantitative complement to the 2D maps. The horizontal profiles are especially sensitive to the Doppler asymmetry between the approaching and receding sides of the disk, whereas the vertical profiles more clearly display the effects of projection and gravitational lensing. On a logarithmic intensity scale, the $n=1$ lensing-ring contribution and the higher-order $n\geq 2$ photon-ring subimages appear as narrow peaks superimposed on the broader direct-emission envelope. The positions and relative amplitudes of these peaks vary with $a$, $B$, and $\theta_O$, suggesting that such profiles can serve as useful diagnostic templates for comparing different disk configurations and spacetime parameters.

Finally, the comparison between prograde and retrograde disks highlights the importance of the disk inner edge. For a rapidly rotating black hole, the retrograde ISCO is shifted to much larger radii than the prograde ISCO. As a result, the direct emission is displaced to larger impact parameters, producing a wider emission-depleted central region. In this case, the $n=1$ lensing-ring contribution and the higher-order photon-ring subimages can be more clearly separated from the direct emission. This demonstrates that the macroscopic brightness distribution is not determined by the critical curve of the spacetime alone, but also by which parts of the strongly lensed photon trajectories are illuminated by the accretion flow.

In summary, the Kerr-BR spacetime provides a useful setting for studying how a magnetic background can affect both photon propagation and disk emission. The magnetically driven synchrotron emissivity proxy adopted here allows us to connect the image morphology with the local electromagnetic structure in a simple analytic framework. The results show that the ISCO modulation, the model-dependent magnetic-dominance cutoff, and the higher-order subimage structure can all leave identifiable imprints on the optical appearance. Future work should extend the present treatment to more realistic plasma models, including non-equatorial and geometrically thick emission regions, absorption and self-absorption effects, non-thermal electron distributions, and full GRMHD-based radiative transfer. Such extensions will be necessary for making more direct comparisons with high-resolution observations by the Event Horizon Telescope (EHT) and the next-generation EHT (ngEHT).

\begin{acknowledgments}
This work was supported by the National Natural Science Foundation of China under Grant No. 12305070, and the Basic Research Program of Shanxi Province under Grant Nos. 202303021222018 and 202303021221033.
\end{acknowledgments}

\bibliography{references}

\end{document}